\documentclass[12pt,a4paper]{article}
\usepackage[utf8]{inputenc}

\usepackage{xr}
\usepackage{amsfonts,amsmath, amssymb}
\usepackage{amsthm}
\usepackage{graphicx}
\usepackage{longtable, rotating,makecell}
\usepackage{float}
\usepackage{array}
\usepackage{enumerate}
\usepackage{threeparttable}
\usepackage{mathrsfs}
\usepackage[text={15cm,24cm},top=2cm,margin=2cm]{geometry} 
\usepackage{listings}
\usepackage{url}
\usepackage{subfigure}
\usepackage{multirow}
\usepackage{textcomp,booktabs}
\usepackage[usenames,dvipsnames]{color}
\usepackage{colortbl}
\usepackage{lscape}
\usepackage{bm}
\usepackage[numbers,square]{natbib}
\usepackage[colorlinks=true,urlcolor=black,citecolor=blue,linkcolor=black,bookmarks=true]{hyperref} 
\usepackage{verbatim}
\usepackage{ulem}
\usepackage{tikz}
\usetikzlibrary{positioning,arrows}
\tikzset{
	block/.style={
		draw, 
		rectangle, 
		minimum height=1cm, 
		minimum width=2cm, align=center
	}, 
	line/.style={->,>=latex'}
}
\usepackage{pdfpages}

\definecolor{mygray}{gray}{.9}
\usepackage[labelfont={bf,it},textfont={bf,it}]{caption}
\DeclareGraphicsExtensions{.jpg,.pdf,.gif,.png}

\newcommand{\PP}{\mathbb{P}}
\newcommand{\FS}[2]{\displaystyle\frac{#1}{#2}}

\newcommand{\R}{\mathbb{R}}

\DeclareMathOperator\tr{tr}
\DeclareMathOperator\skw{skw}
\DeclareMathOperator\sym{sym}

\DeclareMathOperator\corr{corr}

\numberwithin{equation}{section}
\numberwithin{definition}{section}
\numberwithin{remark}{section}
\numberwithin{theorem}{section}
\numberwithin{proposition}{section}
\numberwithin{lemma}{section}
\numberwithin{remark}{section}
\numberwithin{example}{section}
\numberwithin{figure}{section}
\numberwithin{conjecture}{section}
\numberwithin{table}{section}

\begin{document}
\pagestyle{plain}
\title{\bf Physical Confinement and Cell Proximity Increase Cell Migration Rates and Invasiveness: A Mathematical Model of Cancer Cell Invasion through Flexible Channels}
\author{Q. Peng$^{1^*,2}$, F.J. Vermolen$^{2,3}$, D. Weihs$^4$}
\date{\footnotesize 1$^*$. Mathematical Institute, Faculty of Science, Leiden University. Neils Bohrweg 1, 2333 CA, Leiden, The Netherlands. \texttt{q.peng@math.leidenuniv.nl}.\\
2. Computational Mathematics Group, Department of Mathematics and Statistics, Faculty of Science, Hasselt University. Campus Diepenbeek, Agoralaan Gebouw D, 3950 BE, Diepenbeek, Belgium.\\
3. Delft Institute of Applied Mathematics, Delft University of Technology. Mekelweg 4, 2628 CD, Delft, The Netherlands.\\
4. Faculty of Biomedical Engineering, Technion-Israel Institute of Technology, 3200003 Haifa, Israel.
}
\maketitle

\begin{abstract}
	Cancer cell migration between different body parts is the driving force behind cancer metastasis, which is the main cause of mortality of patients. Migration of cancer cells often proceeds by penetration through narrow cavities in locally stiff, yet flexible tissues. In our previous work, we developed a model for cell geometry evolution during invasion, which we extend here to investigate whether leader and follower (cancer) cells that only interact mechanically can benefit from sequential transmigration through narrow micro-channels and cavities.
	 We consider two cases of cells sequentially migrating through a flexible channel: leader and follower cells being closely adjacent or distant. Using Wilcoxon's signed-rank test on the data collected from Monte Carlo simulations, we conclude that the modelled transmigration speed for the follower cell is significantly larger than for the leader cell when cells are distant, i.e. follower cells transmigrate after the leader has completed the crossing. Furthermore, it appears that there exists an optimum with respect to the width of the channel such that cell moves fastest. On the other hand, in the case of closely adjacent cells, effectively performing collective migration, the leader cell moves $12\%$ faster since the follower cell pushes it. This work shows that mechanical interactions between cells can increase the net transmigration speed of cancer cells, resulting in increased invasiveness. In other words, interaction between cancer cells can accelerate metastatic invasion. 
\end{abstract}

\section{Introduction}
\noindent
One of the main reasons of death in cancer patients is metastasis \citep{Massalha2016}. When the cell invades through dense tissues, it often applies forces on its immediate surroundings to migrate through narrow channels and cavities \citep{Rappel2017}. Cancer cells that are more invasive, tend to be very pliable and dynamic, both internally \citep{Gal2012} and externally \citep{Cross2007,Guck2005,Paluch2009, Swaminathan2011}, and hence they are able to adjust their cytoskeleton and geometry. In addition to that, cancer cells have been shown  to apply a significantly larger traction force on substrates, compared to benign cells \citep{Massalha2016};  Importantly, the forceful, invasive interactions of tumor cells with their environments have been used to provide rapid, clinically relevant cancer diagnosis and prognosis \citep{Kortam_2021,Merkher_2020}.


Cells are complex viscoelastic objects, and their shapes are determined by a local balance between retraction and protrusion of its boundaries  \citep{Ebata2018}. The driving force for cell deformation is typically triggered by external stimuli, for example, cell-substrate adhesion, biochemical signaling and forces exerted on the cell \citep{Mogilner2009, Paluch2009}.  Furthermore, cell adhesion and motility are affected by the mechanics of their environment \citep{Yeung_2004} and, for example, cells appear to be more rounded on a softer substrate and more elongated on a stiffer substrate \citep{Ladoux2016,Massalha2016,Zemel2010}.  In other words, cell shape is also an outcome of mechanical equilibrium as a result of external forces \citep{Paluch2009}. However, the mechanisms that determine global cell morphology in relation to its function remain poorly understood \citep{Haupt2018,Keren2008,Mogilner2009}.


Directed cell migration is an active process involving cell polarization that is often driven by chemotaxis or mechanotaxis \citep{Haupt2018}.  During single cell migration, the cell must first polarize for all modes of migration \citep{Ladoux2016,Llense2015,Rappel2017}. Cell polarization into well-defined front (leading edge) and rear parts \cite{Cusseddu2019,Haupt2018}, is one of the most significant responses of animal cells to their environment \citep{Ladoux2016}. Indeed, there are many possible causes of cell polarization: external forces \citep{Verkhovsky1999}, Rho family small guanosine triphosphate (GTP)--binding proteins (Rho GTPases) \citep{Cusseddu2019,Ladoux2016}, chemokines and cytokines \citep{Ji2008, Llense2015} etc. Subsequently, the cell migrates in the direction of the polarity axis, that is, the longitudinal axis of the cell.

Mathematical modelling has been acknowledged as an important tool to help turn blurred concepts and ideas into testable, quantifiable and rigorous hypotheses and to reveal the correlations between various factors, which are otherwise difficult to determine in complex biological phenomena and microscopic experiments. In general, mathematical models in biology can be categorized as agent-based models (for small-scale materials) and continuum models (for large scale materials). Agent-based modelling is extremely beneficial to model cellular activities like cell division, differentiation and migration etc., since the model treats every cell as an individual and hence, it is capable of tracking cell positions and cell-substrate interactions. A further advantage of agent-based models is that most input parameters are expressed in terms of directly measurable quantities. Since in this manuscript, we will focus on cell geometry and single cell metastasis, we utilize agent-based modelling. Agent-based models have been widely used to investigate the evolution of cell geometry. \citet{Rens2019} developed an algorithm approach to model the impact of cellular forces with the cellular Potts model, which is based on the assumption of minimizing the energy configuration; the work of \citet{Zhao2017} mainly focuses on the intracellular environment where the finite-element method is applied to describe the total energy of the cell; the model in \citet{Cusseddu2019} emphasizes the impact of Rho GTPases. 

The current work is an extension of our previous work \citep{Peng2021}, where the Poisson effect, cell polarization and focal adhesion were not considered, yet still provides a basic model to depict the cell shape evolution under external stimuli and dynamic processes that occur during cell invasion.  We use the finite-element method to approximate the solutions to all partial differential equations (PDEs) in the bulk domain. As energy is consumed during cell migration and deformation, energetic effects may also have an impact on the cell migration and invasion; those aspects will be studied in future studies. Furthermore, to the best of our knowledge there are currently no in-vitro results for cells that are migrating through deformable channels, which makes impedes experimental validation of the model. Our simulation study may therefore  serve as a precursor for the development of in-vitro experiments of cells transmigrating through deformable channels.

The manuscript is structured as follows: The biological assumptions and the resulting mathematical model are presented in Section \ref{Sec_maths_modelling}. In Section \ref{Sec_pre_results}, numerical results from the simulations with one set of input parameter values are displayed. Since the model contains randomness and we aim at generalizing the conclusions in Section \ref{Sec_pre_results}, Monte Carlo simulations are presented in Section \ref{Sec_MC_simu}. Finally, in Section \ref{Sec_conclusions}, we summarize our findings and indicate the possible directions to further this research.

\section{Mathematical Modelling}\label{Sec_maths_modelling}
\noindent 
In this manuscript, we extend the phenomenological model in \citet{Peng2021}, where the Poisson effect was not taken into account in the mechanics of the cell. In that study, we were only modeling the interplay between the cell and its immediate extracellular environment. We are aware that to simulate the Poisson effect, the intracellular environment should be modelled as in \citet{Zhao2020,Zhao2017}. However, in the work of \citet{Zhao2017,Zhao2020}, the interaction between the cell and the substrate, in particular, the traction forces exerted by the cell to the substrate were not incorporated. On the other hand, from a computational point of view, the model will be relatively complicated if we model the intracellular and extracellular environment separately, as well as the impact of the traction forces on the substrate. Hence, we are interested in continuing the phenomenological model \citep{Peng2021} to depict the Poisson's effect, in particular, when the cell needs to transmigrate through a small pore or channel. In this problem, we define the intracellular mechanics by means of a spring model with nodal masses on the cell boundary that are connected to their immediate neighbours and to the midpoint of the cell.

\subsection{Biological Assumptions}
\noindent
We are aware that it is impossible and infeasible to include all the cellular activities of viable cells in the model. Hence we simplify reality regarding cell polarization and migration. Cells need to be polarized before migration, which results into leading edges and  non-leading edges of cells. According to \citet{Zhao2017}, leading edges are characterized by having a positive inner product of the outward unit vector, $\boldsymbol{n}_c$, and velocity vector, $\boldsymbol{v}_c$,  i.e. $(\boldsymbol{n}_c, \boldsymbol{v}_c)>0$. Note that $(\cdot, \cdot)$ represents the inner product of two vectors. We consider the combination of cell migration and deformation by dividing the cell boundary into a set of nodal points, of which the coordinates are represented by $\boldsymbol{x}_j$. We consider a two-dimensional model, hence we consider the projection of cells onto a substrate layer. The boundary segments are line segments with two neighboring nodal points as vertices. Since the nodal points define the cell membrane, we use a weighted average to approximate the outward normal unit vector at a nodal point by the two line segments that contain the point as a vertex, hence,
\begin{equation}
\label{Eq_leading_edge}
\boldsymbol{n}_c(\boldsymbol{x}_j) = \frac{\|\boldsymbol{e}^{j-1, j}\|\boldsymbol{n}_c^{j-1,j}+\|\boldsymbol{e}^{j, j+1}\|\boldsymbol{n}_c^{j,j+1}}{\|\|\boldsymbol{e}^{j-1, j}\|\boldsymbol{n}_c^{j-1,j}+\|\boldsymbol{e}^{j, j+1}\|\boldsymbol{n}_c^{j,j+1}\|},
\end{equation}   
where $\boldsymbol{e}^{i,j}$ stands for the line vector, that is, $\boldsymbol{e}^{i,j} = \boldsymbol{x}_j - \boldsymbol{x}_i$, which connects $\boldsymbol{x}_i$ and $\boldsymbol{x}_j$, and $\boldsymbol{n}_c^{i,j}$ is the outward normal unit vector of the line segment $\boldsymbol{e}^{i,j}$, hence $(\boldsymbol{n}_c^{i,j}, \boldsymbol{e}^{i,j}) = 0$. We assume that the leading edges are more sensitive to the signalling molecules.

Since the cytoplasm of cells contains water and polymers, one often models cells as visco-elastic objects. Therewith they are deformable and compressible only up to a certain extent. Therefore, we assume that the difference between the area of the equilibrium status and the area at time $t$ cannot exceed $10\%$, that is,
$$\frac{\left|A(\Omega_C(t)) - A(\Omega_C^{eq})\right|}{A(\Omega_C^{eq})}\leqslant10\%, \ t\geqslant0,$$
where $A(\Omega_C(t))$ is the area of the cell at time $t$ and $A(\Omega_C^{eq})$ represents the area of the cell when it is in its equilibrium shape.
 
We introduce two indicators, namely circularity and aspect ratio to evaluate the evolution of the cell shape quantitatively \citep{Haupt2018,Massalha2016}. For the two-dimensional case, the circularity is a measure of how circular a two dimensional object is. The circularity is defined by 
\begin{equation}
\label{Eq_circularity}
C(\Omega_C(t)) = \frac{4\pi A(\Omega_C(t))}{l^2(\partial\Omega_C(t))}, \mbox{$\Omega_C(t)\subset\R^2$, $t\geqslant0$,}
\end{equation}
where $A(\Omega_C(t))$ represents the area of cell $\Omega_C$  at time $t$ and $l(\partial\Omega_C(t))$ is the circumference of the cell boundary. The circularity value ranges $0$ and $1$, and the circularity value of a circle is $1$, objects degenerating to lines and curves provide circularities tending to zero. Objects with circularity value between $0$ and $1$ may have an elliptic shape, or have an irregular shape that can contain many protrusions and lamellipodia. The aspect ratio is defined by: 
\begin{equation}
\label{Eq_aspect_ratio}
S(\Omega_C(t)) = d_1(\Omega_C(t))/d_2(\Omega_C(t)),  \mbox{$\Omega_C(t)\subset\R^2$, $\ t\geqslant0$,}
\end{equation}
where $d_1(\Omega_C)$ and $d_2(\Omega_C(t))$ represent the length and width of $\Omega_C$ at time $t$, respectively. The aspect ratio indicates the internal polarity axes of the cell, and illustrates how elongated a cell is; the aspect ratio of a circle is $1$.  
 
Due to inhomogeneities in the extracellular domain, the migration of cells is subject to randomness. To model the random (uncertain) part of cellular displacement, a memorized random walk is embedded in the model \citep{Takagi2008}, which is developed from the generalized Langevin model.  Let $d\boldsymbol{W}(t)$ represent a vector Wiener process in which the contributions to the different coordinate directions are treated as independent events, and a normal distribution with zero mean and variance $\delta t$ for all components, with $\delta t$ denoting a time interval.  From statistical principles, we have the following link between random walk described by the Wiener process for the particle position and the evolution of the probability density function of the position of the particle: $\sigma_{rw} = \sqrt{2D_c}$ , where $D_c$ represents the diffusivity of the cell phenotype in this extracellular matrix. This term models the random displacements caused by unpredictable localized microstructural inhomogeneities.  Then, the spontaneous velocity of the cell $\boldsymbol{v}_s(t)$ and the memory of the velocity $\boldsymbol{V}_s(t)$ are described by \citep{Selmeczi2005}
\begin{equation}
	\label{Eq_memorised_rw}
	\left\{
	\begin{aligned}
	d \boldsymbol{v}_s(t) &= -\beta_s(\boldsymbol{v}_s(t)) \boldsymbol{v}_s(t) dt + \alpha_s \boldsymbol{V}_s(t) dt + \sigma_{rw} d \boldsymbol{W}(t), \\ \\
	d \boldsymbol{V}_s(t) &= (\alpha_s \boldsymbol{v}_s(t) - \gamma_s \boldsymbol{V}_s(t)) dt,
	\end{aligned}
	\right.
\end{equation}
where $\beta_s(\boldsymbol{v}_s(t))$ is the velocity decay rate with the velocity dependency, $\alpha_s$ is the memory rate and $\gamma_s$ is the memory decay rate. Since \citet{Takagi2008} suggested that $\beta_s(\boldsymbol{v}_s(t))$ is proportional to $\|\boldsymbol{v}_s(t)\|^2$, we assume that $\displaystyle \beta_s(\boldsymbol{v}_s(t)) = \frac{1}{S(\Omega_C(t))} + 1.7\|\boldsymbol{v}_s(t)\|^2$, where $S(\Omega_C(t))$ is the aspect ratio of the cell defined in Equation (\ref{Eq_aspect_ratio}). The assumption is based on an experimental observation that elongated cells migrate faster, and therefore, we use inverse proportionality of the velocity decay rate on the aspect ratio of the cell \citep{Keren2008}.

\subsection{Cellular Traction Forces}\label{Sec_passive}
\noindent
It has been documented that cancer cells exert forces on their direct environment \citep{Massalha2016}, which results in the deformation of the extracellular environment, which , in turn, has an impact on the cell as well. Furthermore, cells, in particular cancer cells, are known to be able to permanently change their immediate environment. These permanent changes could give rise to shrinkage or growth processes, and to mechanical displacements as a result of forces that are exerted by the cells themselves. Similar to \citet{Peng2021}, we therefore use a morphoelastic approach to model passive convection of the cell, and (permanent) displacements of the extracellular matrix. Morphoelasticity is commonly used to combine the modeling of mechanical stresses and processes like growth and shrinkage  \citep{ ben2015morpho,goriely2011morphoelasticity,koppenol2017biomedical,rudraraju2019computational}. The morphoelastic model reads as \citep{koppenol2017biomedical}:
\begin{equation}
\label{Eq_morpho}
\left\{
\begin{aligned}
&\rho[\FS{D\boldsymbol{v}}{Dt}+\boldsymbol{v}(\nabla\cdot\boldsymbol{v})]-\nabla\cdot\boldsymbol{\sigma}=\boldsymbol{f}(\boldsymbol{x};t) = \sum_{i=1}^{N_c}\boldsymbol{f}^i(\boldsymbol{x};t),\mbox{in $\Omega, t>0$,}\\
&\FS{D\boldsymbol{\epsilon}}{Dt}+\boldsymbol{\epsilon}\skw(\boldsymbol{L})-\skw(\boldsymbol{L})\boldsymbol{\epsilon}+[\tr(\boldsymbol{\epsilon})-1]\sym(\boldsymbol{L})=-\alpha\boldsymbol{\epsilon},\mbox{in $\Omega, t>0$,}\\
&\boldsymbol{v}(\boldsymbol{x},t)=\boldsymbol{0}, \mbox{on $\partial\Omega, t>0$,}	\\
&\boldsymbol{v}(\boldsymbol{x},0)=\boldsymbol{0}, \mbox{in $\Omega, t=0$,}
\end{aligned}
\right.
\end{equation}     
where $\rho$ is the density of the extracellular matrix, $\boldsymbol{\epsilon}$ is the local (Eulerian) effective strain tensor to be solved, $\boldsymbol{L}=\nabla\boldsymbol{v}$ is the deformation rate gradient, and $\alpha$ is a non-negative constant that accounts for the amount of plastic deformation proportional to local (Eulerian) effective strain. This parameter $\alpha$ could have an electro-chemical origin, however, in the current study, we treat it as a constant for the sake of simplicity. Note that if $\alpha=0$, then as soon as the force $\boldsymbol{f}=\boldsymbol{0}$, the tissue will gradually recover to its original shape and volume. Furthermore, let $y$ be a scalar function, then the material derivative is given by $\displaystyle \frac{D{y}}{Dt} = \frac{\partial{y}}{\partial t}+\boldsymbol{v}\nabla\cdot{y}$ where $\boldsymbol{v}$ is the migration velocity of any point within the domain of computation. In order to have a fixed boundary of computational domain $\Omega$, we use a homogeneous Dirichlet boundary condition for the velocity. From a mechanical point of view, we treat the computational domain $\Omega$ as a linear and isotropic medium. Further, as a result of the presence of liquid phases in the tissue, the mechanical balance is also subject to viscous, that is friction, effects. Therefore, we use Kelvin–Voigt’s viscoelastic dashpot model, of which the stress tensor reads as
\begin{equation}
\label{Eq_morpho_visco_sigma}
\begin{aligned}
\boldsymbol{\sigma} &= \boldsymbol{\sigma}_{elas} + \mu_{visco} \boldsymbol{\sigma}_{visco}\\
&= \frac{E_s}{1+\nu_s}\{\boldsymbol{\epsilon}+\tr(\boldsymbol{\epsilon})[\frac{\nu_s}{1-2\nu_s}]\boldsymbol{I}\}+\mu_{visco}\left(\mu_1\sym(\boldsymbol{L})+\mu_2\tr(\sym(\boldsymbol{L}))\boldsymbol{I}\right),
\end{aligned}
\end{equation}
where $E_s$ is the stiffness of the substrate, $\nu_s$ is the Possion ratio of the substrate, $\boldsymbol{\epsilon}$ is the strain tensor, $\mu_{visco}$ is the weight of the viscosity in viscoelasticity, $\mu_1$ and $\mu_2$ are the dimensionless shear and bulk viscosity, respectively. The morphoelasticity model provides a set of nonlinear partial differential equations for the displacement velocity and the strain tensor. Since the initial displacement is zero everywhere, the displacement of the domain can be approximated by integrating the velocity over time: $\displaystyle \boldsymbol{u}(\boldsymbol{x}(t), t)=\int_{0}^{t}\boldsymbol{v}(\boldsymbol{x}(\tau), \tau)d\tau$. 

We assume that the traction forces applied by the cell are modelled by Dirac Delta distributions. Let $\Omega\subset\R^d$ be an open region, then the Dirac Delta distribution is given by
\begin{enumerate}
	\item $\delta(\boldsymbol{x}) = 0$, for all $\boldsymbol{x}\in\R^d\backslash\{\boldsymbol{0}\}$;
	\item $\int_{\Omega}\delta(\boldsymbol{x}) d\Omega = 1$, if $\boldsymbol{0}\in\Omega$.
\end{enumerate}
We currently model (cancer) cells that are migrating through narrow channels and we incorporate the pushing forces that are exerted by the cancer cells on their immediate environment. These forces are directed in the outward normal direction of the cell boundary and they are only applied on the cell boundary segments that are in mechanical contact with the extracellular obstacles. These obstacles can be other cells or the wall of the narrow channel. Following the same form of traction forces as in our previous work \citep{Peng2021}, we denote $N^i_m = \{j_1,\dots,j_m\}$ as the set of line segments of the cell membrane that are in mechanical  contact with the obstacles, then the traction forces for cell $i$ are expressed as 
\begin{equation}
	\label{Eq_force}
	\boldsymbol{f}^i(\boldsymbol{x}(t), t) = \sum_{j\in N^i_m}Q(d(\boldsymbol{s}^i_j),t)\boldsymbol{n}(\boldsymbol{s}^i_j(t))\delta(\boldsymbol{x}(t)-\boldsymbol{s}^i_j(t))\Delta\Gamma^{i,j},
\end{equation}
where $\boldsymbol{n}(\boldsymbol{x}(t))$ is the unit outward pointing normal vector at  $\boldsymbol{x}(t)$, $\boldsymbol{s}^i_j(t)$ is the midpoint of line segment $j$ of cell $i$, and $\Delta\Gamma^{i,j}$ is the length of line segment $j$. Note that the cell index is marked as a superscript in the notations. Most of the time, for the simplification, we neglect it unless we explicitly describe the case for multiple cells. Here, $Q(d(\boldsymbol{x}), t)$ is defined analogously as in \citep{Peng2021}:
$$Q(d(\boldsymbol{x}), t) = \frac{\pi}{4}d(\boldsymbol{x})E^*/\|l_m\|.$$
Here, $E^*$ is the total equivalent Young's modulus defined by \citep{popov2019handbook} $$\displaystyle \frac{1}{E^*} = \frac{1-\nu_c^2}{E_c} + \frac{1-\nu_s^2}{E_s},$$
where $E_c$ and $E_s$ are the Young's modulus of the cell and the ECM, respectively, and $\nu_c$ and $\nu_s$ are the Poisson ratio of the cell and the ECM, respectively. Furthermore, $d(\boldsymbol{x})$ is the invagination depth of the cell due to the compression by the obstacle and $\|l_m\|$ is the total curve length of the portion of the membrane of the cell that is in mechanical contact with the obstacles. Note that the division by $\|l_m\|$ is necessary for normalisation since $Q$ represents the force per unit of peripheral length of the cell. The total force follows from integration of Q over the boundary portion that is in physical contact with the obstacle.

\subsection{Cell Cytoskeleton}
\begin{figure}
	\centering
	\subfigure[Cell cytoskeleton]{
	\includegraphics[width=0.51\textwidth]{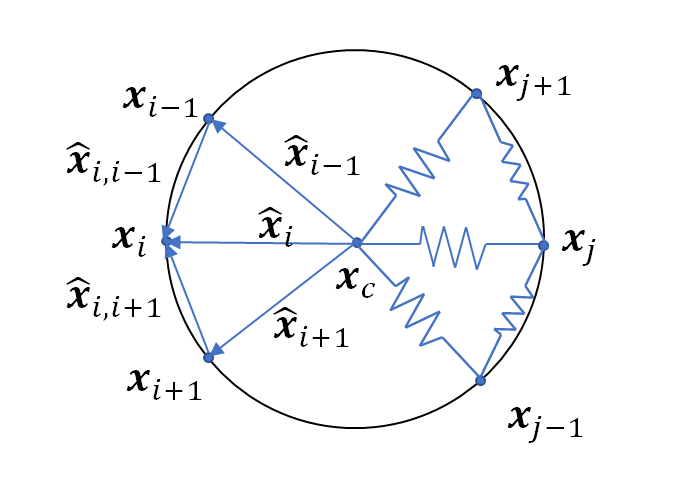}
    \label{fig_cytoskeleton}}
	\subfigure[Cell collides with the other cell]{
	\includegraphics[width=0.42\textwidth]{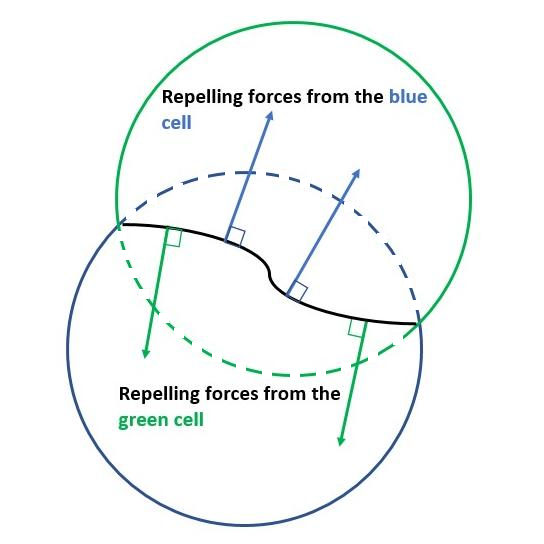}
	\label{fig_collide}}
	\caption{A schematic of the cell cytoskeleton and the repelling (pushing) forces when the cell has mechanical contact with an obstacle. \subref{fig_cytoskeleton} Cell cytoskeleton is built up by a series of elastic springs. For every nodal point on the cell boundary, it is connected to the cell centre and two neighbouring points by a elastic spring separately. Here, the arrows indicate the vectors. \subref{fig_collide} Repelling forces are exerted by the two cells when they collide with each other and they are compressed. Dashed curves represent the equilibrium shape of the cells, and the black curve between two cells is the contacting surface. Note that the forces, indicated by the arrows, are perpendicular to the black curve.}
	\label{Fig_cytoskeleton}
\end{figure}

\noindent
In \citet{Peng2021}, the cell boundary is modeled by a set of points on the cell boundary where all the points are connected to the cell center by springs. In the current study, we extend this principle by connecting adjacent cell boundary points by springs. Using this formalism, it becomes possible to accommodate for surface tension.  Here, we assume that the elasticity of the springs that connect the node to the center and to its neighboring nodes can have different values. The cell cytoskeleton is schematically represented in Figure \ref{Fig_cytoskeleton}.
Similar to  \citet{Peng2021,Chen2018,Vermolen2012cellshape}, without any extracellular stimulus (i.e. here we {\it only} present the elasticity part of the cell deformation; the spontaneous displacement of the cell that is described in Equation (\ref{Eq_memorised_rw}) will be incorporated later into the equation of motion), the position of nodal point $\boldsymbol{x}_j$ over the cell boundary is determined by
\begin{equation}
\label{Eq_cyto}
	d\boldsymbol{x} _j= E_c(\boldsymbol{x}_c(t)+\hat{\boldsymbol{x}}_j-\boldsymbol{x}_j(t))dt + E_m(\boldsymbol{x}_{j-1}(t)+\hat{\boldsymbol{x}}_{j,j-1}-\boldsymbol{x}_j(t))dt + E_m(\boldsymbol{x}_{j+1}(t)+\hat{\boldsymbol{x}}_{j,j+1}-\boldsymbol{x}_j(t))dt,
\end{equation}  
where $E_c$ and $E_m$ represent the stiffness of the spring connecting $\boldsymbol{x}_j$ to the centre point and neighbouring points respectively; $\hat{\boldsymbol{x}}_j = \tilde{\boldsymbol{x}}_j(t)-\boldsymbol{x}_c(t)$, $\hat{\boldsymbol{x}}_{j, j\pm1} = \tilde{\boldsymbol{x}}_{j}(t)-\tilde{\boldsymbol{x}}_{j\pm1}(t)$ are the vectors connecting the nodal point $j$ to the centre point and the neighbouring points when the cell is in equilibrium shape, and  $\tilde{\boldsymbol{x}}_{j}(t)$ represents the equilibrium position of the nodal point $j$ relative to the centre $\boldsymbol{x}_c(t)$.

To maintain the right orientation of the cell, we consider the rotation matrix defined by
\begin{equation}
\label{Eq_Bmatrix}
\boldsymbol{B}(\phi) = \begin{pmatrix}
\cos(\phi) & -\sin(\phi) \\
\sin(\phi) & \cos(\phi)
\end{pmatrix},
\end{equation}
such that the angle of orientation, $\phi$, can be computed from
\begin{equation}
\label{Eq_phi}
\tilde{\phi} = \arg\min_{\phi\in[0,2\pi)}\left(\sum_{j=1}^{N}\|\boldsymbol{B}(\phi)\tilde{\boldsymbol{x}}_j(t) - \boldsymbol{x}_j(t)\|^2\right).
\end{equation}
The orientation of the cell is represented by the angle of the vector connecting the ‘front and tail’ of the cell. The overall displacement of the nodes of the cell boundary are determined by translation and rotation. This matrix $\boldsymbol{B}(\phi)$ monitors the angle of rotation of the cell with respect to the cell position (and hence boundary nodes) at the previous timestep. This angle of rotation is important for the determination of the equilibrium points of the cell boundary nodes. The equilibrium points reflect the position and shape to which the cell boundary nodes will converge to if the cell is not subject to any external cue for migration, that is, the cell no longer migrates. Without any external cues, the cell will always return to its initial orientation. Subsequently, Equation (\ref{Eq_cyto}) is rewritten as 
\begin{equation}
	\label{Eq_cytoskeleton}
	\begin{aligned}
	d\boldsymbol{x}_j &= E_c(\boldsymbol{x}_c(t)+\boldsymbol{B}(\tilde{\phi})\hat{\boldsymbol{x}}_j-\boldsymbol{x}_j(t))dt + E_m(\boldsymbol{x}_{j-1}(t)+\boldsymbol{B}(\tilde{\phi})\hat{\boldsymbol{x}}_{j,j-1}-\boldsymbol{x}_j(t))dt \\
	&+ E_m(\boldsymbol{x}_{j+1}(t)+\boldsymbol{B}(\tilde{\phi})\hat{\boldsymbol{x}}_{j,j+1}-\boldsymbol{x}_j(t))dt.
	\end{aligned}
\end{equation}
Note that here we have not yet incorporated the memorized random walk of the cell that is described in Equation (\ref{Eq_memorised_rw}). The memorized random walk will be added to the equation of motion in the next section.

\subsection{Concentration of Generic Signal}
\noindent
For cancer cells, examples of such cues for migration are oxygen and nutrients (such as sugar-like chemicals)  \citep{Koumoutsakos2013, Roussos2011}. We consider a generic chemical, which originates from an artificial point source. Since we only want to scrutinize the model's ability to predict collective cell migration that is impacted by external mechanical factors, we keep this cue the same for all cells (both leader cells and follower cells). Therefore, contrary to our previous work, we use a steady-state diffusion equation (i.e. $\displaystyle\frac{\partial c}{\partial t } = 0$):
\begin{equation}
\label{Eq_adr_concent}
\begin{aligned}
-\nabla\cdot(D\nabla c_s(\boldsymbol{x}))=k\delta(\boldsymbol{x}(t)-\boldsymbol{x}_s), \mbox{$\boldsymbol{x}\in\Omega, t>0$,}
\end{aligned}
\end{equation}
where $c_s(\boldsymbol{x})$ is the concentration of the signalling molecule, $D$ is the diffusion coefficient which has been taken constant in the current study, $k$ is the secretion rate of the signal source, $\boldsymbol{x}_s$ is the position of the source. Furthermore, $\Omega$ is an open bounded simply connected subset in $\mathbb{R}^2$ with boundary $\partial \Omega$. The steady-state problem is closed by the following Robin condition $$\frac{\partial c_s}{\partial\boldsymbol{n}}+\kappa_s c_s = 0, \mbox{on $\partial\Omega,t>0$,}$$ which deals with a balance between the diffusive flux across the boundary and the flux between the boundary and the region far away from the domain of computation. The symbol $\kappa_s$, which is non-negative, represents the mass transfer coefficient. Note that as $\kappa_s\rightarrow0$ then the Robin condition tends to a homogeneous Neumann condition, which represents no flux (hence isolation), which in the current steady-state case, yields an ill-posed problem in terms of existence (and uniqueness). Whereas  $\kappa_s\rightarrow\infty$ represents the case that $c_s\rightarrow0$ on the boundary, which, physically, is reminiscent to having an infinite mass flow rate at the boundary into the surroundings. The Robin condition, also referred to as a mixing boundary condition, is able to deal with both these two limits and all cases between these limits. From a mathematical perspective, the solution $c_s$ to Equation (\ref{Eq_adr_concent}) does not live in the right finite element space ($H^1(\Omega)$) \citep{Peng_2022_JCAM}. Therefore one observes issues regarding accuracy and convergence of the finite element approximation in regions close to the point where the Dirac distribution is acting, that is, ${\bf x}_s$. However, away from $\boldsymbol{x}_s$, the accuracy of the finite element solution is as good as the classical accuracy in the $L^2$--norm of the solution \citep{K_ppl_2014}. We note that in our application, we are only using the solution away from the point of action $\boldsymbol{x}_s$.

\subsection{Focal Adhesions}\label{Subsec_FA}
\noindent 
Focal adhesions play an important role in cell migration. The movement of a cell is driven by continuous remodeling of the cytoskeleton and is mediated by the lamellipodia. In general, a cell moves in three steps: (1) Protrusion: the rear part of the cell is attached on the substrate while the membrane of the front (leading) part is extended; (2) Adhesion: new adhesion is generated at the newly extended membrane of the front part; (3) De-adhesion and retraction: the rear part of the cell is detached from the substrate and due to the cytoskeleton, the rear part retracts to maintain the cell geometry. Once cell contraction has completed, the cycle of movement turns back to Step (1). A more detailed description of cell migration, or crawling, over the substrate can be found in \citet{Ananthakrishnan2007, Bershadsky2011,Mitchison1996}. 

For the implementation, we simplified the aforementioned three steps into two steps, that is, the leading part firstly detaches from the substrate and moves forward while the non-leading part is attached, then the leading part is attached to the substrate and the non-leading part detaches and migrate due to cell contraction. For each nodal point on the cell membrane, we consider a stochastic Markov Chain model, such that the impact of the previous time step (i.e. the previous status) is also accounted for. We assume that the probability that the nodal point is detached from the substrate is related to the local shear force and the local gradient of the concentration of the signalling molecules. In other words, if there is larger local shear force and/or higher gradient of concentration of the signalling molecules, then it is more likely that the nodal point detaches from the substrate. Denoting $1$ for the attached status and $0$ for the detached status of the nodal point, respectively, then the probability of the occurrence of status $s$ when the previous status is $s'$ is given by
\begin{equation}
\label{Eq_FA_prob}
	\PP(X_{n} = s|X_{n - 1} = s') = s'(1-\exp\{-\lambda_1(\boldsymbol{x})dt\}) + (1 - s')(1 - \exp\{-\lambda_2(\boldsymbol{x})dt\}),
\end{equation}  
where $X_n$ is the next status and $X_{n-1}$ is the present status; $\lambda_1 = \lambda_1(\boldsymbol{\sigma}_{12}(\boldsymbol{x},t),c_s(\boldsymbol{x},t))= |\boldsymbol{\sigma}_{12}(\boldsymbol{x})| + 100\|\nabla c_s(\boldsymbol{x})\| + 5$, and $\lambda_2 = \lambda_2(\boldsymbol{\sigma}_{12}(\boldsymbol{x})) =  |\boldsymbol{\sigma}_{12}(\boldsymbol{x})| + 5$ respectively, and $\boldsymbol{\sigma}_{12}$ is given by Equation (\ref{Eq_morpho_visco_sigma}), where the matrix is symmetric and the non-diagonal element is related to the shear stress. Equation (\ref{Eq_FA_prob}) can be rephrased as a Markov Chain matrix:
\begin{equation}
\label{Eq_FA_MC_matrix}
\boldsymbol{P}(\boldsymbol{x}) = \begin{pmatrix}
1 - \exp\{-\lambda_2(\boldsymbol{x})dt\} & \exp\{-\lambda_2(\boldsymbol{x})dt\} \\
1 - \exp\{-\lambda_1(\boldsymbol{x})dt\} & \exp\{-\lambda_1(\boldsymbol{x})dt\}
\end{pmatrix}.
\end{equation}
The initial status is randomly assigned to the nodal points on the cell membrane. To determine the status of such a point, a random value from a uniform distribution between $0$ and $1$ is drawn and tested with the transition probabilities.

\subsection{Cell Geometry Evolution}
\noindent
The cell shape is determined by the positions of the nodal points on the cell boundary. The movement of the nodal points is determined by various processes, which are the retraction towards the equilibrium shape, fluid flow, spontaneous displacement (random walk), chemotaxis and possible external mechanical forces. In Section \ref{Subsec_FA}, the focal adhesion part of cell migration was formulated in terms of a stochastic model where the front and rear parts of the cell detach and attach onto the substrate by chance. Nodal points on the cell boundary that adhere onto the solid substrate are only subject to movement that is determined by the local displacement of the substrate. On the contrary, if a nodal point on the cell boundary is detached, then its movement is determined by all the other aforementioned processes, and hence
\begin{equation}
\label{Eq_disp_chemo}
\begin{aligned}
&d\boldsymbol{x}_j = \mu_s(S(\Omega_C), \boldsymbol{x}_j)\frac{\nabla c_s(\boldsymbol{x},t)}{\|\nabla c_s(\boldsymbol{x},t)\|+\gamma}dt + E_c(\boldsymbol{x}_c(t)+\boldsymbol{B}(\tilde{\phi})\hat{\boldsymbol{x}}_j-\boldsymbol{x}_j(t))dt \\ 
&+E_m(\boldsymbol{x}_{j-1}(t)+\boldsymbol{B}(\tilde{\phi})\hat{\boldsymbol{x}}_{j,j-1}-\boldsymbol{x}_j(t))dt + E_m(\boldsymbol{x}_{j+1}(t)+\boldsymbol{B}(\tilde{\phi})\hat{\boldsymbol{x}}_{j,j+1}-\boldsymbol{x}_j(t))dt\\
&+\boldsymbol{v}dt+\boldsymbol{v}_sdt,
\end{aligned}
\end{equation}
where $\gamma$ is a small positive constant to prevent the denominator being zero and $\boldsymbol{v}$ is the velocity of the substrate determined by Equation (\ref{Eq_morpho}). Here, we take $E_c$ and $E_m$ as constant. Furthermore, $\mu_s(S(\Omega_C),\boldsymbol{x}_j)$ is the weight of chemotaxis/mechanotaxis that is related to the aspect ratio of the cell and whether the nodal point in on the leading part. It has been found that the receptors are distributed differently of the polarized cell, that is, the leading edges and non-leading edges have a different response to the chemotaxis \citep{Devreotes1988,Wang2009}. Hence, we assume $\mu_s(S(\Omega_C), \boldsymbol{x})$ is defined by
\begin{equation}
	\label{Eq_mu_s_chemo}
	\mu_s(S(\Omega_C), \boldsymbol{x}_j) = \exp\left\{4\left(\frac{\boldsymbol{v}_c}{\|\boldsymbol{v}_c\|}, \boldsymbol{n}_c(\boldsymbol{x}_j)\right)\right\} + \frac{1}{d_1(\Omega_C)} \left((\boldsymbol{x}_j - \boldsymbol{x}_c), \frac{\boldsymbol{v}_c}{\|\boldsymbol{v}_c\|}\right) + S(\Omega_C) +5, 
\end{equation}
where $\boldsymbol{v}_c$ is the cell velocity at time $t$ that is defined by the centre position of the cell: 
$$\boldsymbol{v}_c(t) \approx \frac{\boldsymbol{x}_c(t) - \boldsymbol{x}_c(t-dt)}{dt},$$
and here, $d_1(\Omega_C)$ is the length of the cell and the initial velocity is zero. The first term in Equation (\ref{Eq_mu_s_chemo}) indicates that the points on the leading edge react more sensitively to the signalling molecules. Since it is assumed that the cell migrates in the direction of its longitude-axis,  $\displaystyle\left((\boldsymbol{x}_j - \boldsymbol{x}_c), \frac{\boldsymbol{v}_c}{\|\boldsymbol{v}_c\|}\right)$ in the second term computes the projection of the vector $(\boldsymbol{x}_j - \boldsymbol{x}_c)$ onto the cell velocity vector. We divide this by the diameter of the cell for the sake of normalization. The second term in Equation (\ref{Eq_mu_s_chemo}) is meant to emphasize that the more to the front the point is located, the more sensitive it responses to the signalling molecules. Furthermore, when a cell is elongated, it is more sensitive to the signalling molecules. In summary, the cell geometry is obtained by connecting these nodal points on the cell membrane in the predefined (anti-)clockwise order.

Cells that penetrate through narrow channels and narrow blood vessels are subject to friction with the channel or vessel walls. This friction causes a reduction of the migration speed. Denote $\partial\Omega_{ob}$ as the boundary of the obstacle (for a microtube, this represents the walls). Then if a nodal point collides with an obstacle, then both the repelling force, exerted by the cell, and the force that results by the compression of the springs are taken into account. Note that the repelling force {\it does not} contribute to the friction if the springs are stretched or in their equilibrium length. For every nodal point on the cell membrane, there are three springs connected with the centre and the two adjacent nodal points. Hence, the force that is caused by these springs are given by
\begin{equation}
\label{Eq_force_springs}
\begin{aligned}
\boldsymbol{f}_s(\boldsymbol{x};t) &= E_c\left(\left(\boldsymbol{x}_j(t) - \boldsymbol{B}(\tilde{\phi})\tilde{\boldsymbol{x}}_j\right), \frac{\boldsymbol{B}(\tilde{\phi})\hat{\boldsymbol{x}}_j}{\|\boldsymbol{B}(\tilde{\phi})\hat{\boldsymbol{x}}_j\|}\right)\frac{\boldsymbol{B}(\tilde{\phi})\hat{\boldsymbol{x}}_j}{\|\boldsymbol{B}(\tilde{\phi})\hat{\boldsymbol{x}}_j\|}\\
&+ E_m\left(\left(\boldsymbol{x}_j(t) - \boldsymbol{B}(\tilde{\phi})\tilde{\boldsymbol{x}}_j\right), \frac{\boldsymbol{B}(\tilde{\phi})\hat{\boldsymbol{x}}_{j,j-1}}{\|\boldsymbol{B}(\tilde{\phi})\hat{\boldsymbol{x}}_{j,j-1}\|}\right)\frac{\boldsymbol{B}(\tilde{\phi})\hat{\boldsymbol{x}}_{j,j-1}}{\|\boldsymbol{B}(\tilde{\phi})\hat{\boldsymbol{x}}_{j,j-1}\|}\\
&+ E_m\left(\left(\boldsymbol{x}_j(t) - \boldsymbol{B}(\tilde{\phi})\tilde{\boldsymbol{x}}_j\right), \frac{\boldsymbol{B}(\tilde{\phi})\hat{\boldsymbol{x}}_{j,j+1}}{\|\boldsymbol{B}(\tilde{\phi})\hat{\boldsymbol{x}}_{j,j+1}\|}\right)\frac{\boldsymbol{B}(\tilde{\phi})\hat{\boldsymbol{x}}_{j,j+1}}{\|\boldsymbol{B}(\tilde{\phi})\hat{\boldsymbol{x}}_{j,j+1}\|},
\end{aligned}
\end{equation}
where $\boldsymbol{B}(\tilde{\phi})\tilde{\boldsymbol{x}}_j$ is the equilibrium position of $\boldsymbol{x}_j$ if there is only orientation but no other mechanics, $\displaystyle\frac{\boldsymbol{B}(\tilde{\phi})\hat{\boldsymbol{x}}_j}{\|\boldsymbol{B}(\tilde{\phi})\hat{\boldsymbol{x}}_j\|}$, $\displaystyle\frac{\boldsymbol{B}(\tilde{\phi})\hat{\boldsymbol{x}}_{j,j-1}}{\|\boldsymbol{B}(\tilde{\phi})\hat{\boldsymbol{x}}_{j,j-1}\|}$ and $\displaystyle\frac{\boldsymbol{B}(\tilde{\phi})\hat{\boldsymbol{x}}_{j,j+1}}{\|\boldsymbol{B}(\tilde{\phi})\hat{\boldsymbol{x}}_{j,j+1}\|}$ represent the unit vector of the springs, respectively when they are compressed. In this modeling framework, we consider the force that is exerted by the cell in order to retain the equilibrium shape configuration. One may compare this spring force to the relaxation of skin after a force has been applied. The model also considers a second force, which is the force that the cell actively exerts in order to open the channel further so that the cell can continue its migration. The current formalism superposes these two forces. Then, the friction force orthogonal to the obstacle that has an impact on the migration of the cell is determined by
\begin{equation}
\label{Eq_friction}
\boldsymbol{f}_f(\boldsymbol{x};t) = ((\boldsymbol{f}_s(\boldsymbol{x};t) + \boldsymbol{f}(\boldsymbol{x};t)) ,\boldsymbol{n}_{ob}(\boldsymbol{x}(t))) \boldsymbol{n}_{ob}(\boldsymbol{x}(t)), \mbox{if $\boldsymbol{x}(t)\in \partial\Omega_{ob}$,}
\end{equation}
where $\boldsymbol{f}_s(\boldsymbol{x};t)$ is the force from the springs defined in Equation (\ref{Eq_force_springs}) and $\boldsymbol{f}(\boldsymbol{x};t)$ is the repelling forces actively exerted by the cell that is defined in Equations (\ref{Eq_morpho}) and (\ref{Eq_force}). It is assumed that the magnitude of the friction is proportional to the repelling force. Furthermore, $\boldsymbol{n}_{ob}$ is the unit orthogonal vector pointing outward the cell centre, hence the vector points into the substrate. In our model, we simply subtract a friction part of the velocity in the tangential direction of the obstacle. Hence, the displacement of the nodal point which collides the wall of the microtube is given by
\begin{equation}
\label{Eq_disp_friction}
\begin{aligned}
d\boldsymbol{x}_j(t)&\leftarrow d\boldsymbol{x}_j(t)-\mu_f\|\boldsymbol{f}_f(\boldsymbol{x}_j(t))\|\times (d\boldsymbol{x}_j(t), \boldsymbol{\tau}_{ob}(\boldsymbol{x}_j(t)))\boldsymbol{\tau}_{ob}(\boldsymbol{x}_j(t)), \mbox{if $\boldsymbol{x}_j(t)\in\partial\Omega_{ob}$,}
\end{aligned}
\end{equation}
where $\mu_f$ is the cell friction coefficient, $\boldsymbol{f}(\boldsymbol{x}_j(t))$ is the repelling force exerted by the cell and $\boldsymbol{\tau}_{ob}(\boldsymbol{x})$ is the tangential direction (also unit vector) of the obstacle boundary $\partial\Omega_{ob}$.

Furthermore, the cell is not allowed to penetrate into the wall itself and hence velocity components normal to the wall are subtracted. In other words, the displacement of the nodal point that collides the obstacle is rephrased as 
\begin{equation}
\label{Eq_disp_repel}
d\boldsymbol{x}_j(t)\leftarrow d\boldsymbol{x}_j(t)-(d\boldsymbol{x}_j(t), \boldsymbol{n}_{ob}(\boldsymbol{x}_j(t)))\boldsymbol{n}_{ob}(\boldsymbol{x}_j(t)),\mbox{if $\boldsymbol{x}_j(t)\in\partial\Omega_{ob}$.}
\end{equation}

\section{Numerical Results}\label{Sec_pre_results}
\noindent
In this section, we investigate how cells influence each other's migration when they migrate through a narrow pore or channel. In the in-vitro experiments which contain microtubes, the width of the microtube is often set a bit smaller than the cell diameter, which experimentally models the transmigration of cells through narrow channels in real tissues. However, as we mentioned earlier, in the  in-vitro experiments, the microtubes are undeformable \citep{Mak2013ASerial,Mak2013, Zhang2021-yu}, which is contrary to real tissues, where the immediate environment of cells, consists of viscoelastic material. Therefore, inspired by the microtube experiments,  we adjust the experimental setting such that the channel can deform due to the forces exerted by cells in our simulations.

We consider two different settings, which are distinguished by how many cells are migrating through the channel at the same time. In Case (1) (see Figure \ref{Fig_Case_1_Snapshots}\subref{fig_seperate_initial}), there is only one cell in the computational domain penetrating through the channel. In other words, the next cell is only allowed to enter the channel when the previous one has completely exited. We are aware that this setting does not probably reflect reality since cancer cells mostly migrate collectively. However, this setting is helpful to determine and to quantify whether the follower cell benefits from the leader cell, which possesses the ability to permanently widen the channel. The permanent nature of the deformation has been incorporated by the use of the morphoelastic component in the solid mechanical part of the model. In Case (2) (see Figure \ref{Fig_Case_2_Snapshots}\subref{fig_repel_initial}), we consider two cells that simultaneously migrate through the narrow channel. This case reflects the mechanism of collective migration somewhat better. In this case, we consider the forces that the cell exert in order to repel and push each other. Note that here we assume that there is only mechanical contact between the two cells, as likely occurs in cadherin deficient cells, such as the commonly used metastatic, breast cancer cell line MDA-MB-231 \citep{Zhang2021-yu}.


In both cases, we have similar experimental settings and in this manuscript, all the simulations are carried out in two dimensions. In the computational domain $\Omega =(-100, 100)\times(50,50)$, a sinusoidal deformable channel that starts at $X_l = -64$ and ends at $X_r = 60$, is given by the following top and bottom part of the channel walls
\begin{equation}
	\label{Eq_sine_channel}
	y = \pm(b + A\sin(wx)), \mbox{ $b, A, w>0$.}
\end{equation}
Note that $2(b+A)$ is set to be smaller than the cell diameter to ensure that the cell is confined in the channel. The advantages of a sinusoidal channel are: (1) it is a well-defined geometry, where the channel width ranged between $2(b-A)$ and $2(b+A)$; (2) it resembles the walls of a channel in a tissue, where the wall is formed by cells  \citep{Heuz__2011}. We assume that the cells are located initially at the left side of the channel and at the other side of the channel, there is a point source of signalling molecule. The parameter values that are used in the simulations in this section are shown in Table \ref{Tbl_ParaValue_All}. Note that we also define the initial positions of the most left and right point of the channel, since the channel will deform due to the forces exerted by the cells. As a result, the position of the channel is altering in the same time: we define $\{\boldsymbol{X}_{channel}\}$ as the set of initial positions of the channel, then the new positions of the channel $\{\boldsymbol{x}_{channel}\}$ can be tracked by $$\boldsymbol{x}(t) = \boldsymbol{X}+\boldsymbol{u}(\boldsymbol{X}, t), \mbox{ $\forall \boldsymbol{X}\in\{\boldsymbol{X}_{channel}\}$},$$ where $\boldsymbol{u}(\boldsymbol{X},t)$ is the displacement of the computational domain, post processed by solving the morphoelasticity model in Equation (\ref{Eq_morpho}), and $\boldsymbol{X}$ is the initial position. We say that the cell enters the channel when there is more than one nodal point of the cell membrane in the channel (the moment is denoted by $t^i_{in}$ for cell $i$) and the cell leaves the channel completely when there is no nodal point of cell membrane in the channel (the moment is denoted by $t^i_{out}$ for cell $i$). To rephrase it, the entering time and exiting time are defined by 
\begin{align*}
	t^i_{in} = \arg\min_{t\geqslant 0}\{\boldsymbol{x}^i_j\in\partial\Omega_{C^i}: \exists\boldsymbol{x}^i_j \mbox{ enters the channel}\};\\
	t^i_{out} = \arg\min_{t>t^i_{in}}\{\boldsymbol{x}^i_j\in\partial\Omega_{C^i}: \forall\boldsymbol{x}^i_j \mbox{ exits the channel}\},
\end{align*}
for the cell $i$ and $\partial\Omega_{C^i}$, and we remind the reader that $\boldsymbol{x}^i_j$ denotes the cell membrane and $j$-th nodal point on the cell membrane of the cell $i$. Subsequently, the transmigration time of cell $i$ is given by $$T^i = t^i_{out} - t^i_{in},$$ and the (average) speed of the cell migration reads as $$\bar{v}^i = \frac{\|\boldsymbol{x}^i_c(t^i_{out}) - \boldsymbol{x}^i_c(t^i_{in})\|}{T^i},$$
where $\boldsymbol{x}^i_c(t)$ represents the central position of cell $i$ at time $t$. 
\begin{table}\footnotesize
	\centering
	\caption{Parameter values of the cell invasion model used in Section \ref{Sec_pre_results}}
	\begin{tabular}{m{2cm}<{\centering}m{6cm}<{\centering}m{2cm}<{\centering}m{2.5cm}<{\centering}m{2.5cm}<{\centering}}
		\toprule
		{\bf Parameter}& {\bf Description} & {\bf Value} & {\bf Units} & {\bf Source}\\
		\toprule
		$E_s$ & Substrate elasticity & $50$ & $kg/(\mu m\cdot min^2)$ & \citet{chen2017model}\\
		$E_c$ & Stiffness of the springs that connect the cell centre and the nodal point on the cell membrane & $5$ & $kg/(\mu m\cdot min^2)$ & \citet{chen2017model}\\
		$R$ & Cell radius & $5$ & $\mu m$ & \citet{chen2017model} \\
		$\mu_f$ & Cell friction coefficient & $0.03$ & $-$ & \citet{Angelini2012} \\
		$\nu_s$ & Poisson's ratio of the ECM & $0.48$ & $-$ & \citet{koppenol2017biomedical} \\
		$\nu_c$ & Poisson's ratio of cells & $0.38$ & $-$ & \citet{Trickey2006}\\
		$\kappa_s$ & Parameter in Robin’s boundary condition to solve Equation (\ref{Eq_adr_concent}) & $1$ & $1/\mu m$ & \citet{Peng2020} \\
		$\mu_{visco}$ & Weight of viscosity in viscoelasticity in Equation (\ref{Eq_morpho_visco_sigma}) & $1$ & $-$ & \citet{PengGorter2022}\\
		$\mu_1$ & Shear viscosity of the ECM & $33.783$ & $-$ & \citet{Peng2020}\\
		$\mu_2$ & Bulk viscosity of the ECM & $22.523$ & $-$ & \citet{Peng2020}\\
		$\alpha_s$ & Memory rate of spontaneous displacement & $0.098$ & - & \citet{Takagi2008}\\ 
		$\gamma_s$ & Decay rate of spontaneous displacement & $0.098$ & - & \citet{Takagi2008}\\ 
		\bottomrule
		&&&&\\
		\toprule
		\multicolumn{5}{l}{\bf Estimated Parameter Value in this study}\\
		\toprule
		$E_m$ & Stiffness of the springs that connect the neighboring nodal points on the cell membrane & $0.5$ & $kg/(\mu m\cdot min^2)$ &\\
		$k$ & Secrete rate of the signal & $100$ & $kg/(\mu m^3\cdot min)$ & \\
		$D$ & Diffusion rate of the signal & $233.2$ & $\mu m^2/ min$ & \\
		$N$ & Number of nodal points on the cell membrane & $40$ & $-$ & \\
		$\sigma_{rw}$ & Weight of random walk & $1$ & $-$ &  \\
		$\alpha$ & Degree of permanent deformation in Equation (\ref{Eq_morpho}) & $1.5$ & $-$ &  \\
		$\rho$ & Density of ECM in Equation (\ref{Eq_morpho}) & $1$ & $kg/\mu m^3$ &  \\
		$x0$ & Length of the computational domain in the x-coordinate & $200$ & $\mu m$ &  \\
		$y0$ & Length of the computational domain in the y-coordinate & $100$ & $\mu m$ &  \\
		$X_l$ & The x coordinate of the most left point of the channel & $-64$ & $\mu m$ &  \\
		$X_r$ & The x coordinate of the most right point of the channel & $60$ & $\mu m$ &  \\
		$A$ & Amplitude of the initial sinusoidal flexible channel in Equation (\ref{Eq_sine_channel}) & $1$ & $-$ &  \\
		$w$ & Angular frequency of the initial sinusoidal flexible channel in Equation (\ref{Eq_sine_channel}) & $0.7$ & $-$ &  \\
		$b$ & Midline of the initial sinusoidal flexible channel in Equation (\ref{Eq_sine_channel})& $2.5$ & $-$ &  \\
		\bottomrule
	\end{tabular}
	\label{Tbl_ParaValue_All}
\end{table}

\subsection{Case (1): Single Cell Migration}\label{Subsec_pre_Case_1}
\begin{sidewaysfigure}
	\centering
	\subfigure[t = 0]{
		\includegraphics[width=0.23\textwidth]{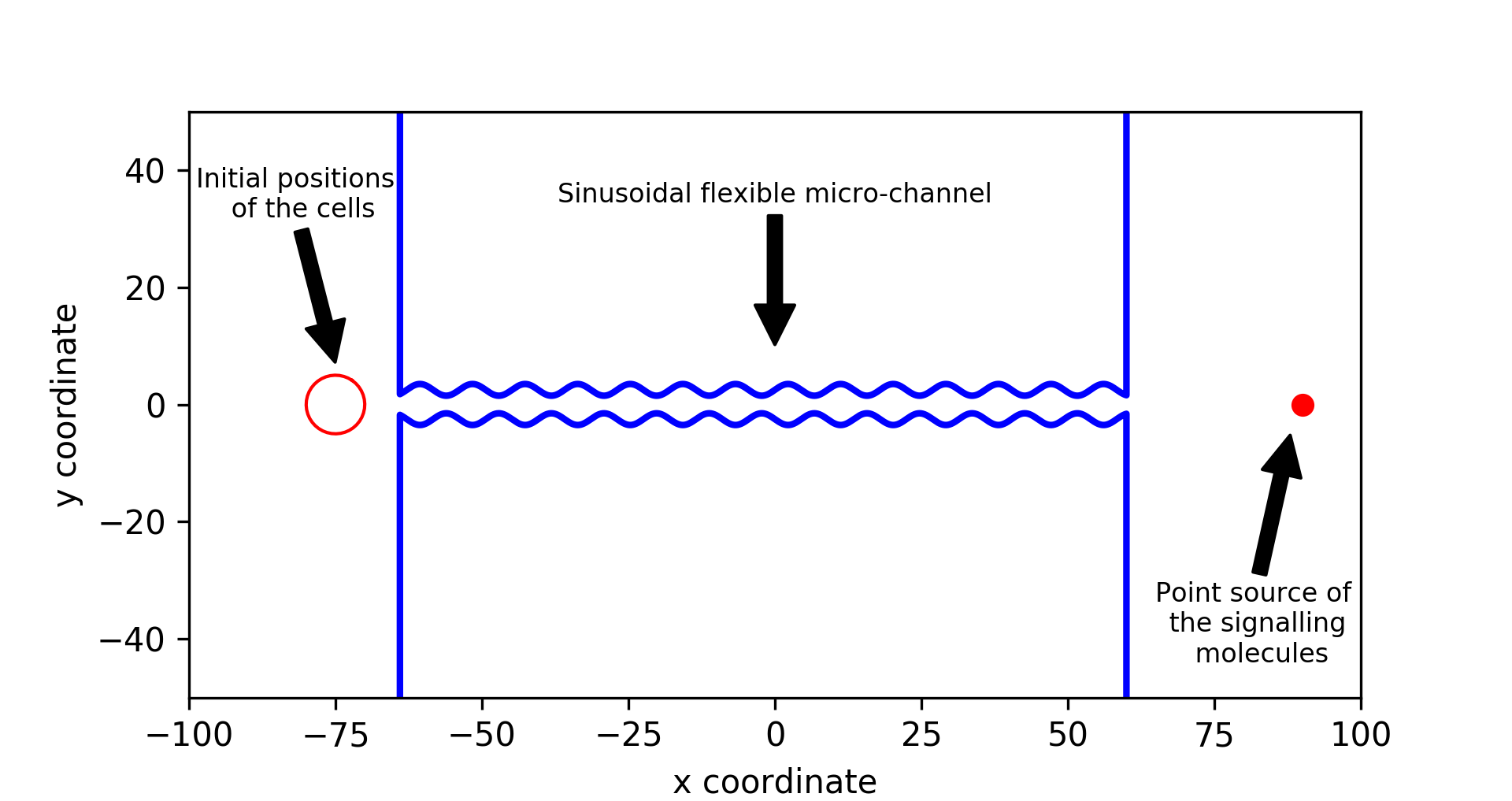}
		\label{fig_seperate_initial}}
	\subfigure[t = 2.1 min]{
		\includegraphics[width=0.23\textwidth]{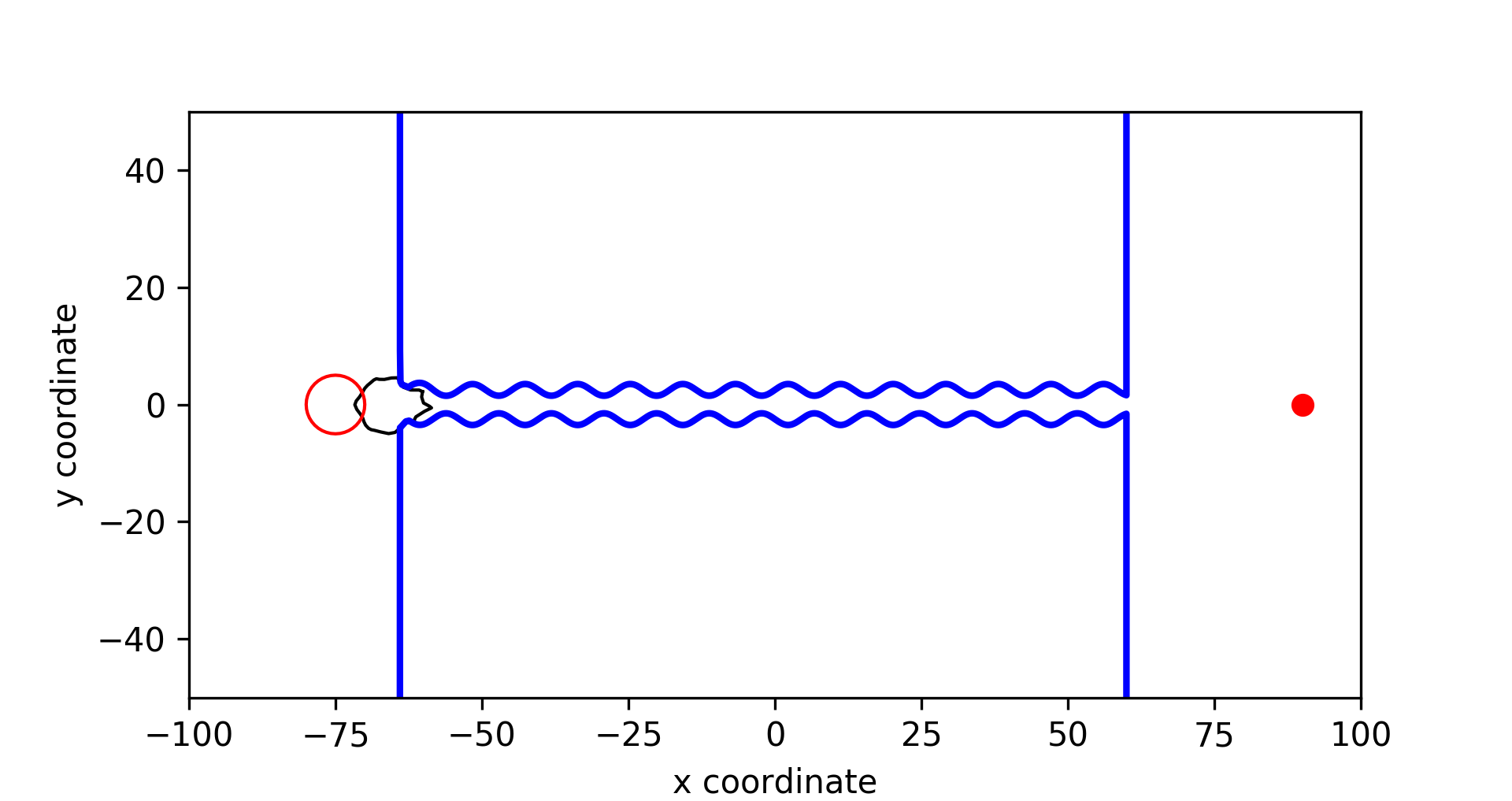}}
	\subfigure[t = 3.5 min]{
		\includegraphics[width=0.23\textwidth]{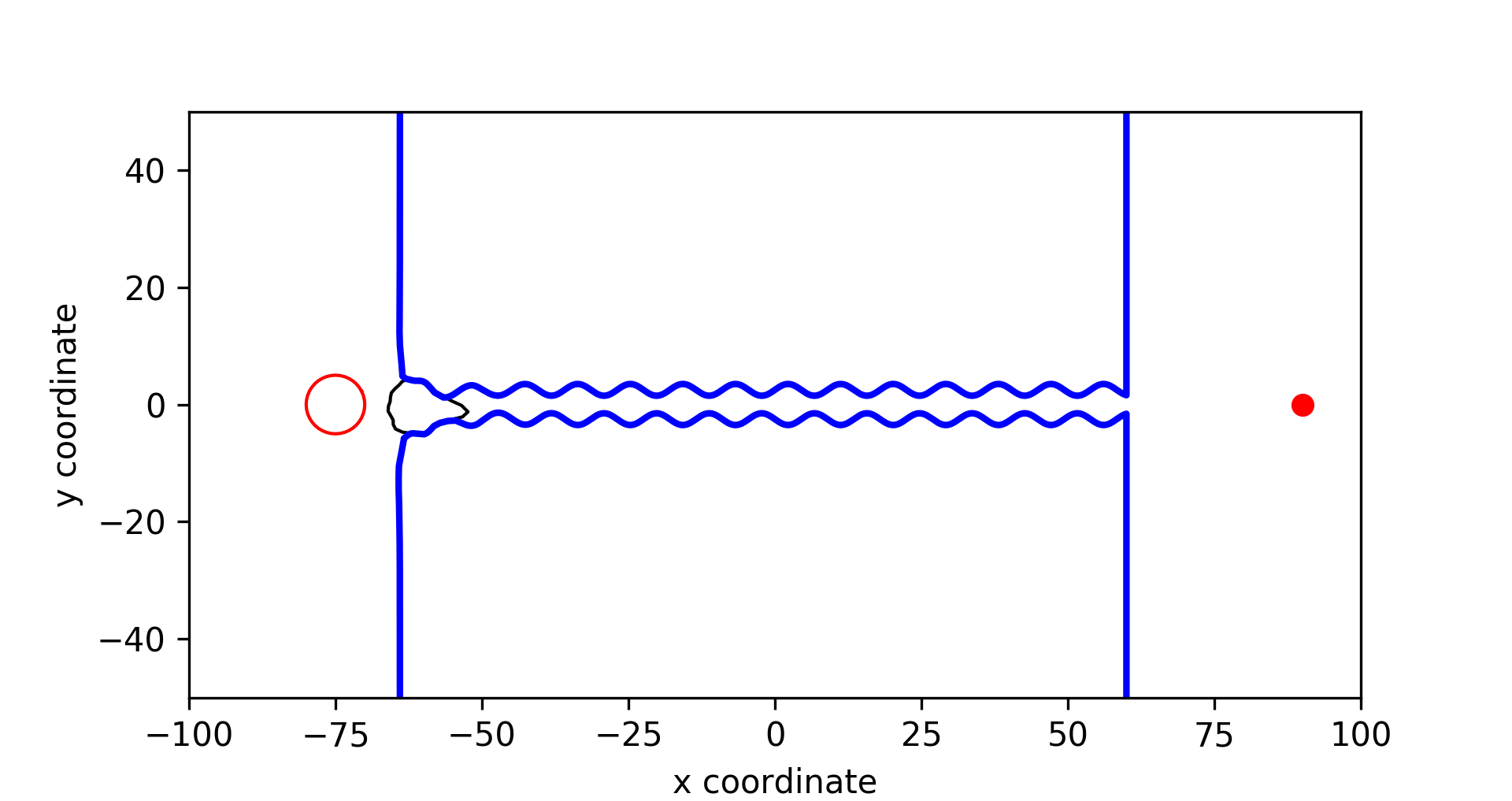}}
	\subfigure[t = 14.0 min]{
		\includegraphics[width=0.23\textwidth]{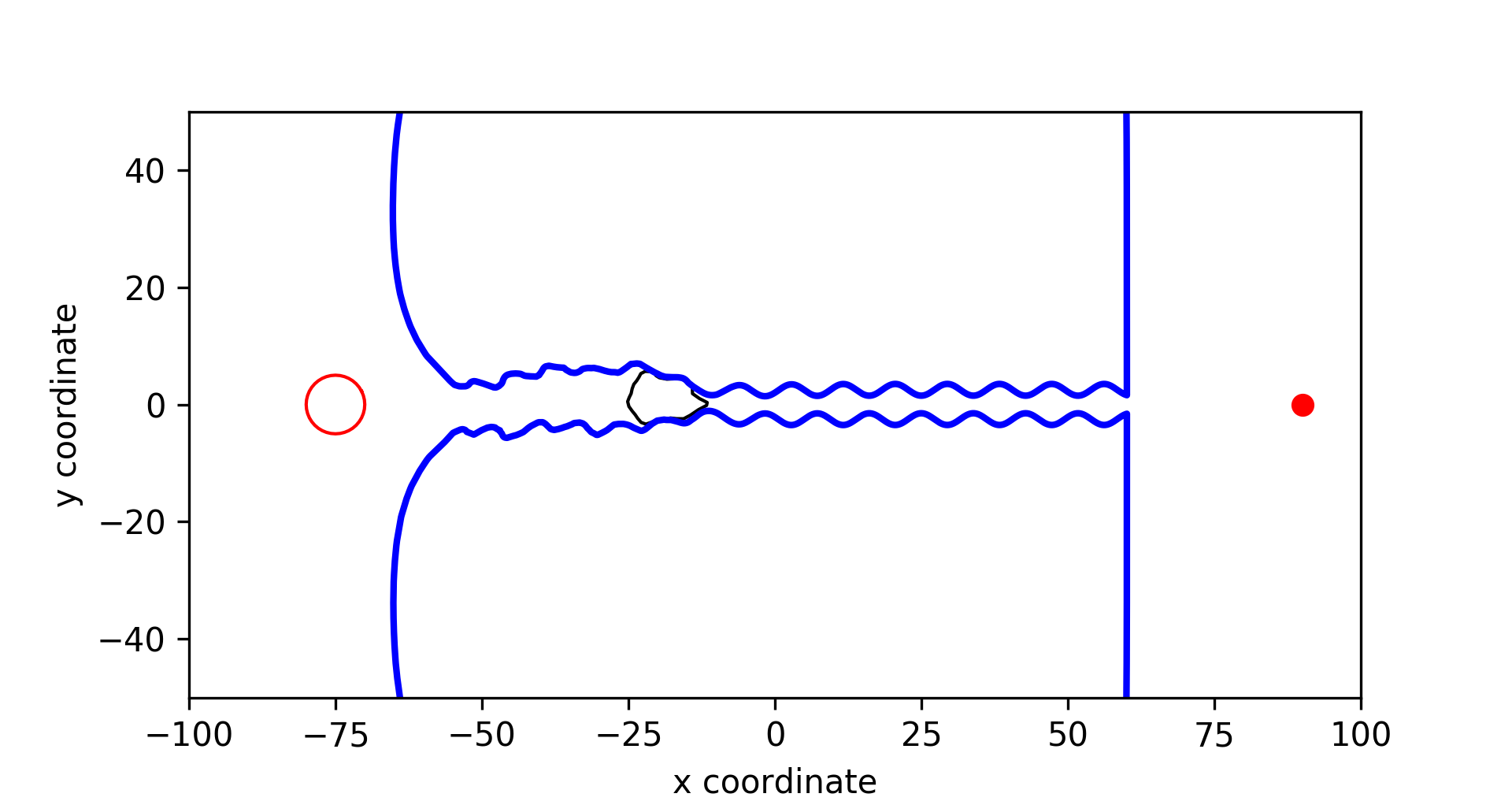}}
	\subfigure[t = 35.63 min (the first cell leaves the channel completely)]{
		\includegraphics[width=0.23\textwidth]{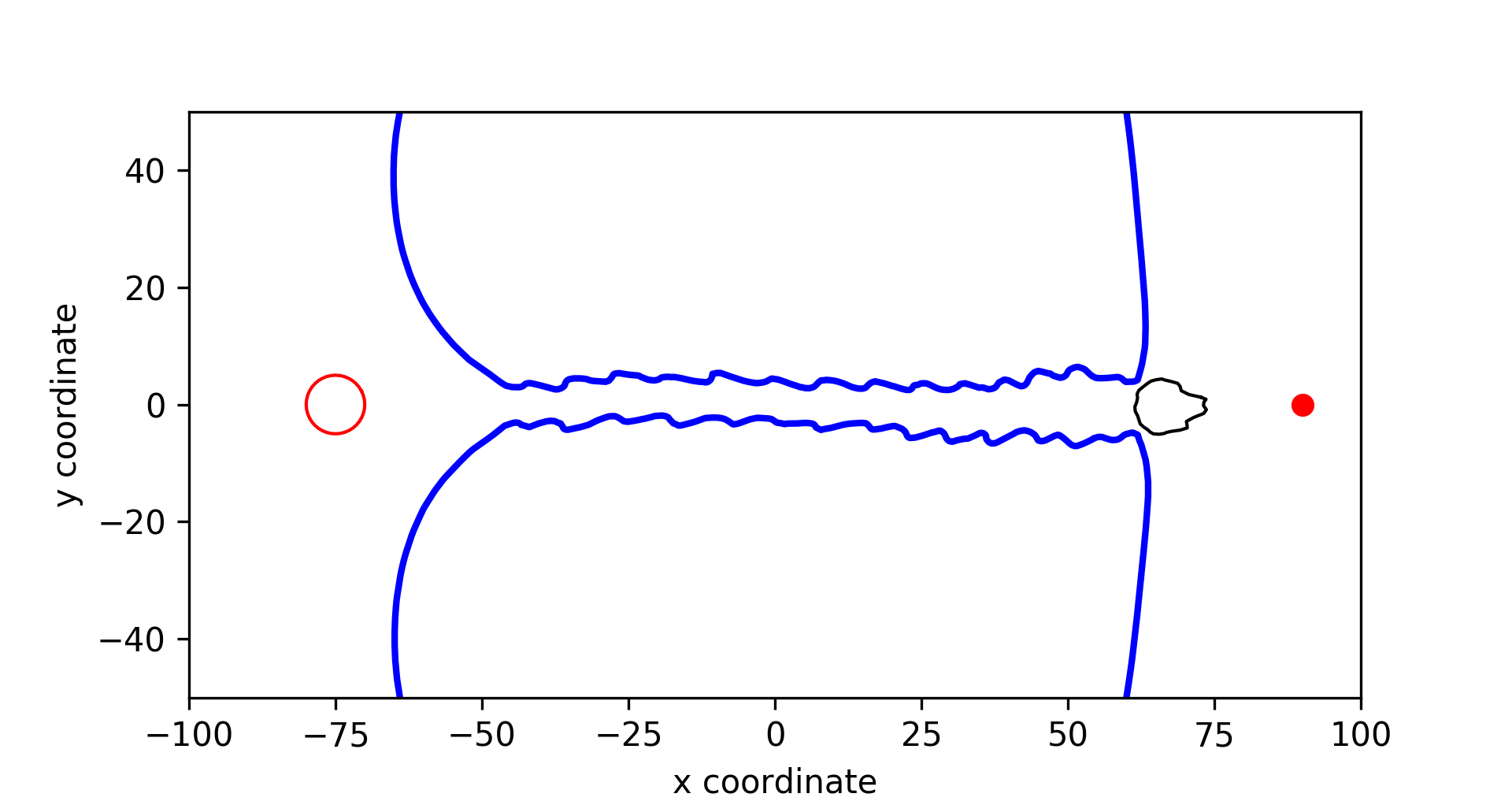}}
	\subfigure[t = 35.70 min (the second cell appears in the computational domain)]{
		\includegraphics[width=0.23\textwidth]{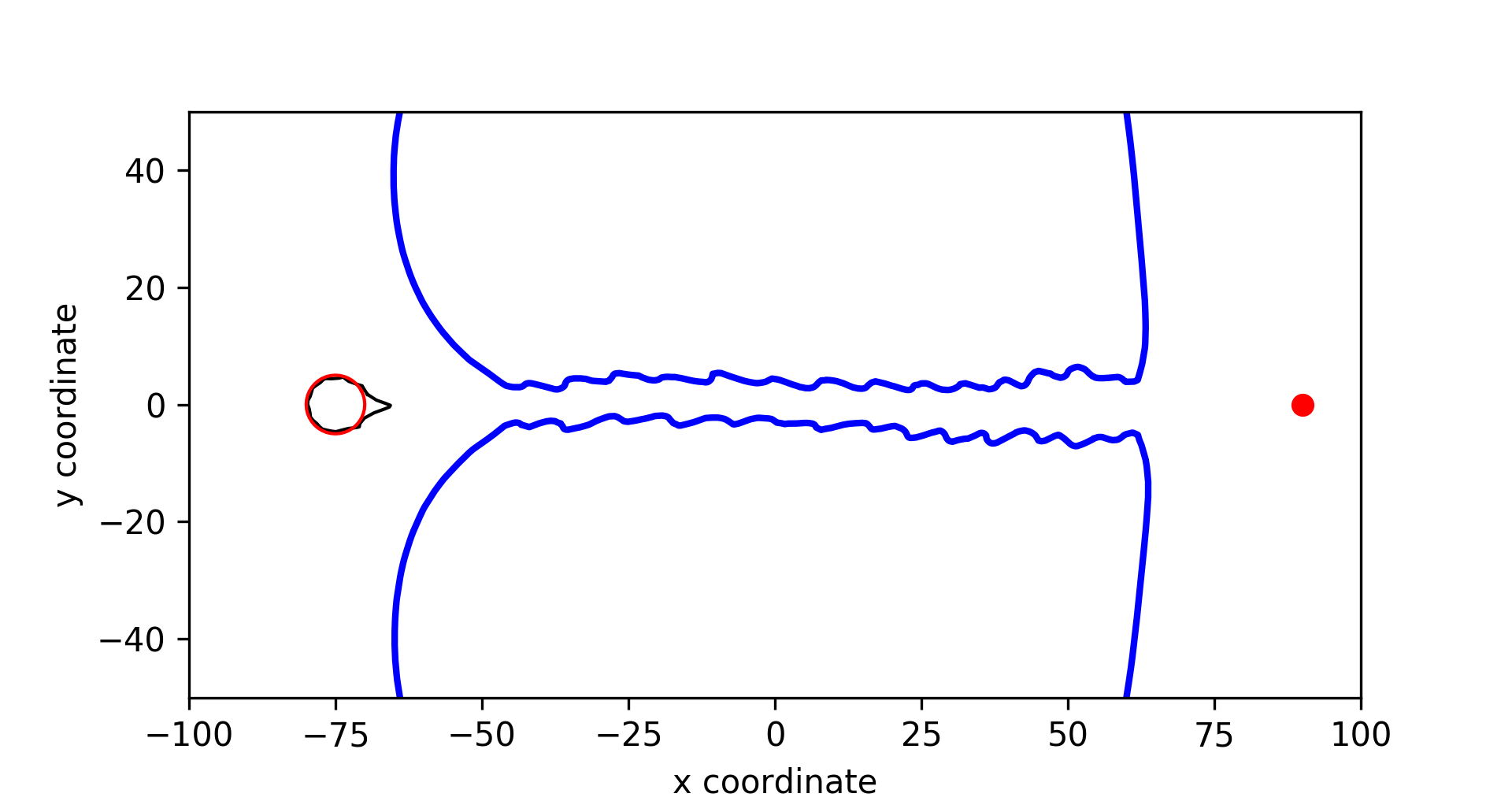}}
	\subfigure[t = 41.02 min]{
		\includegraphics[width=0.23\textwidth]{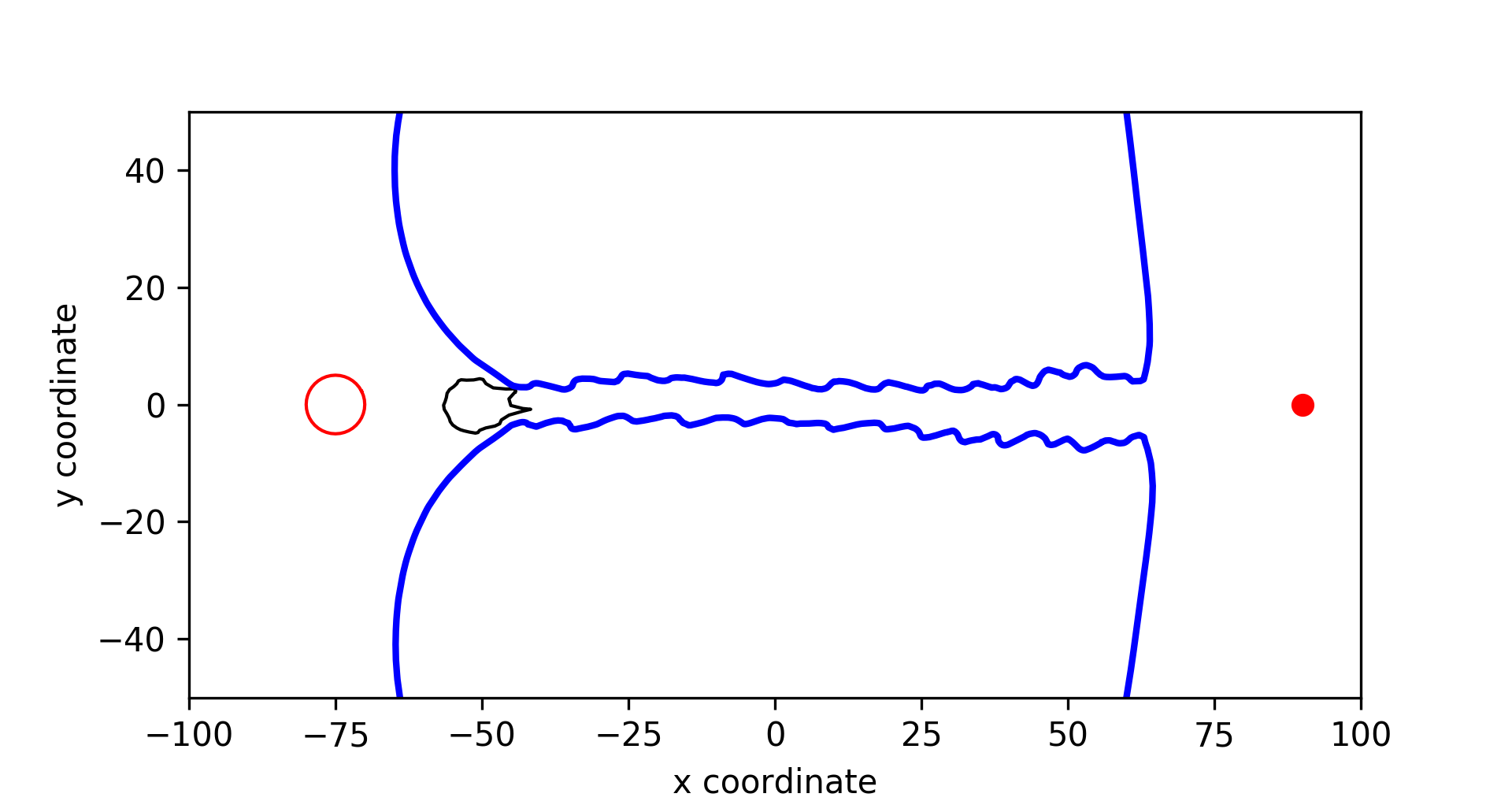}}
	\subfigure[t = 48.02 min]{
		\includegraphics[width=0.23\textwidth]{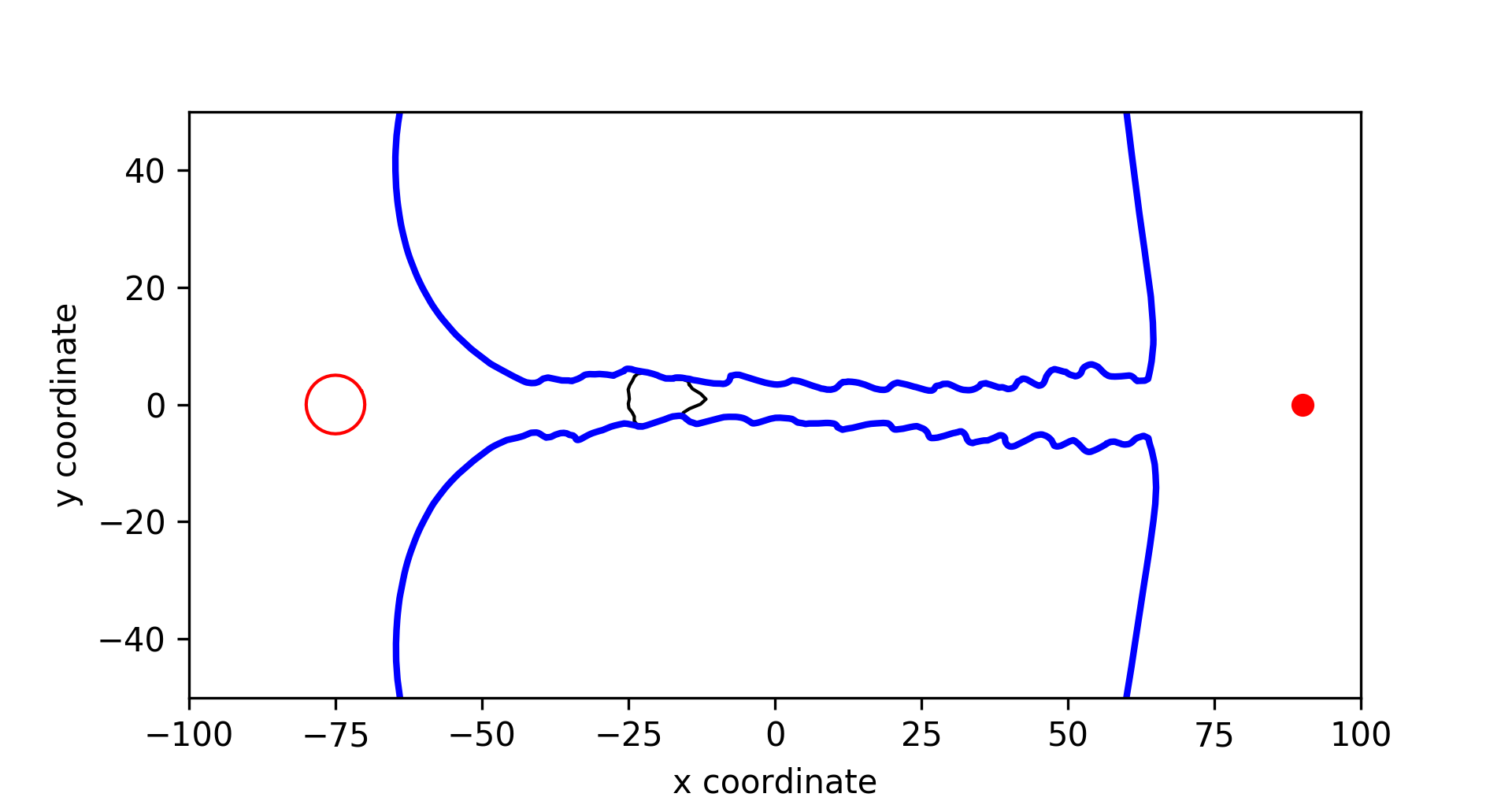}}
	\subfigure[t = 67.27 min (the second cell leaves the channel completely)]{
		\includegraphics[width=0.23\textwidth]{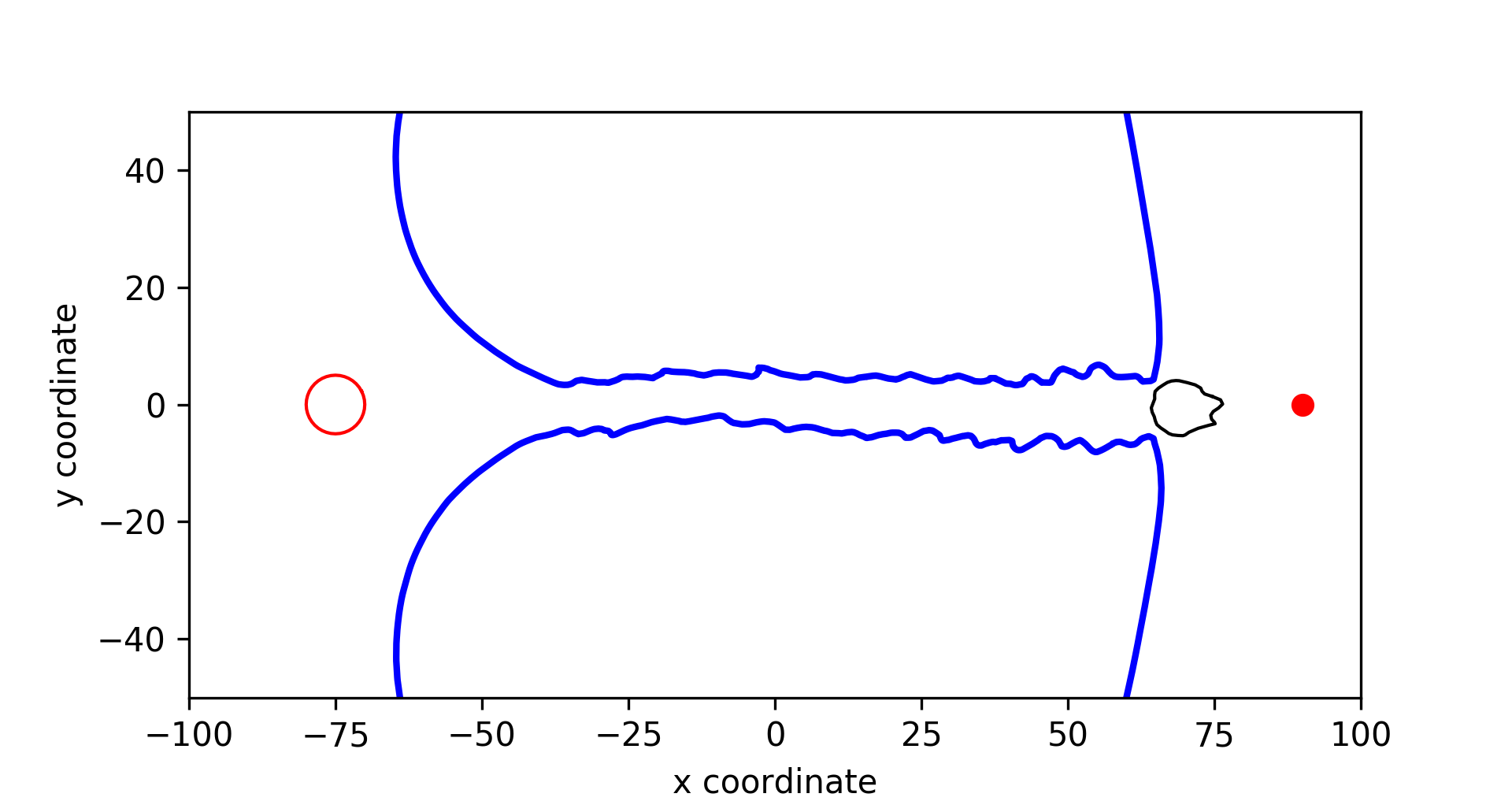}}
	\subfigure[t = 67.34 min (the third cell appears in the computational domain)]{
		\includegraphics[width=0.23\textwidth]{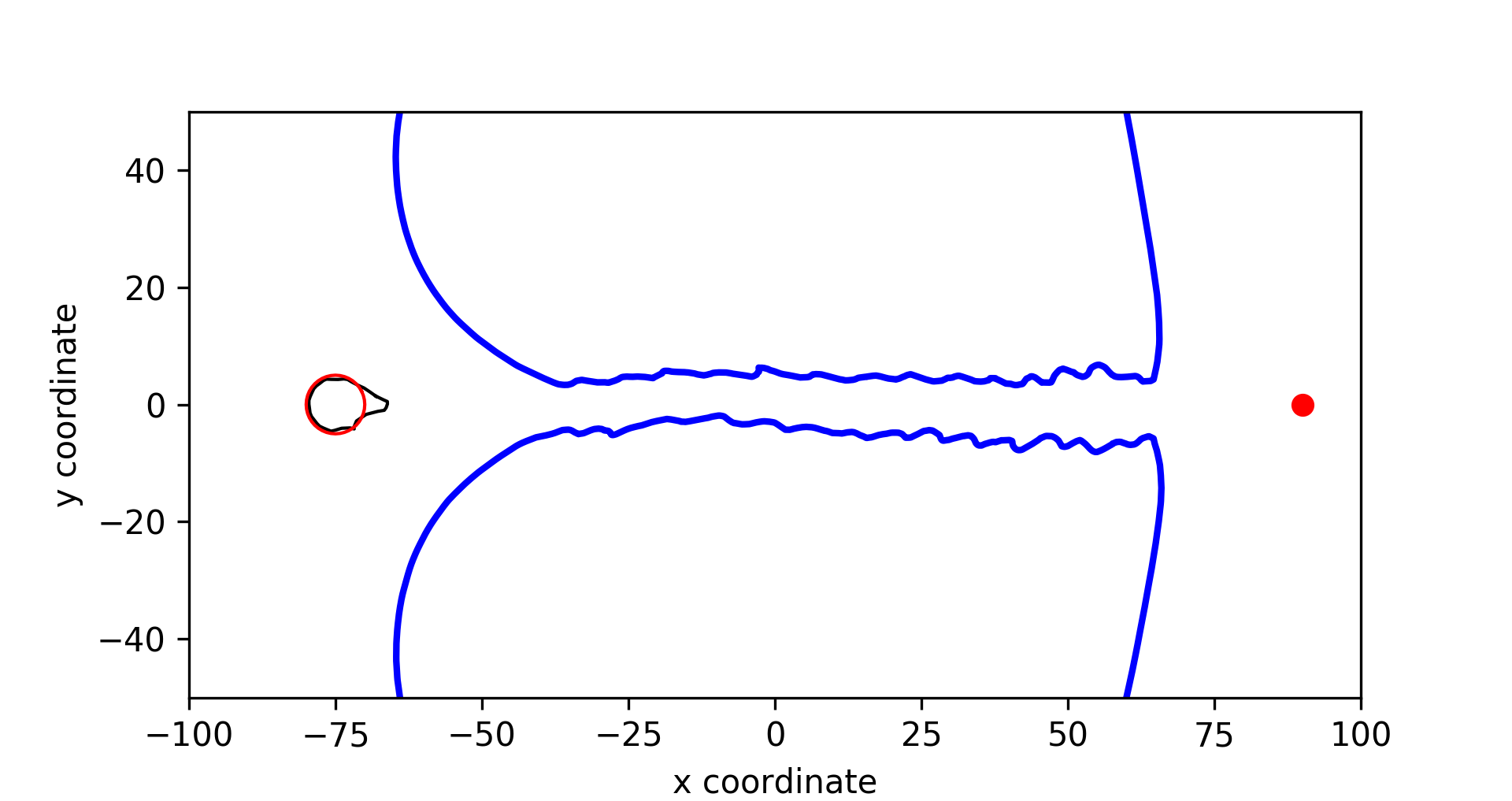}}
	\subfigure[t = 81.20 min]{
		\includegraphics[width=0.23\textwidth]{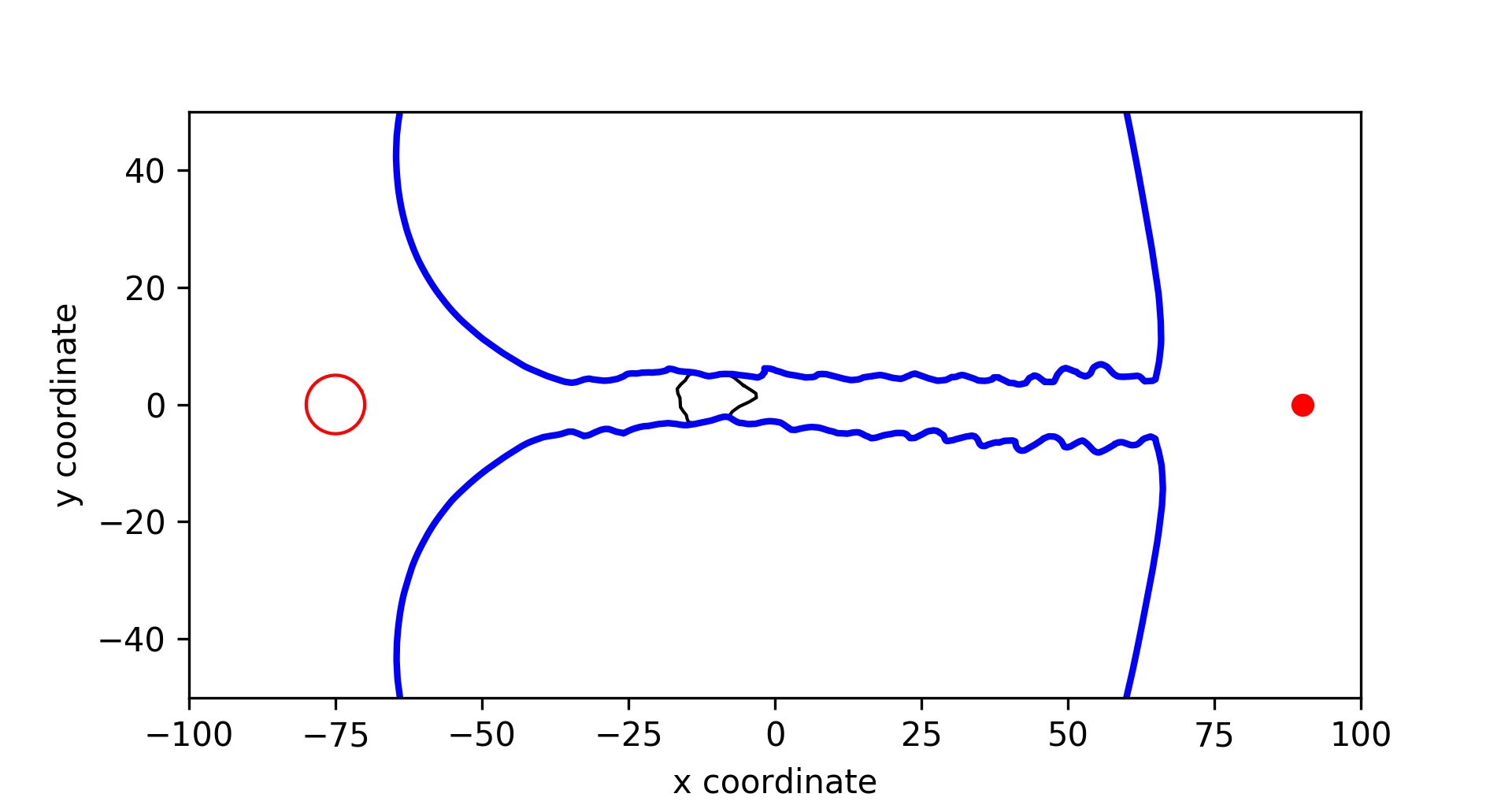}}
	\subfigure[t = 99.05 min]{
		\includegraphics[width=0.23\textwidth]{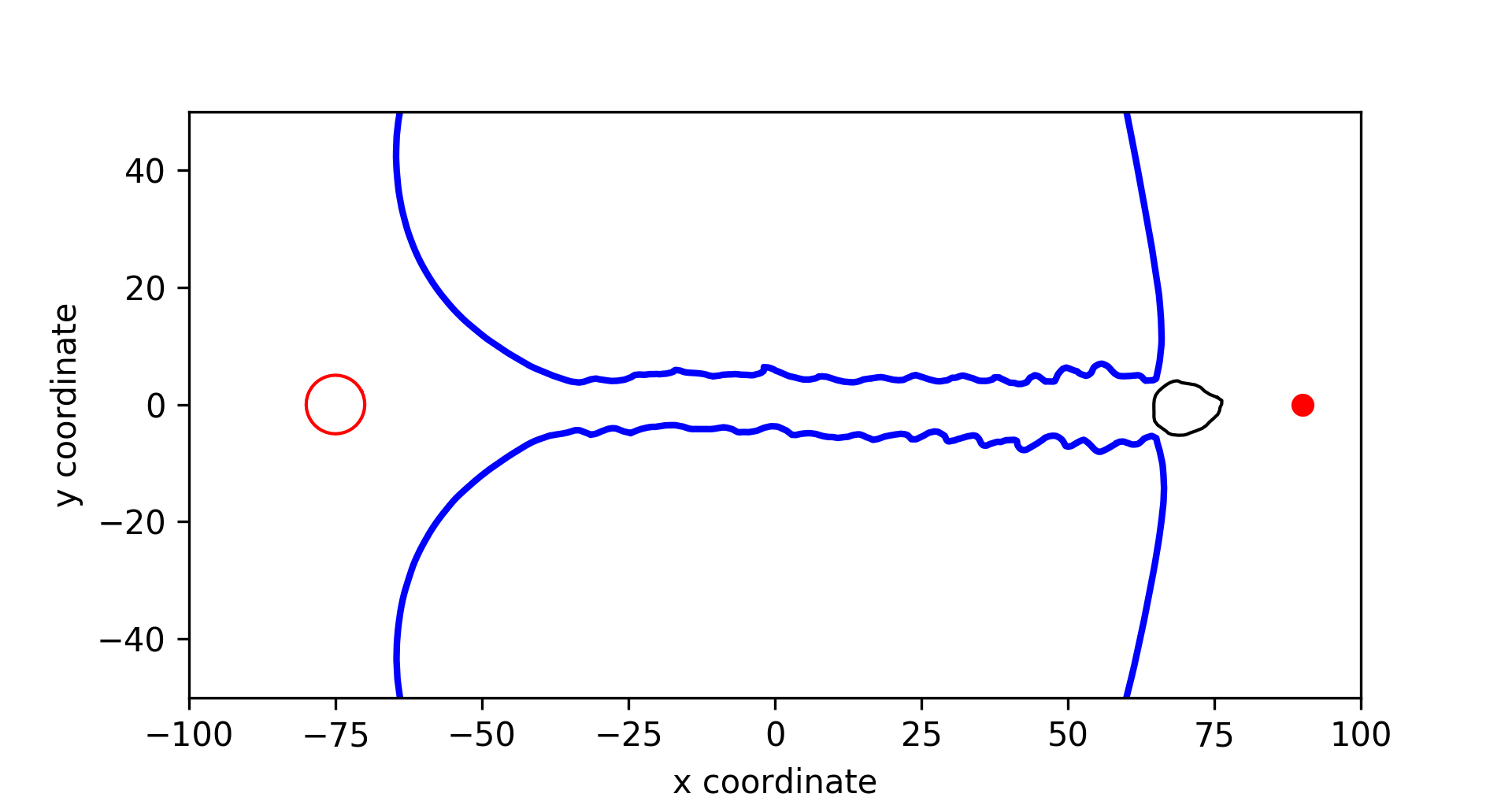}}
	\caption{The screenshots of the simulation for Case (1), where the next cell appears only when the previous cell leaves the channel completely. }
	\label{Fig_Case_1_Snapshots}
\end{sidewaysfigure}

\noindent
The initial setting of the simulation is shown in Figure \ref{Fig_Case_1_Snapshots}\subref{fig_seperate_initial}. We consider a deformable sinusoidal micro-channel that separates the left and right parts of the domain. The cells can migrate through the channel. To ensure most settings are the same for every cell, the initial positions of cells are the same, that is, all the cells start moving from location where the centre position is $(-75,0)$. 

Several snapshots at consecutive times of the simulation are shown in Figure  \ref{Fig_Case_1_Snapshots}. Furthermore, to have a better visualization of the simulation, a short animation is attached (see Supplementary Material Video 1). Figure \ref{Fig_avg_all}\subref{fig_Case_1_avg_vel} shows the cells' speed when they are migrating through the channel. Note that here $t=0\min$ in the figure is the moment when the cell enters the channel. In general, it can be seen that the curve of the first cell has the largest amplitude of the oscillation, while the migration speed profiles of the other cells exhibit smaller amplitudes. Furthermore, it is clear that the first cell moves the slowest and takes the longest time in the channel compared to the follower cells. However, regarding the cell speed, it is hard to see the difference between the second and the third cell, except that the third cell spends less time in the channel (since the black curve disappears the earliest in the figure). Furthermore, in Figure \ref{Fig_avg_all}\subref{fig_Case_1_avg_CSI} and \subref{fig_Case_1_avg_as}, cells are in an elongated shape when they migrate, as the circularity and the aspect ratio never recover to $1$, which is the case for the equilibrium (and initial) shape. However, it seems that the difference in the circularity is not that significant, while regarding the aspect ratio, the follower cells are more elongated, which accelerated their velocity according to our model assumptions.

For the sake of quantification, we show the transmigration time and average speed of three cells in Table \ref{Tbl_Case_1_time_vel}, and the change of the width of the channel is shown in Table \ref{Tbl_Case_1_width}. Since cells are migrating to the right-hand side, the channel is stretched to the right as well, as it is seen in Figure \ref{Fig_Case_1_Snapshots}. Furthermore, we also noticed that the sinusoidal channel becomes more and more flat after the transmigration of several cells, which is mainly caused by the fact that the repulsion force is proportional to the penetration depth of the cell. From Table \ref{Tbl_Case_1_time_vel}, it can be concluded that, in general, the follower cells move faster and move over a shorter distance, compared to the leader cell. Furthermore, every cell contributes to making the channel wider and after all three cells penetrating through the channel, even the minimal width of the channel is comparable with the cell size; see Table \ref{Tbl_Case_1_width}. As a result, the follower cells benefit from the leader cell that widens the channel such that the follower cells deform less and there is also less friction from the wall of the channel. However, we observed that the speed of Cell 3 is $1.53\%$ smaller than the speed of Cell 2, which may imply an optimum  channel width to maximize the cell speed. Such an optimum could easily exist since the wall is also expanded by the cell in order to migrate. Furthermore, there is a significant difference of the cell aspect ratio between Cell 2 and Cell 3, which results from the fact that before Cell 3 enters the channel, the channel has been expanded significantly such that Cell 3 is not required to change shape as much; hence, cell polarization and the cell aspect ratio are less compared with Cell 2. Consequently, Cell 3 transmigrates more slowly than Cell 2.

\begin{figure}
	\centering
	\subfigure[Average velocity of Cells in Case (1)]{
	\includegraphics[width=0.48\textwidth]{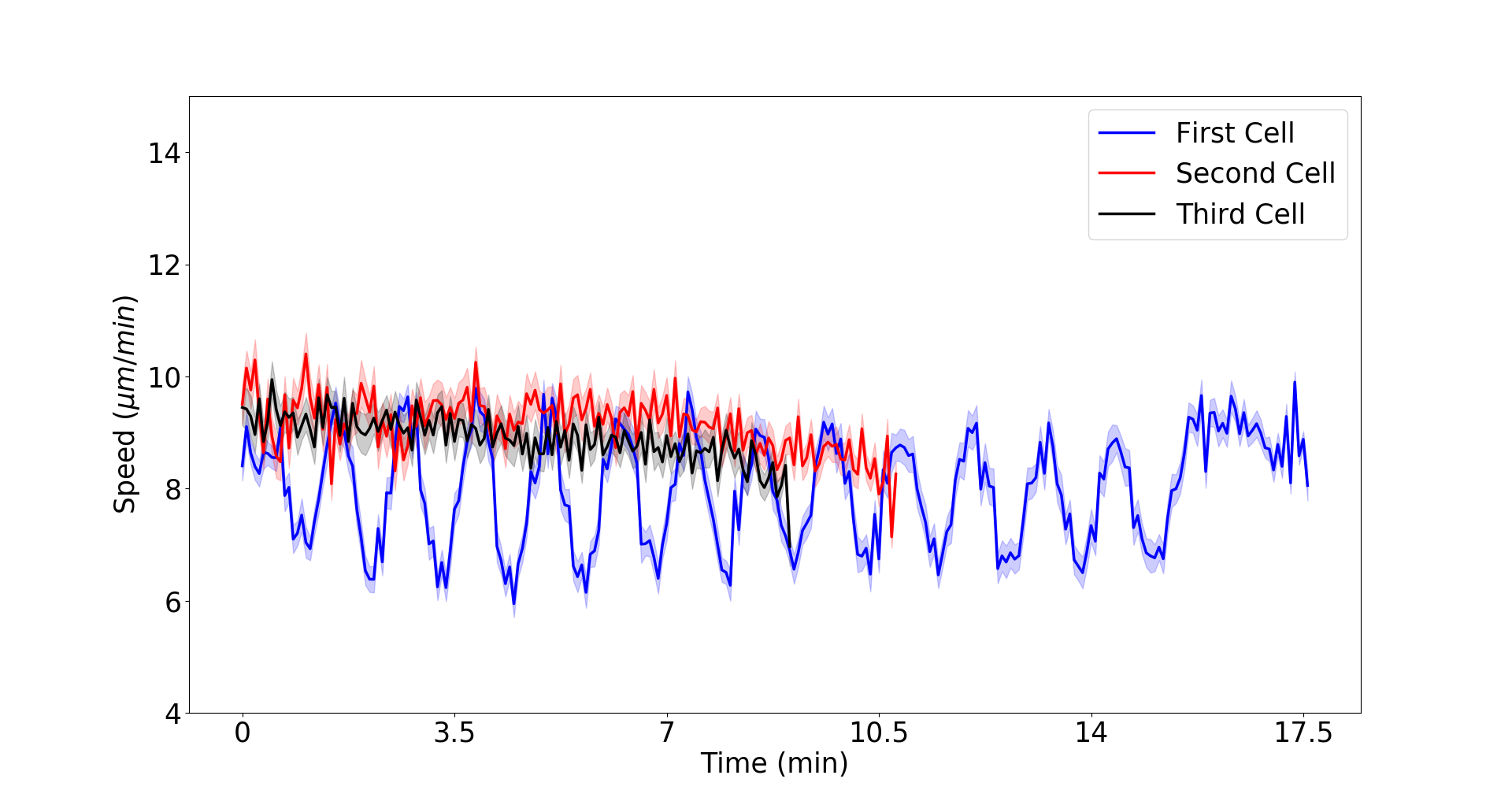}
	\label{fig_Case_1_avg_vel}}
	\subfigure[Average velocity of Cells in Case (2)]{
	\includegraphics[width=0.48\textwidth]{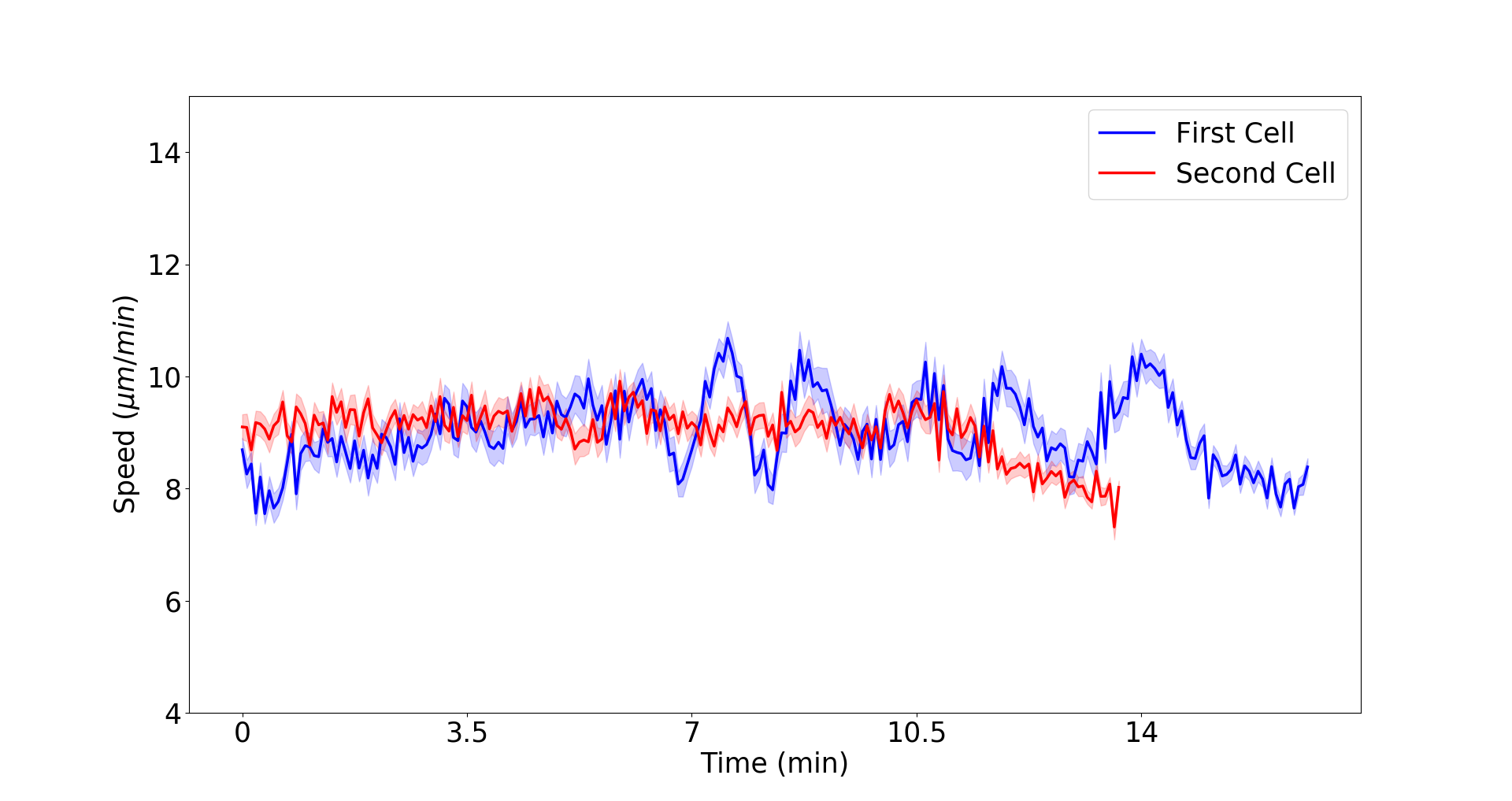}
	\label{fig_Case_2_avg_vel}}
%
	\subfigure[Circularity of Case (1)]{
		\includegraphics[width=0.48\textwidth]{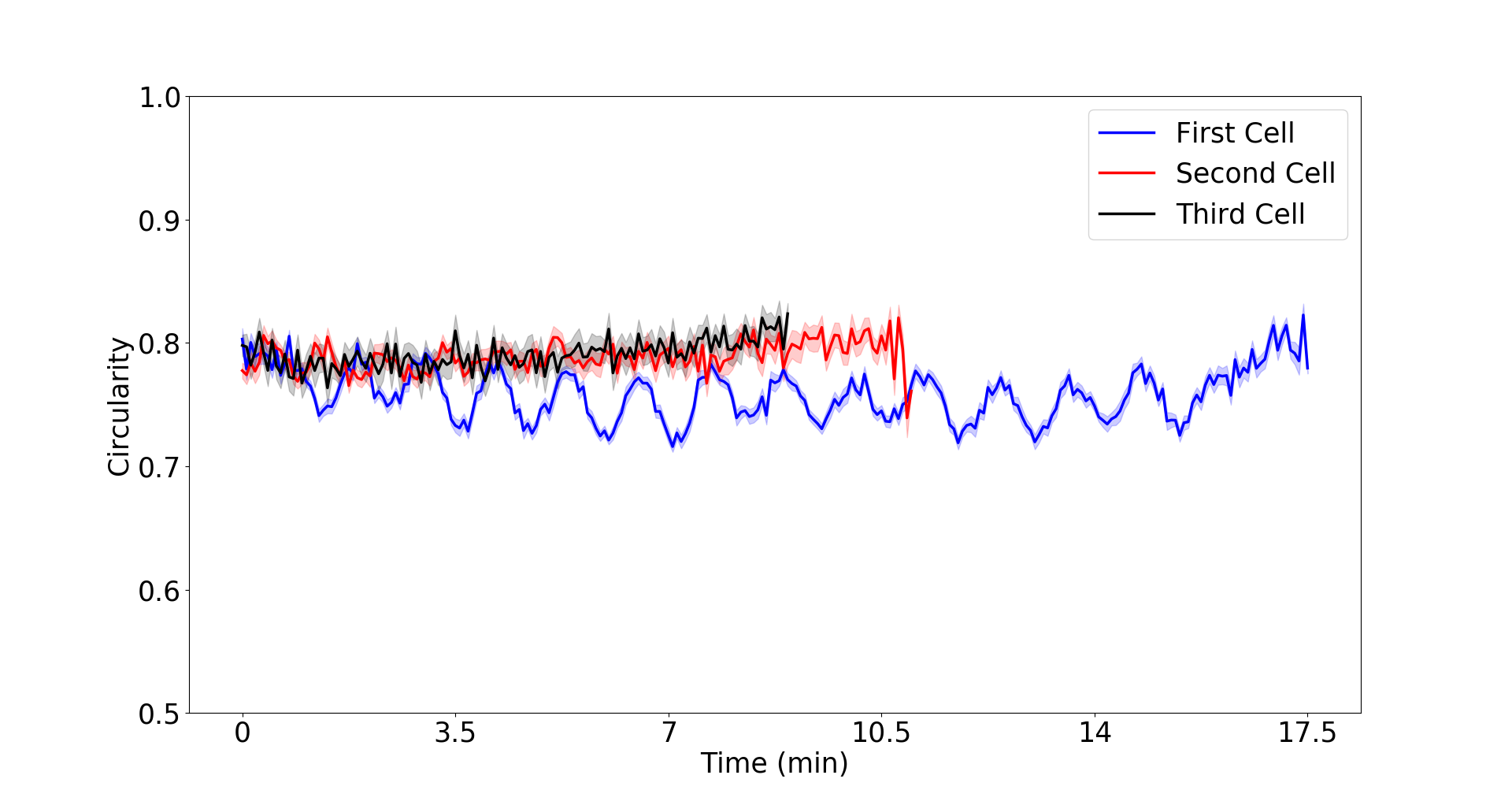}
		\label{fig_Case_1_avg_CSI}}
	\subfigure[Circularity of Case (2)]{
		\includegraphics[width=0.48\textwidth]{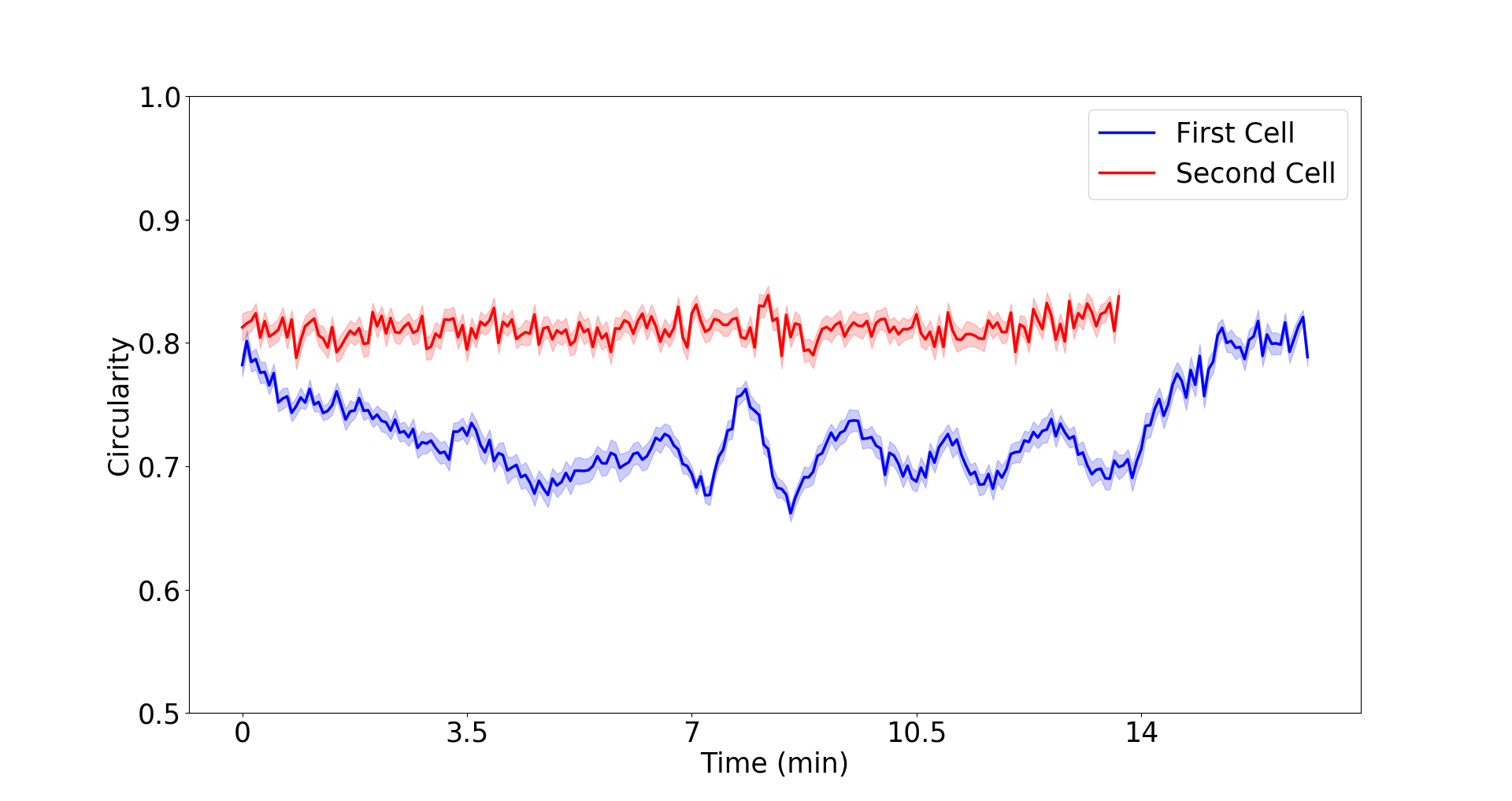}
		\label{fig_Case_2_avg_CSI}}
	\subfigure[Aspect Ratio of Case (1)]{
		\includegraphics[width=0.48\textwidth]{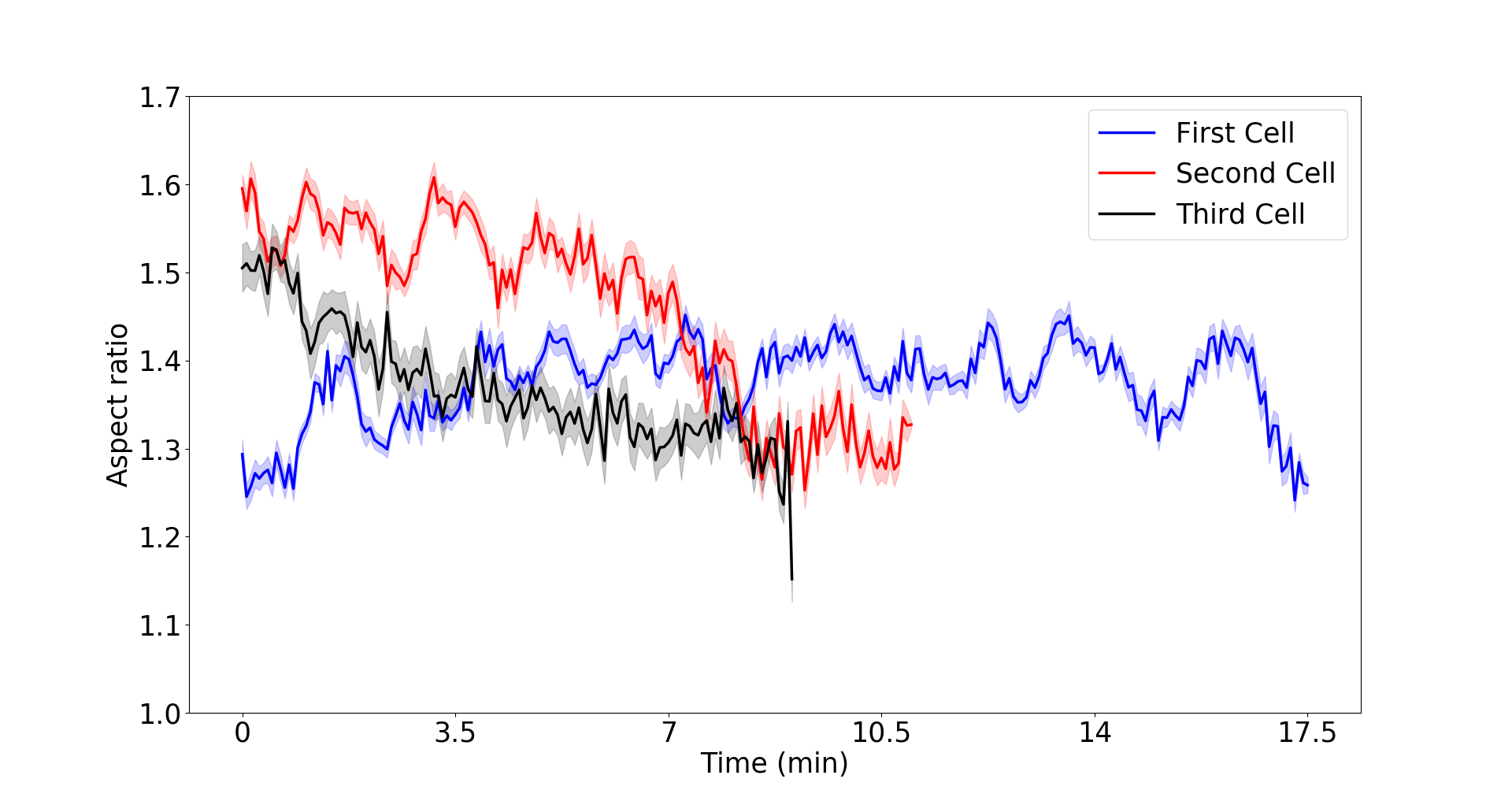}
		\label{fig_Case_1_avg_as}}
	\subfigure[Aspect Ratio of Case (2)]{
		\includegraphics[width=0.48\textwidth]{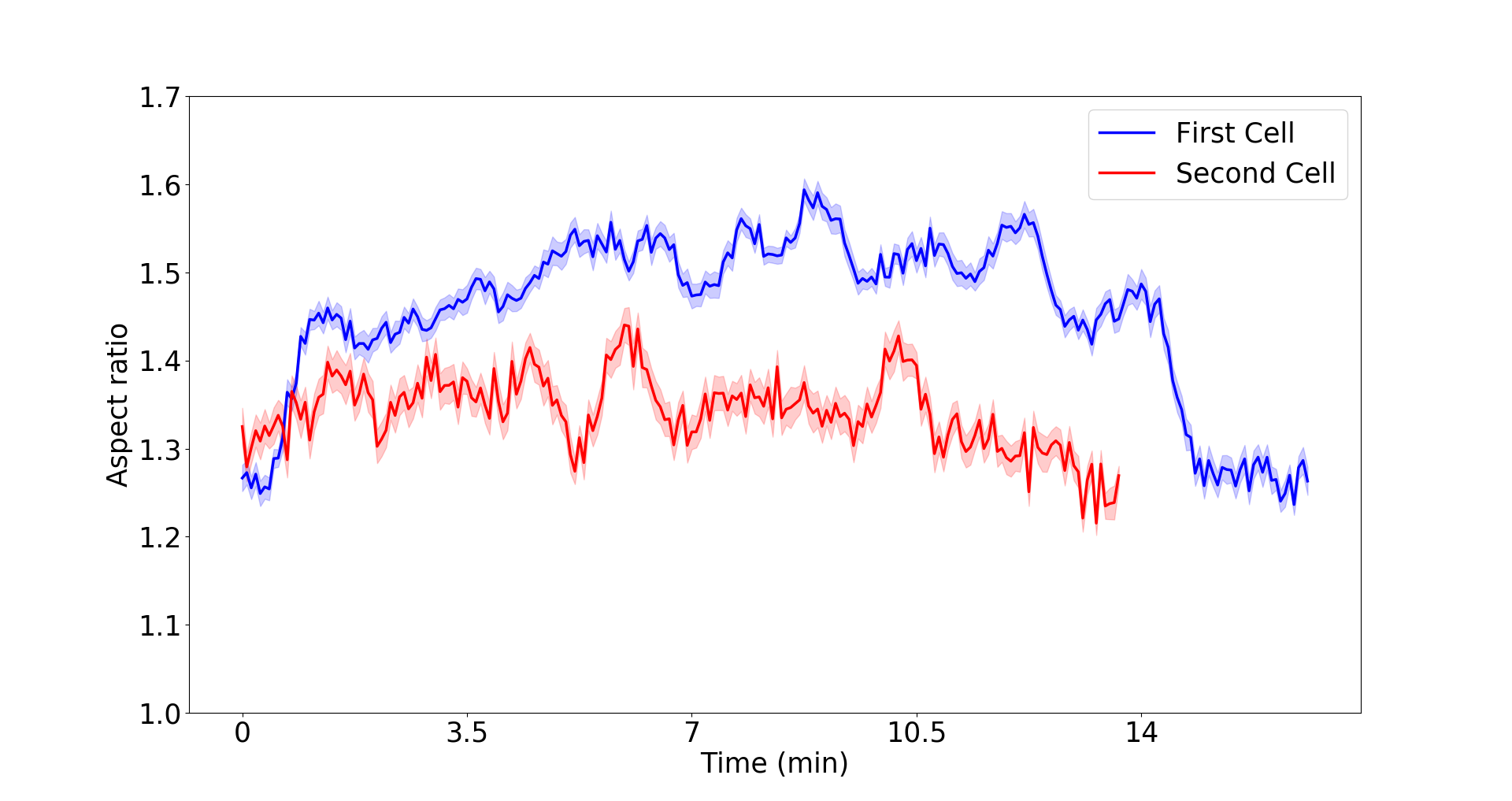}
		\label{fig_Case_2_avg_as}}
	\caption{In each case, we ran $50$ simulations in total and we take the average speed (subfigures (a) and (b)), circularity (subfigures (c) and (d)) and aspect ratio ((subfigures (e) and (f)) of every time step for all the cells to obtain the curves shown above. The 95\% confidence interval is also indicated as an envelop in both figures with different colors accordingly. Blue, red and black curves represent the first (leader cell), the second (follower cell) and the third cell (only existed in Case (1)), respectively.}
	\label{Fig_avg_all}
\end{figure}

\begin{table}\footnotesize
\centering
\caption{Numerical results (i.e. transmigration time and average speed of three cells) of the simulation shown in Figure \ref{Fig_Case_1_Snapshots} for Case (1).}
\begin{tabular}{m{4cm}<{\centering}m{3.5cm}<{\centering}m{3.5cm}<{\centering}m{3.5cm}<{\centering}}
	\toprule
	{\bf Cell Index $i$}& {\bf transmigration time $T^i$ $(min)$} & {\bf Average Speed $\bar{v}^i$ $(\mu m/min)$} & {\bf Migration Distance of Cell Centre $(\mu m)$} \\
	\toprule
	{\bf Cell 1} & $34.93$ & $3.99$ & $139.46$ \\
	{\bf Cell 2} & $26.53$ & $4.59$ & $121.74$ \\
	{\bf Cell 3} & $24.85$ & $4.52$ & $112.34$ \\
	\bottomrule
\end{tabular}
\label{Tbl_Case_1_time_vel}
\end{table}

\begin{table}\footnotesize
	\centering
	\caption{The maximal and minimal width of the channel after different cells exit the channel of the simulation shown in Figure \ref{Fig_Case_1_Snapshots} for Case (1).}
	\begin{tabular}{m{7cm}<{\centering}m{3.5cm}<{\centering}m{3.5cm}<{\centering}}
		\toprule
		{\bf Description}& {\bf Minimal width of the channel $(\mu m)$} & {\bf Maximal width of the channel $(\mu m)$}  \\
		\toprule
		{\bf Initial Condition} & $3.00$ & $7.00$\\
		{\bf After the first cell exits and before the second cell enters the channel} & $5.75$ & $13.09$  \\
		{\bf After the second cell exits and before the third cell enters the channel} & $6.81$ & $14.08$  \\
		{\bf After the third cell exits the channel} & $8.47$ & $14.26$  \\
		\bottomrule
	\end{tabular}
	\label{Tbl_Case_1_width}
\end{table}

\subsection{Case (2): Cell Doublet Migration}\label{Subsec_pre_Case_2}
\begin{figure}
	\centering 
		\subfigure[t = 0 min]{
		\includegraphics[width=0.45\textwidth]{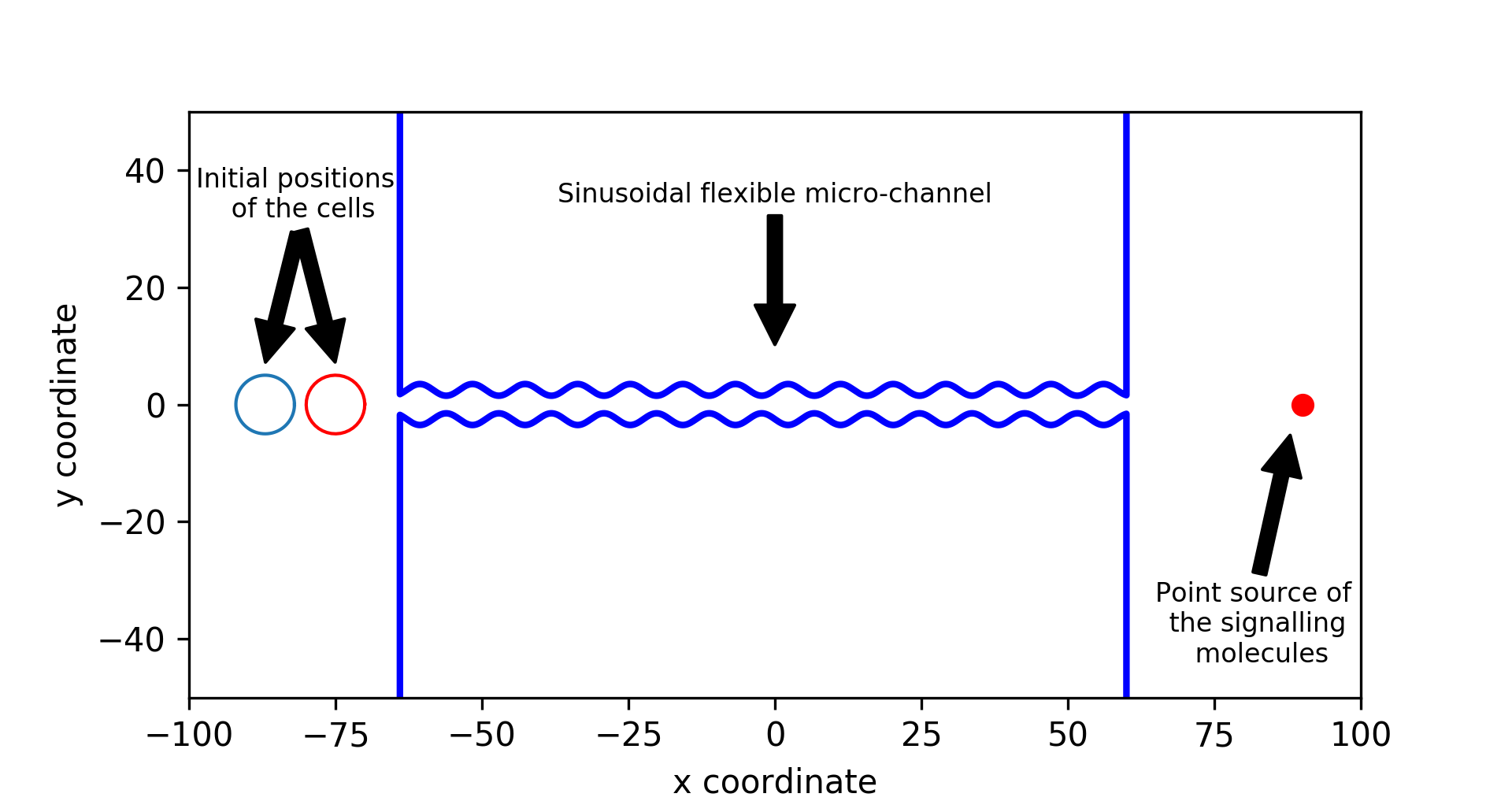}
		\label{fig_repel_initial}}
	\subfigure[t = 2.1 min]{
			\includegraphics[width=0.45\textwidth]{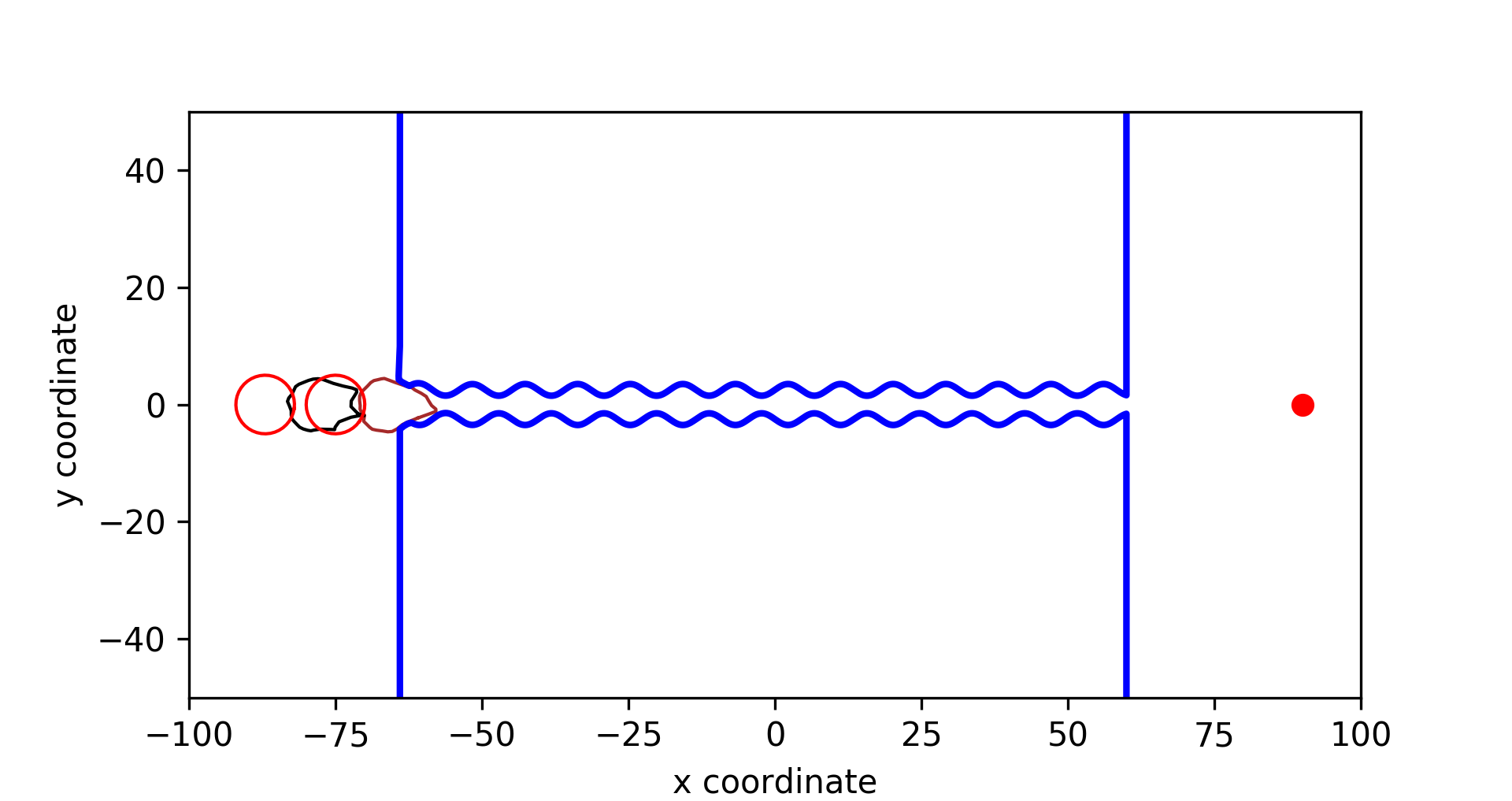}}
	\subfigure[t = 3.5 min]{
			\includegraphics[width=0.45\textwidth]{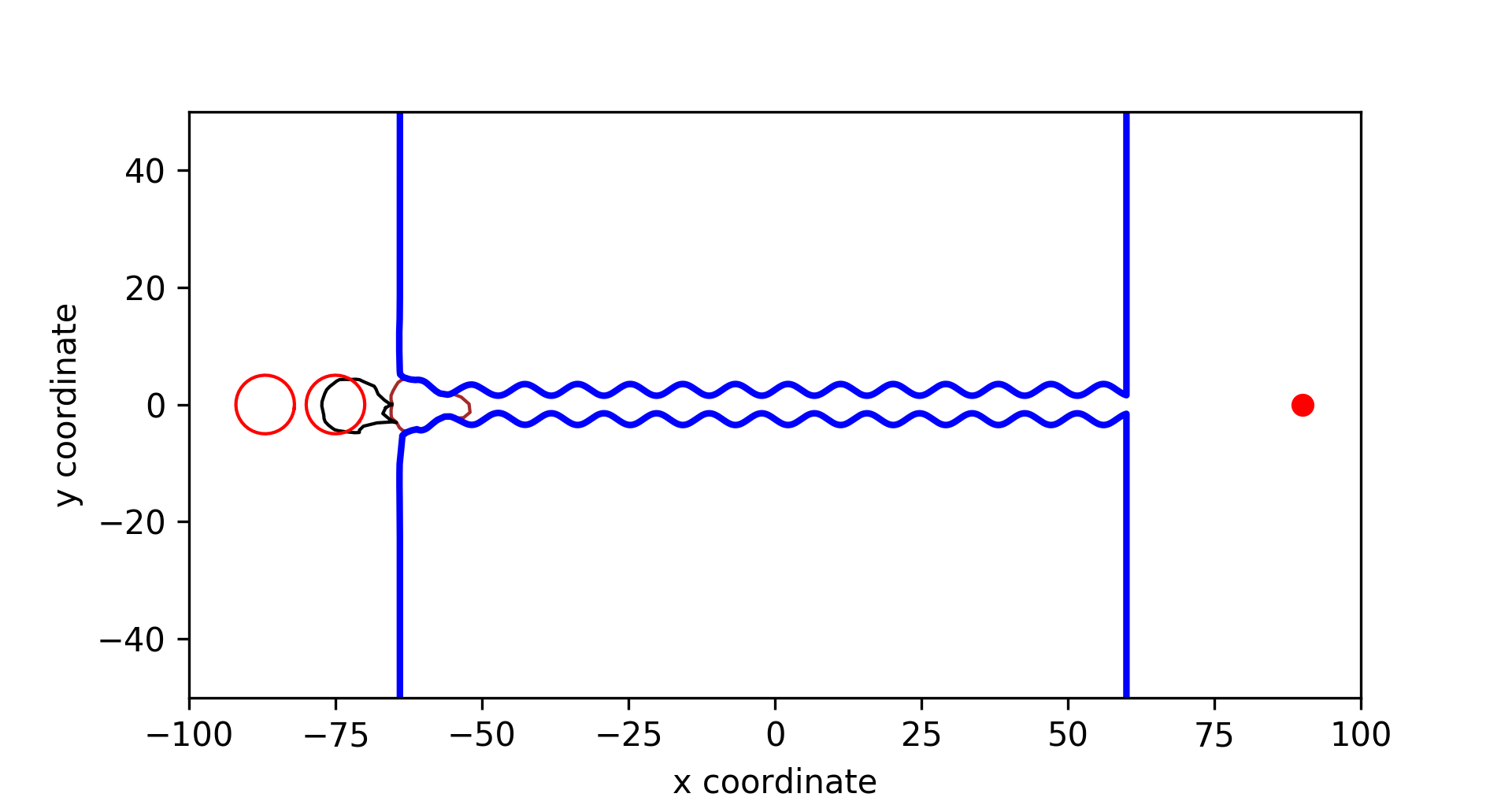}}
	\subfigure[t = 7.0 min]{
			\includegraphics[width=0.45\textwidth]{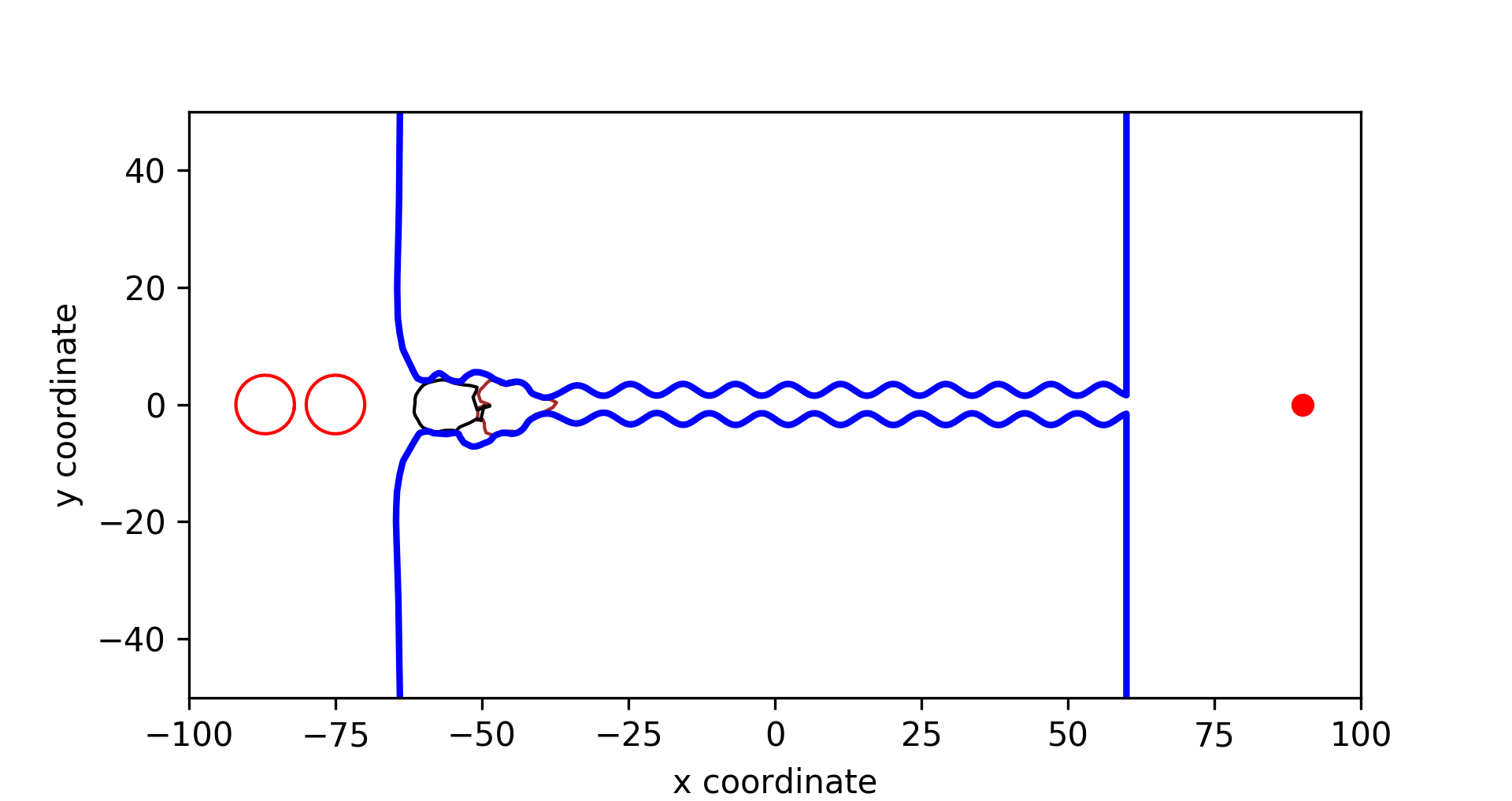}}
	\subfigure[t = 10.92 min]{
			\includegraphics[width=0.45\textwidth]{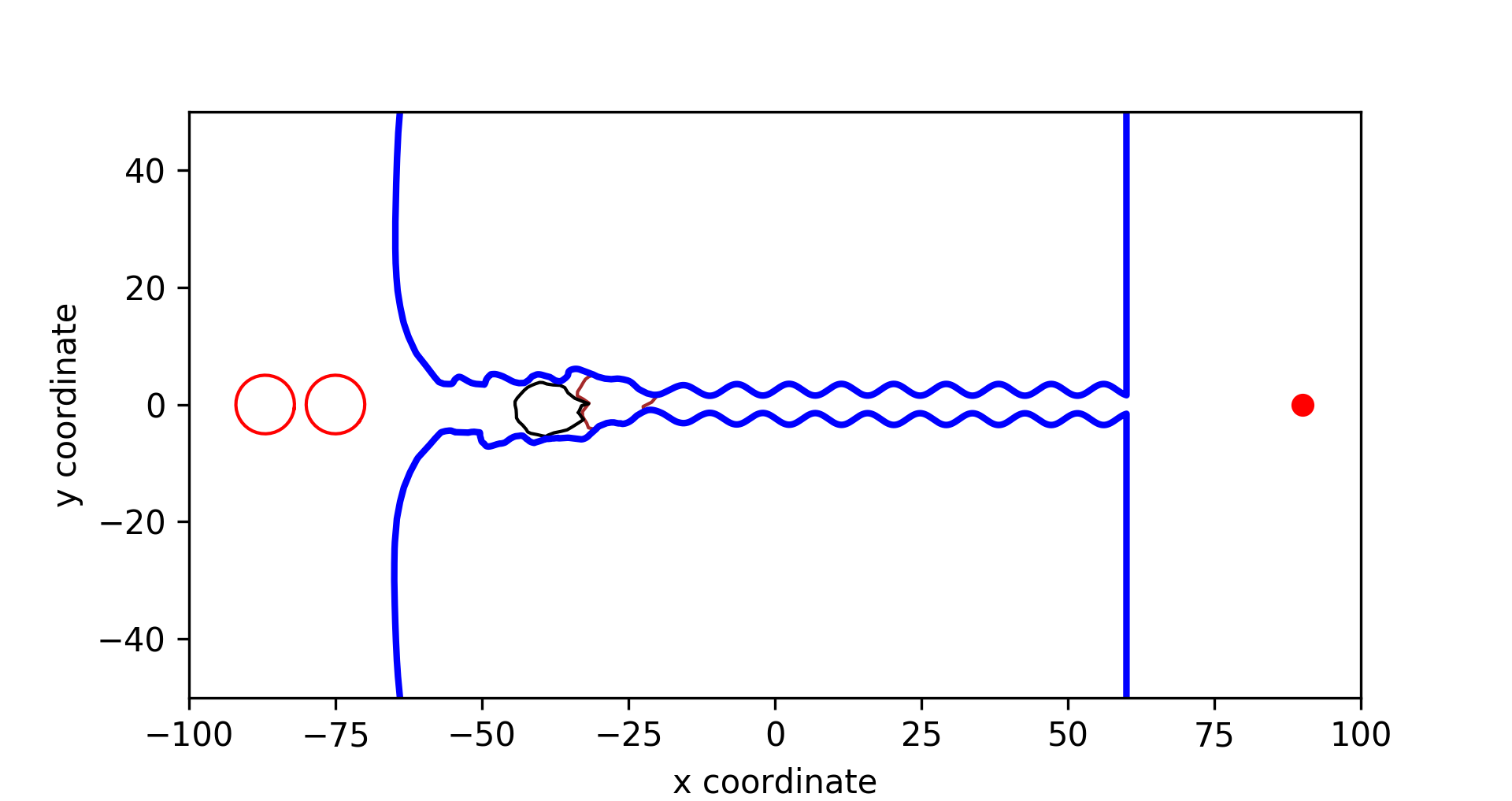}}
	\subfigure[t = 14.14 min]{
			\includegraphics[width=0.45\textwidth]{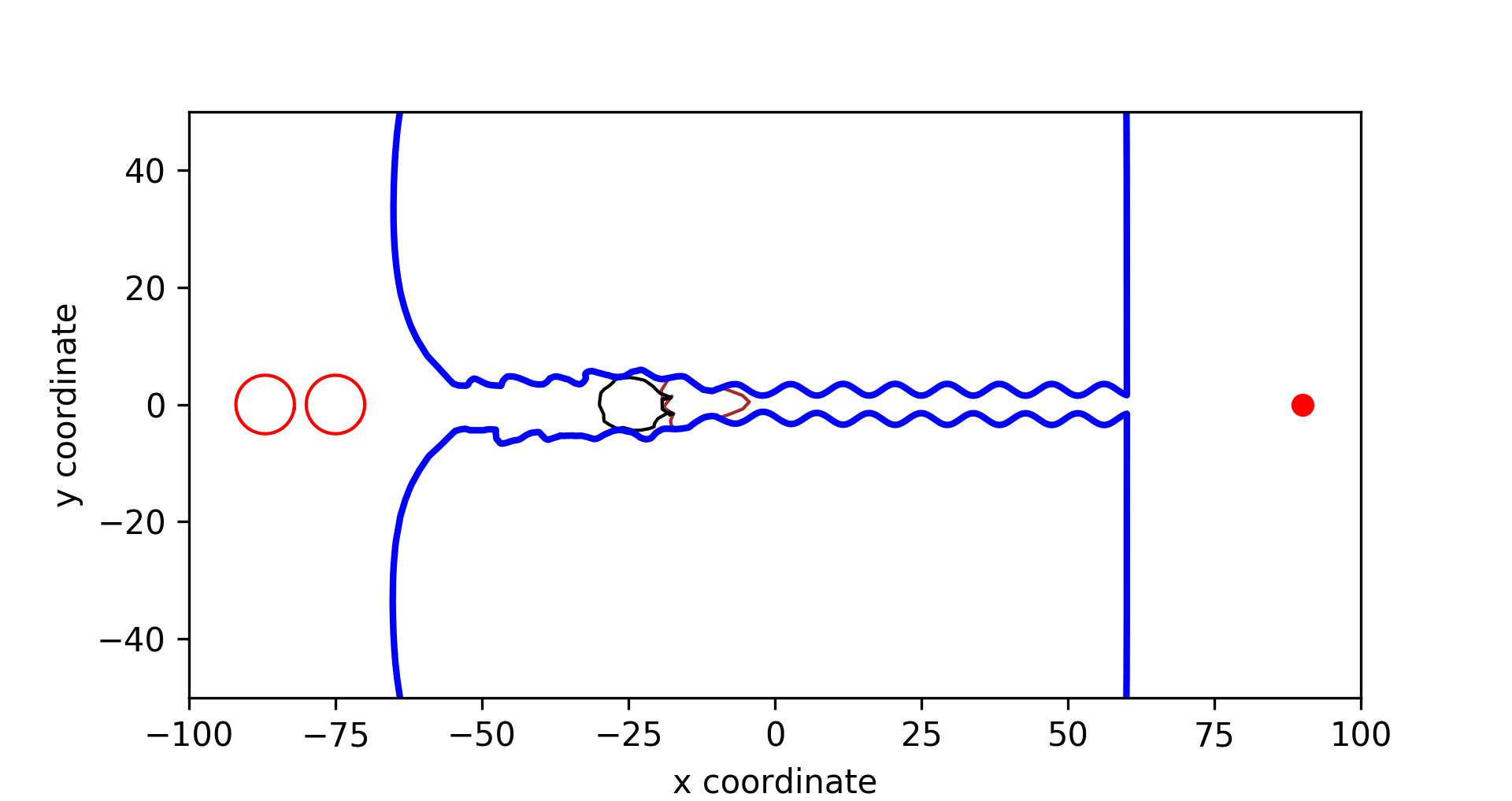}}
	\subfigure[t = 21.0 min]{
			\includegraphics[width=0.45\textwidth]{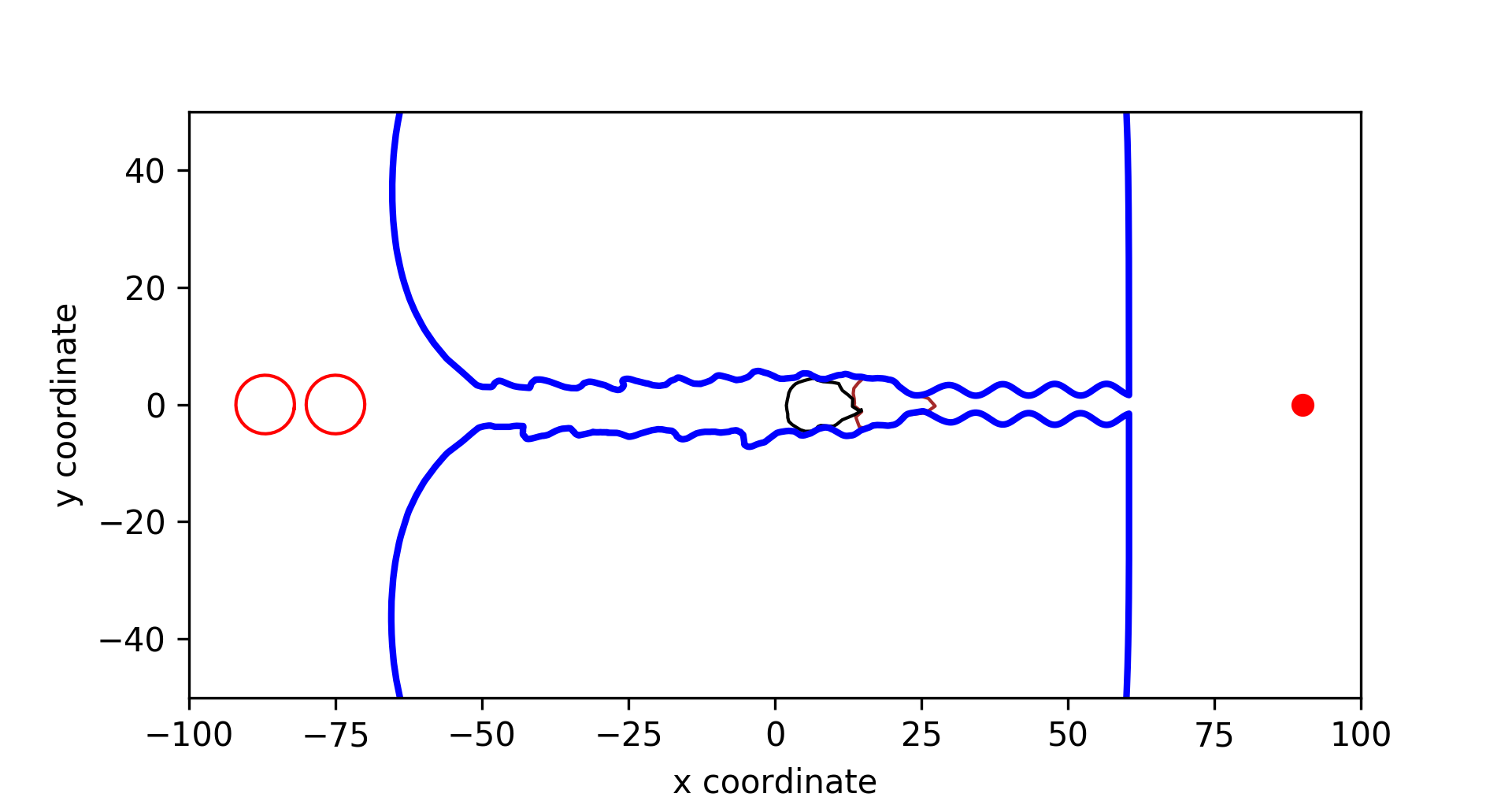}}
	\subfigure[t = 24.50 min]{
			\includegraphics[width=0.45\textwidth]{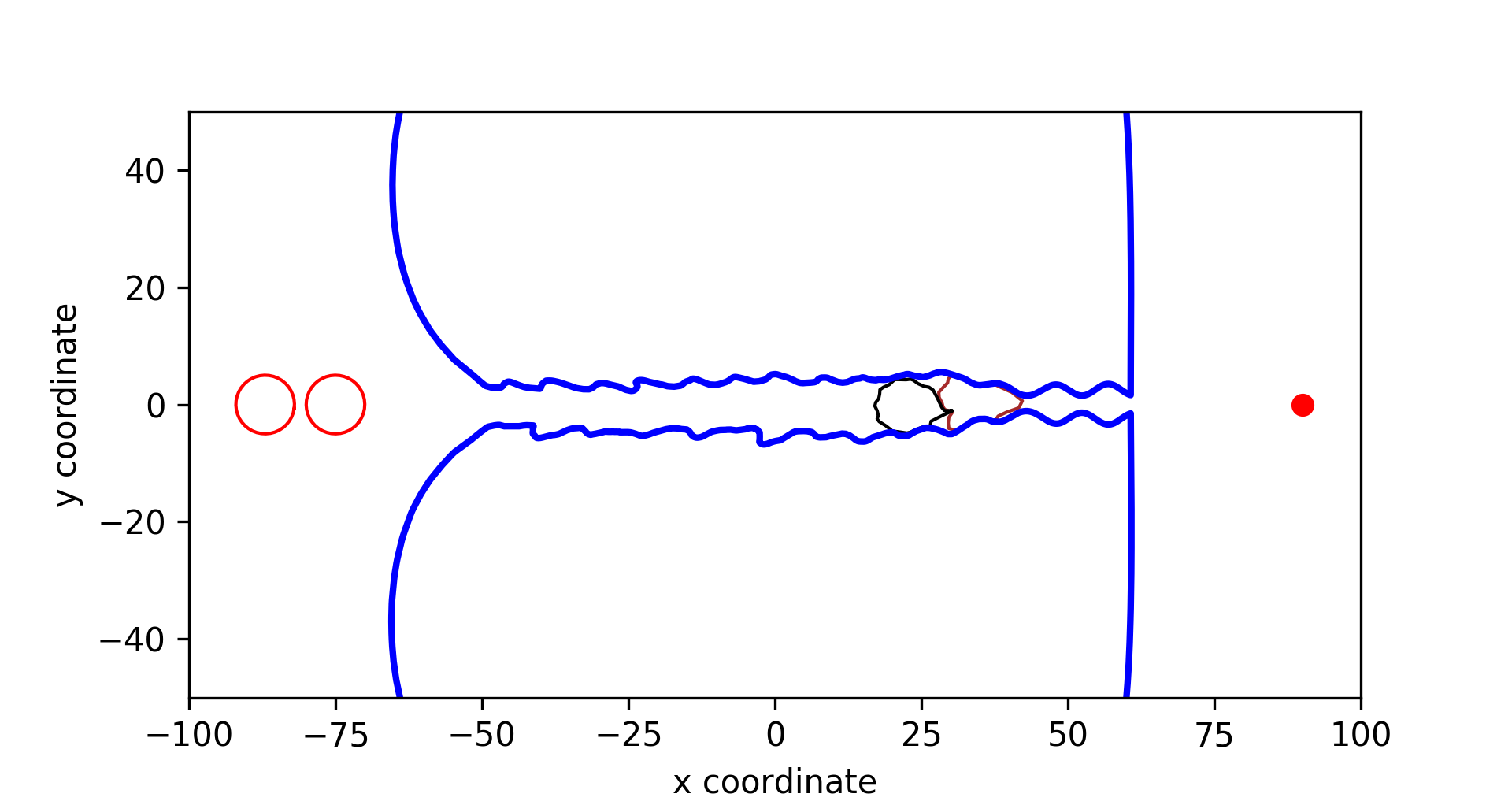}}	
	\subfigure[t = 31.50 min]{
			\includegraphics[width=0.45\textwidth]{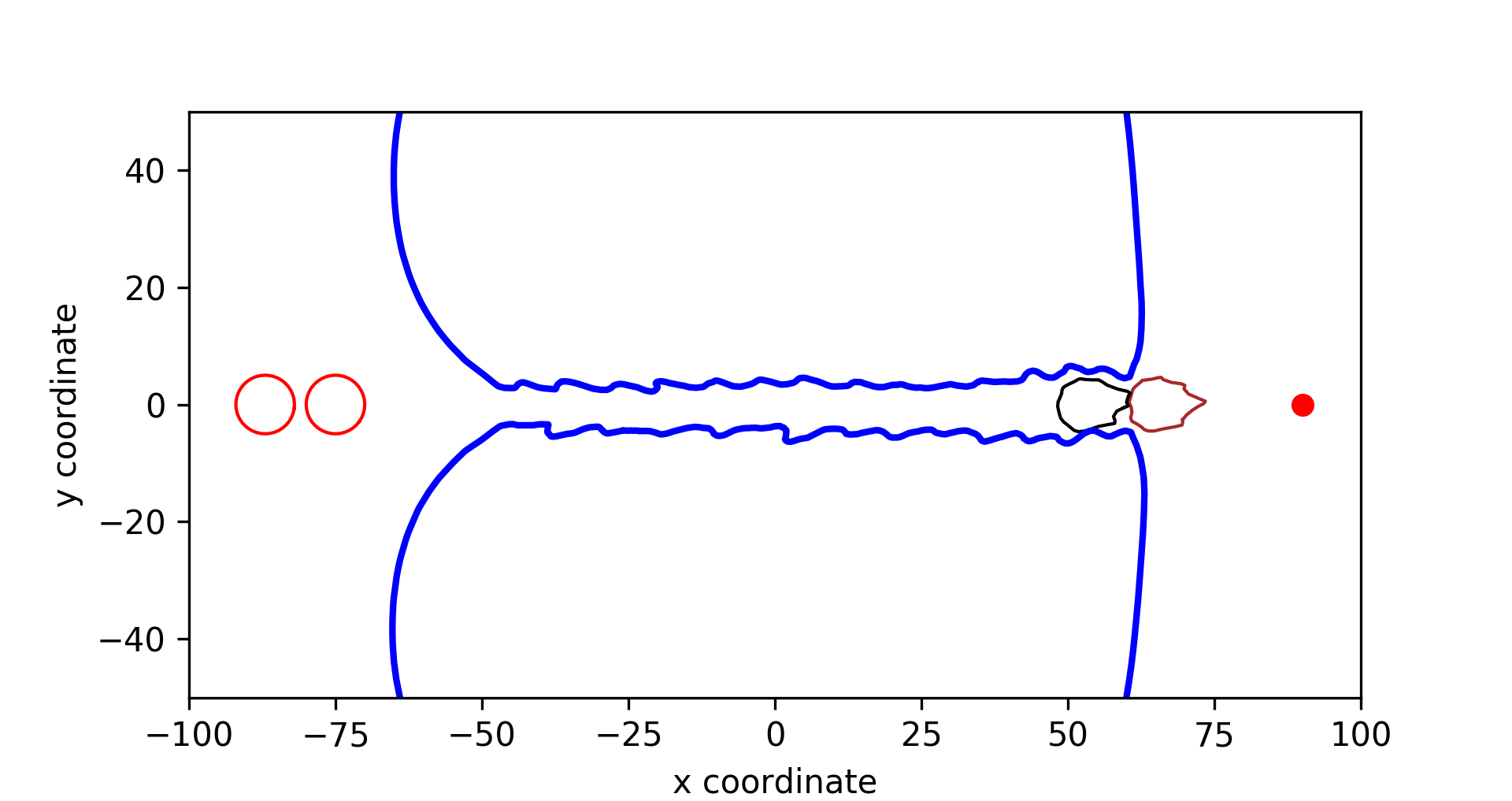}}
	\subfigure[t = 34.37 min]{
		    \includegraphics[width=0.45\textwidth]{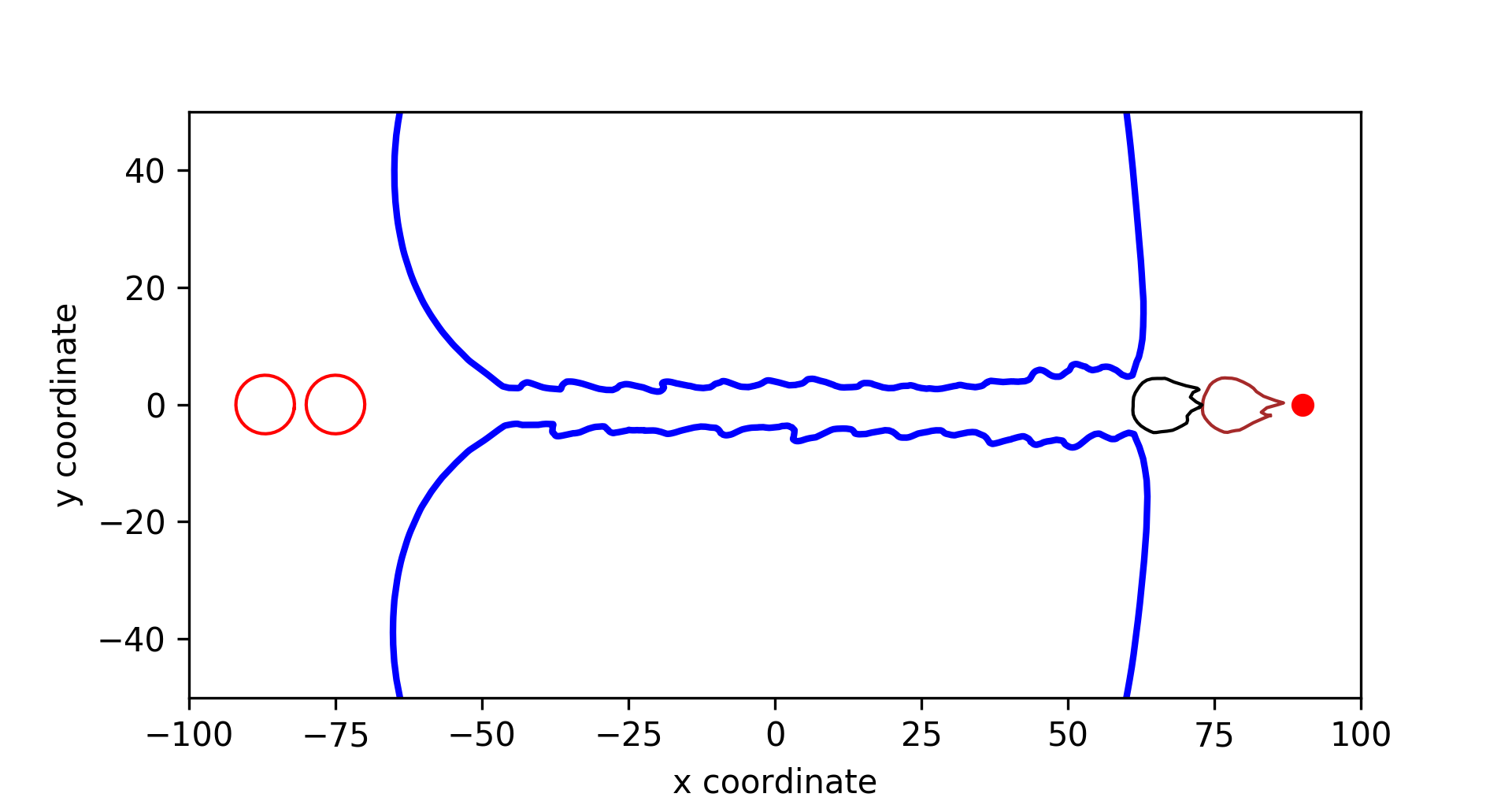}}
	 \caption{The screenshots of the simulation for Case (2), where cells follow each other to enter the channel together.}
	 \label{Fig_Case_2_Snapshots}
\end{figure}

\noindent
We consider the second case, where the follower cell follows the leader cell immediately entering the channel. That is, there are more than one cell in the computational domain at the same time, which is so-called cell doublets in in-vitro experiments \citep{Zhang2021-yu}. The setting of cell doublets is a simplified type of collective migration. Further, this experimental setting (see Figure \ref{Fig_Case_2_Snapshots}\subref{fig_repel_initial}) results into the possibility that cells collide against each other. We assume here that cells can also exert force on the other cell as long as they are compressed. For a better visualization, we refer to the Supplementary Material Video 2, attached to this manuscript. 

Figure \ref{Fig_Case_2_Snapshots} shows several snapshots at consecutive times of the simulation, and Figure \ref{Fig_avg_all}\subref{fig_Case_2_avg_vel} illustrates the speed of the two cells inside the channel. The results of the simulations are summarized in Tables \ref{Tbl_Case_2_time_vel} and \ref{Tbl_Case_2_width}. The difference between the average speeds of the follower cell and the leader cell is very small, and due to the randomness involved in the model, these differences are not significant from a statistical point of view. 
Another, more important, effect that we see is that the average speeds of both the follower and leader cells are about the same as the average of the follower cell in Case (1), which has increased compared to the average speed of the original leader cell. This fact clearly shows that collective cell migration is beneficial for the cells. This also demonstrates that cancer metastasis benefits from collective cell migration. Similarly to the conclusion we drew in the previous section, Figure \ref{Fig_avg_all}\subref{fig_Case_2_avg_CSI} and \subref{fig_Case_2_avg_as} display the circularity and the aspect ratio of both cells. Different from what we saw in Case (1), there is a significant difference regarding the circularity and the aspect ratio of the follower cell and the leader cell: the follower cell alters its equilibrium shape much less than the leader cell, as both indices of the follower cell are closer to $1$ than for the leader cell. Although, we do not explicitly take energy into account in the current model, we expect from these simulations that the follower cell consumes less energy than the leader cell. 

\begin{table}\footnotesize
	\centering
	\caption{Numerical results (i.e. transmigration time and average speed of two cells) of the simulation shown in Figure \ref{Fig_Case_2_Snapshots} for Case (2).}
	\begin{tabular}{m{4cm}<{\centering}m{3.5cm}<{\centering}m{3.5cm}<{\centering}m{3.5cm}<{\centering}}
		\toprule
		{\bf Cell Index $i$}& {\bf transmigration time $T^i$ $(min)$} & {\bf Average Speed $\bar{v}^i$ $(\mu m/min)$} & {\bf Migration Distance of Cell Centre $(\mu m)$} \\
		\toprule
		{\bf Cell 1} & $30.87$ & $4.48$ & $138.21$ \\
		{\bf Cell 2} & $30.73$ & $4.46$ & $137.12$ \\
		\bottomrule
	\end{tabular}
	\label{Tbl_Case_2_time_vel}
\end{table}

\begin{table}\footnotesize
	\centering
	\caption{The maximal and minimal width of the channel after different cells exit the channel of the simulation shown in Figure \ref{Fig_Case_2_Snapshots} for Case (2).}
	\begin{tabular}{m{7cm}<{\centering}m{3.5cm}<{\centering}m{3.5cm}<{\centering}}
		\toprule
		{\bf Description}& {\bf Minimal width of the channel $(\mu m)$} & {\bf Maximal width of the channel $(\mu m)$}  \\
		\toprule
		{\bf Initial Condition} & $3.00$ & $7.00$\\
		{\bf After cell doublets exist the channel} & $5.82$ & $13.72$  \\
		\bottomrule
	\end{tabular}
	\label{Tbl_Case_2_width}
\end{table}

\subsection{Comparison between the Two Cases}\label{Subsec_pre_Case12}
\begin{figure}
	\centering 
	\includegraphics[width=0.85\textwidth]{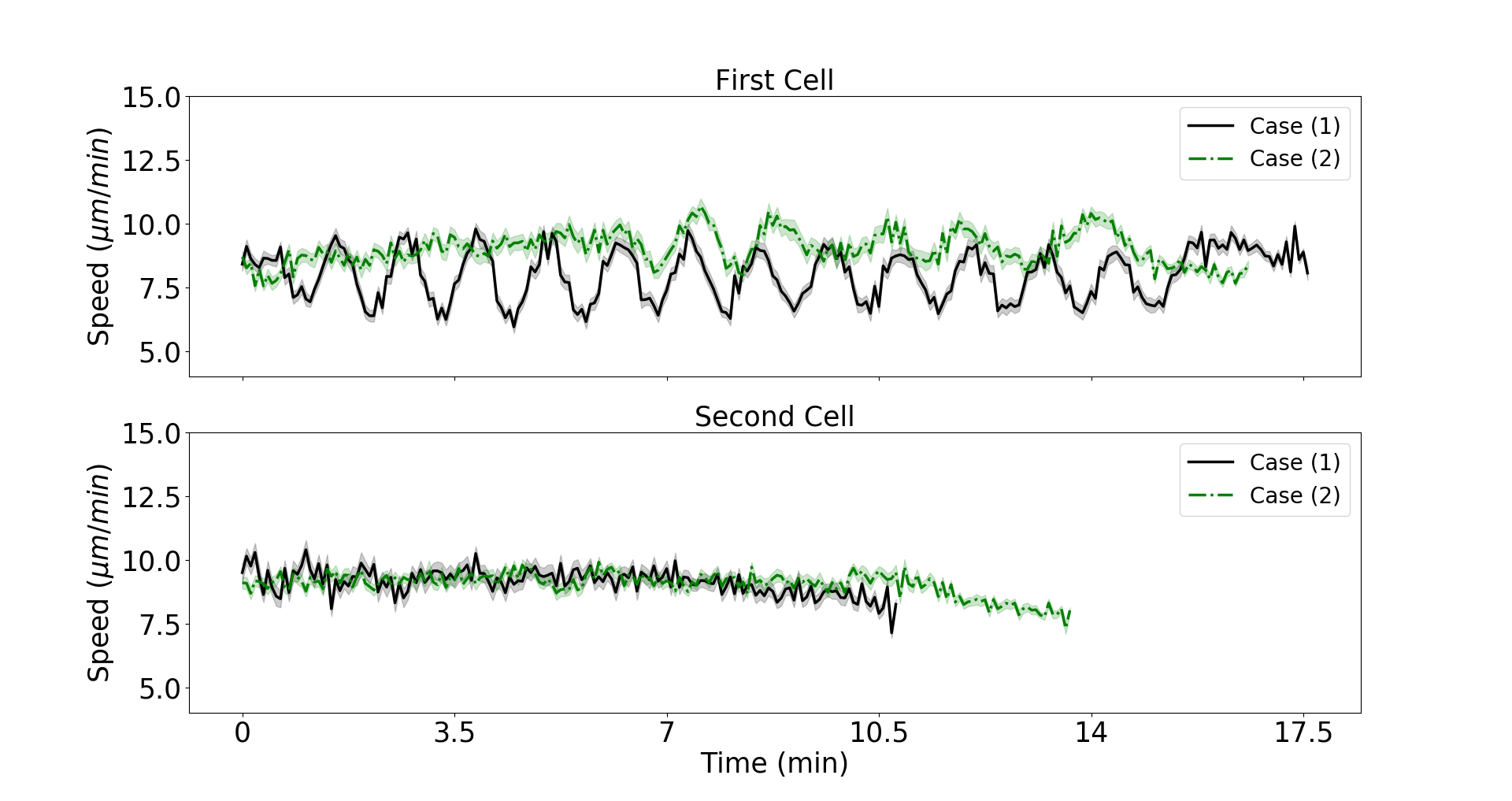}
	\caption{The time series of the average cell speed of Cell $1$ and Cell $2$ in Case (1) and Case (2). Black solid curve and green dashed curve represent Case (1) and Case (2), respectively. The translucent envelops around the curves are the corresponding $95\%$ confidence interval.}
	\label{Fig_both_avg_vel}
\end{figure}

\noindent
From Table \ref{Tbl_Case_1_time_vel} and \ref{Tbl_Case_2_time_vel}, we observe that there is a significant increase in the average speed of Cell 1 in Case (2). To have a clearer view, we plot the average of the time series of the velocity of Cell $1$ and Cell $2$ that are collected from $50$ simulations with the same input parameters in Table \ref{Tbl_ParaValue_All}; see Figure \ref{Fig_both_avg_vel} for the results. It can be concluded that for the selected parameter set, the leader cell (Cell $1$) is moving faster in Case (2) than Case (1), as the follower cell in Case (2) exerts the repelling force, while for the follower cell (Cell $2$), there is hardly any difference with respect to the speed, however, in Case (2), Cell $2$ needs more time to transmigrate through the channel completely since the leader cell is blocking the channel. However, the most important conclusion so far is that in Case (2), both the leading and follower cells transmigrate faster than the leader cell in Case (1). This shows that collective cell movement can increase metastatic  migration rates, where we illustrated this behavior for two cell-interaction modes: (1) the leader cell 'paves' the way through for the follower cell (Case (1)), and (2) the cells interact by pushing each other thereby producing a net larger joint force onto the channel walls (Case (2)).

Regarding the invasiveness, we focus on the channel width {\it after} the leader cell (Cell 1 in Case (1)) or the cell doublet (both Cell 1 and Cell 2 in Case (2)) have exited the channel completely. The results can be found in Table \ref{Tbl_Case_1_width} and Table \ref{Tbl_Case_2_width}. Both minimal and maximal width of the channel in Case (2) ($5.82$ and $13.72\ \mu m$, respectively) are wider than in Case (1) ($5.75$ and $13.09\ \mu m$, respectively), since in Case (2), both cells exert forces on the channel walls at the same time and within a certain distance. This implies that cell doublets enhance the invasiveness and increase its speed as compared with single cell, which is in line with experimental observations by \citet{Merkher2017-va,Tulchinsky2022-vf}.

The aforementioned observations hold for the current set of input values. Next, we will vary the input parameters according to statistical distributions and see whether the above conclusions are general.

\section{Monte Carlo Simulations}\label{Sec_MC_simu}
\noindent
In Section \ref{Sec_pre_results}, we considered series of $50$ simulations with the same input parameters. To investigate whether the conclusions (the potential existence of an optimal width of the channel that increases migration speed under for interacting cells) from Section 3 hold in general, Monte Carlo simulations are performed where the input parameters are subject to the statistical distributions that are given in Table \ref{Tbl_MC_all} and the other constant parameters are the same as in Table \ref{Tbl_ParaValue_All}. In this section, we will show box-plots of the average cell speeds, and conduct the Wilcoxon test to statistically investigate whether the follower cell and leader cells benefit from each other for both Case (1) and Case (2) for the parametric ranges that we use. Since the output parameters do not necessarily follow normal distributions, we compute the Spearman correlation coefficients \citep{Weaver2017} to determine the correlation between input-and output parameters. 
\begin{table}\footnotesize
	\centering
	\caption{Input parameters and outputs of the Monte Carlo simulations for both cases}
	\begin{tabular}{p{2.5cm}<{\centering}p{7cm}<{\centering}p{4.5cm}<{\centering}}
		\toprule
		\multicolumn{3}{l}{\bf Input variables}\\
		\midrule
		{\bf Parameters} & {\bf Description} & {\bf Predefined Distributions}\\
		\midrule
		$E_s$ & Substrate elasticity & $\text{log-normal}(\log(50), 0.5)$\\
		$E_c$ & Stiffness of the springs that connect the cell centre and the nodal point on the cell membrane & $\text{log-normal}(\log(10), 0.1)$\\
		$E_m$ & Stiffness of the springs that connect the neighboring nodal points on the cell membrane & $\text{log-normal}(\log(5), 0.05)$\\
		$\nu_s$ & Poisson's ratio of the ECM & $U(0.4, 0.5)$  \\
		$\nu_c$ & Poisson's ratio of the cells & $U(0.3, 0.4)$\\
		$\mu_f$ & Cell friction coefficient against the channel wall & $U(0.03, 0.06)$ \\
		$w$ & Angular frequency of the initial sinusoidal flexible channel in Equation (\ref{Eq_sine_channel}) & $U(0.3, 0.7)$\\
		$b$ & Midline of the initial sinusoidal flexible channel in Equation (\ref{Eq_sine_channel})& $U(2, 4)$ \\
		$\mu_{visco}$ & Weight of viscosity in viscoelasticity in Equation (\ref{Eq_morpho_visco_sigma}) & $U(0, 1)$ \\
		$\alpha$ & Degree of permanent deformation in Equation (\ref{Eq_morpho}) & $U(0,2)$\\	
		\midrule
		\multicolumn{3}{l}{\bf Outputs with respect to cell $i$, $i = 1, 2, 3^\dag$}\\
		\midrule
		$T^i$ & \multicolumn{2}{c}{\parbox{11.5cm}{The transmigration time of cell $i$ through the channel}}\\  
		$\bar{v}^i$ & \multicolumn{2}{c}{\parbox{11.5cm}{The average speed of cell $i$ penetrating through the channel}}\\  
		cell\_dis(i) & \multicolumn{2}{c}{\parbox{11.5cm}{The length of the channel that cell $i$ has migrated, as the channel is deformed due to the cellular traction forces}}\\
		cell\_max\_as(i) & \multicolumn{2}{c}{\parbox{11.5cm}{The maximal aspect ratio of cell $i$ during the penetration}}\\
		cell\_min\_c(i) & \multicolumn{2}{c}{\parbox{11.5cm}{The minimal circularity of cell $i$ during the penetration}}\\  
		\midrule
		\multicolumn{3}{l}{\bf Outputs with respect to width of the channel, where $i = 1,2,3^\dag$}\\
		\midrule
		min\_width\_0 & \multicolumn{2}{c}{\parbox{11.5cm}{The minimal width of the initial channel}}\\
		min\_width(i) & \multicolumn{2}{c}{\parbox{11.5cm}{The minimal width of the channel \textit{after} cell $i$ completely exits the channel, that is, \textit{before} cell $i+1$ enters the channel}}\\
		max\_width\_0 & \multicolumn{2}{c}{\parbox{11.5cm}{The maximal width of the initial channel}}\\
		max\_width(i) & \multicolumn{2}{c}{\parbox{11.5cm}{The maximal width of the channel \textit{after} cell $i$ completely exits the channel, that is, \textit{before} cell $i+1$ enters the channel}}\\
		\bottomrule
		\multicolumn{3}{l}{\scriptsize $^\dag i =3$ is only applicable in Case (1) in the Monte Carlo simulations but not in Case(2).}
	\end{tabular}
	\label{Tbl_MC_all}
\end{table}
In this section, we compute the Spearman correlation \citep{Weaver2017} as it does not request the data to obey the normal distribution. Furthermore, we did the Shapiro test \citep{shapiro1990test} , which evidenced that most output data does not follow a normal distribution. 

\subsection{Case (1): Single Cell Migration}\label{Subsec_MC_Case_1}
\begin{figure}
	\centering 
	\includegraphics[width=0.7\textwidth]{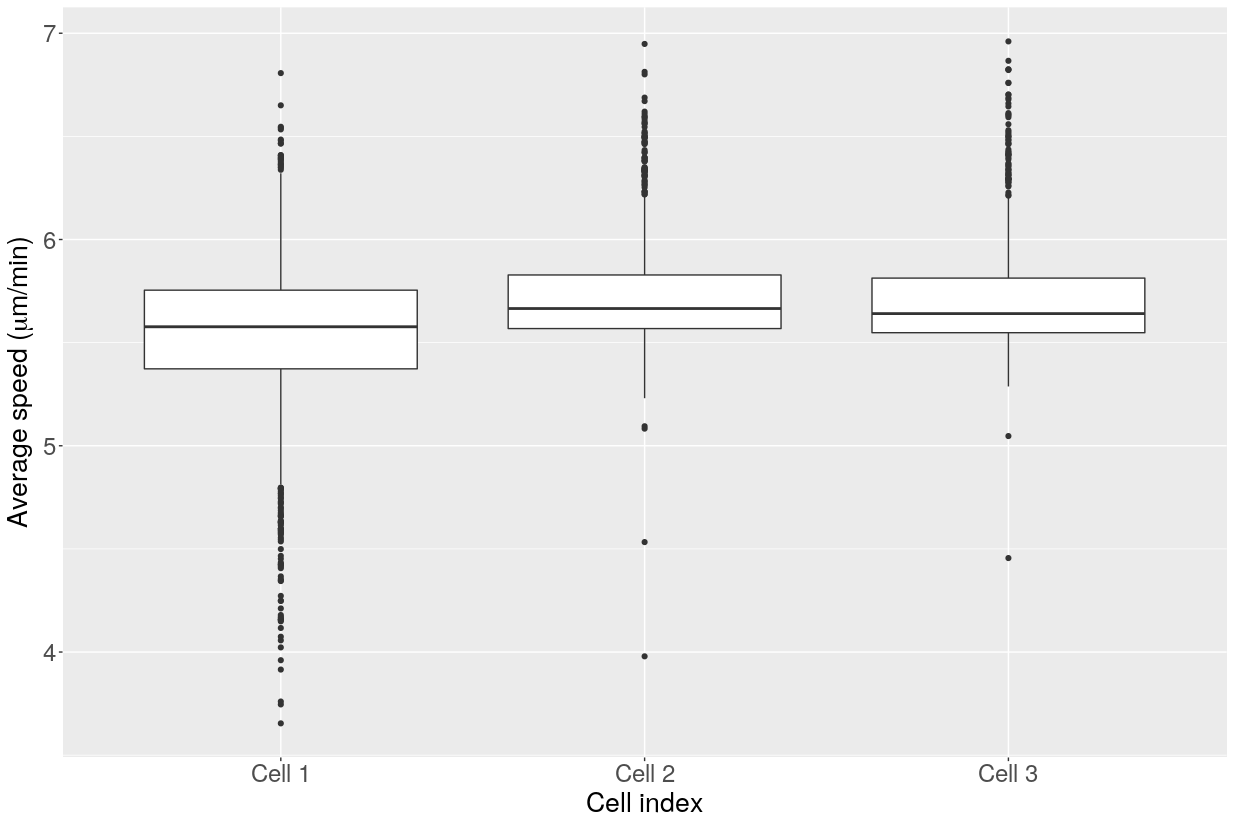}
	\caption{The figure shows the Monte Carlo results of Case (1), in particular, the results of the average velocity of the three cells. The box-plot of the average velocity of the two cells, where the median, 25\% (Q1) and 75\% (Q3) percentile (lower and upper bound of the box), the minimum and the maximum of the data excluding the outliers and the outliers are shown. }
	\label{Fig_MC_Case1_vel}
\end{figure}

\begin{figure}
	\centering
	\includegraphics[width=0.65\textwidth]{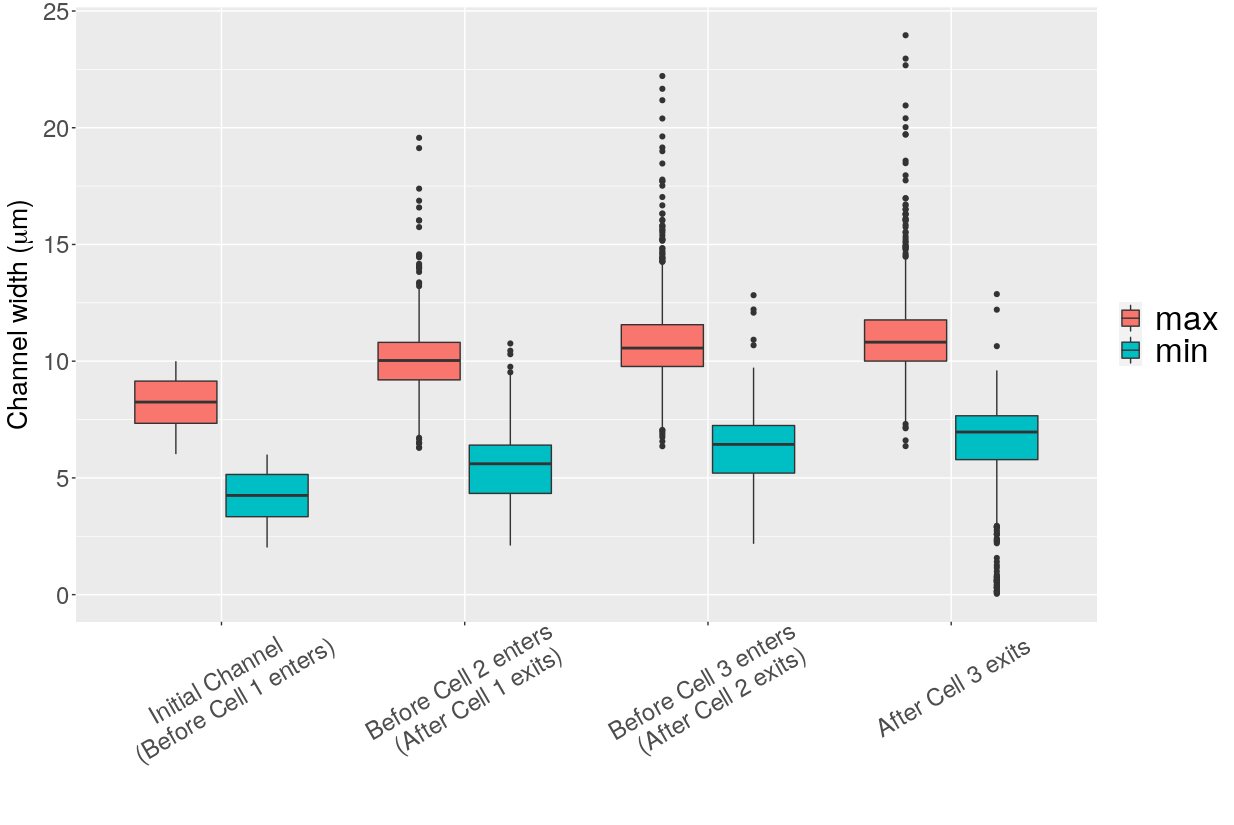}
	\caption{The figures show the Monte Carlo results of Case (1), in particular, the results regarding the maximal and the minimal width of the channel after each cell exits the channel completely. The maximal and minimal width of the flexible channel are shown as orange and green boxes, respectively, at certain moments.  }
	\label{Fig_MC_Case1_width_cell_vel}
\end{figure}

\noindent
In total, $1019$ samples are collected with all the input and output variables mentioned in Table \ref{Tbl_MC_all}. As the channel is deforming continuously, the average speed of the cell during the penetration is more appropriate to investigate. Figure \ref{Fig_MC_Case1_vel} illustrates the average speed of all the cells and the scatter plot between the first and second cells. It is hard to conclude from the box-plot (Figure \ref{Fig_MC_Case1_vel}) whether there is a significant difference between cells. Therefore, Wilcoxon's test is used and the results are shown in Table \ref{Tbl_Wilcox_Case1_cell_vel}. As the leader cell, Cell 1 does help the follower cells (Cell 2 and Cell 3) move faster since the p-value from the Wilcoxon's test is almost $0$. However, Cell 3 moves more slowly than Cell 2, which can be explained by the energy consumption and the confinement: from Figure \ref{Fig_MC_Case1_width_cell_vel}, it can be seen that after Cell 2 has transmigrated through the channel, the channel is expanded more, which results in less confinement for the cell and then the aspect ratio of the cell is smaller; we speculate from the perspective of the energy consumption, that a cell may prefer to stay mostly in its original equilibrium rather than migrate faster, to some extent. Furthermore, we see a relatively strong correlation between the speeds of each cell; see Figure \ref{Fig_MC_Case1_vel}. As a result, to investigate how the properties of the substrate and the flexible channel influence the cell displacement, we will mainly focus on Cell 1.

In Figure \ref{Fig_MC_Case1_width_cell_vel}, we investigate how the width of the flexible channel before the cell enters, affects the average velocity of the cell respectively. As it is expected, once every cell leaves the channel, it contributes to widening the channel in general. In Figure \ref{Fig_scatter_min_width_cell_Case_1}\subref{fig_min_width0_cell1_vel} - \subref{fig_min_width2_cell3_vel}, the correlation for Cell 1 is opposite for the other two cells. Before Cell 1 enters the channel, the minimal width of the channel can be much smaller than the cell size. According to our model assumptions that the cell area can only alter at most 10\%, the cell has to become more elongated, which results into a larger aspect ratio. Therefore, the cell should move faster. However, in the meantime, a narrow channel also increases the friction from the channel on the cell, and it requires more energy from the cell to deform more, which actually slows down the cell --- that is why Figure \ref{Fig_scatter_min_width_cell_Case_1}\subref{fig_min_width0_cell1_vel} shows a parabolic behavior and it shows an opposite tendency compared to Figure \ref{Fig_scatter_min_width_cell_Case_1}\subref{fig_min_width1_cell2_vel} and \subref{fig_min_width2_cell3_vel}. Once the leader cell (Cell 1) expands the channel to a width that is comparable to the cell size (see Figure \ref{Fig_MC_Case1_width_cell_vel}), the follower cells (Cell 2 and Cell 3) do not need to deform as much as before, thus, they stay more akin the equilibrium shape and less elongated since cells are not confined. Subsequently, a decreasing tendency appears in Figure \ref{Fig_scatter_min_width_cell_Case_1}\subref{fig_min_width1_cell2_vel} and \subref{fig_min_width2_cell3_vel}. In other words, there should exist an optimum of the width of the channel, such that the cell moves fastest with a combination of a relatively large aspect ratio and a relatively small deformation--- this is verified by the fact that Cell 2 moves significantly faster than Cell 3 in the Wilcoxon's test ($\text{p-value} = 8.28\times10^{-47}$) in Table \ref{Tbl_Wilcox_Case1_cell_vel}. 

%

\subsection{Case (2): Cell Doublet Migration}\label{Subsec_MC_Case_2}
\begin{figure}
	\centering 
	\includegraphics[width=0.7\textwidth]{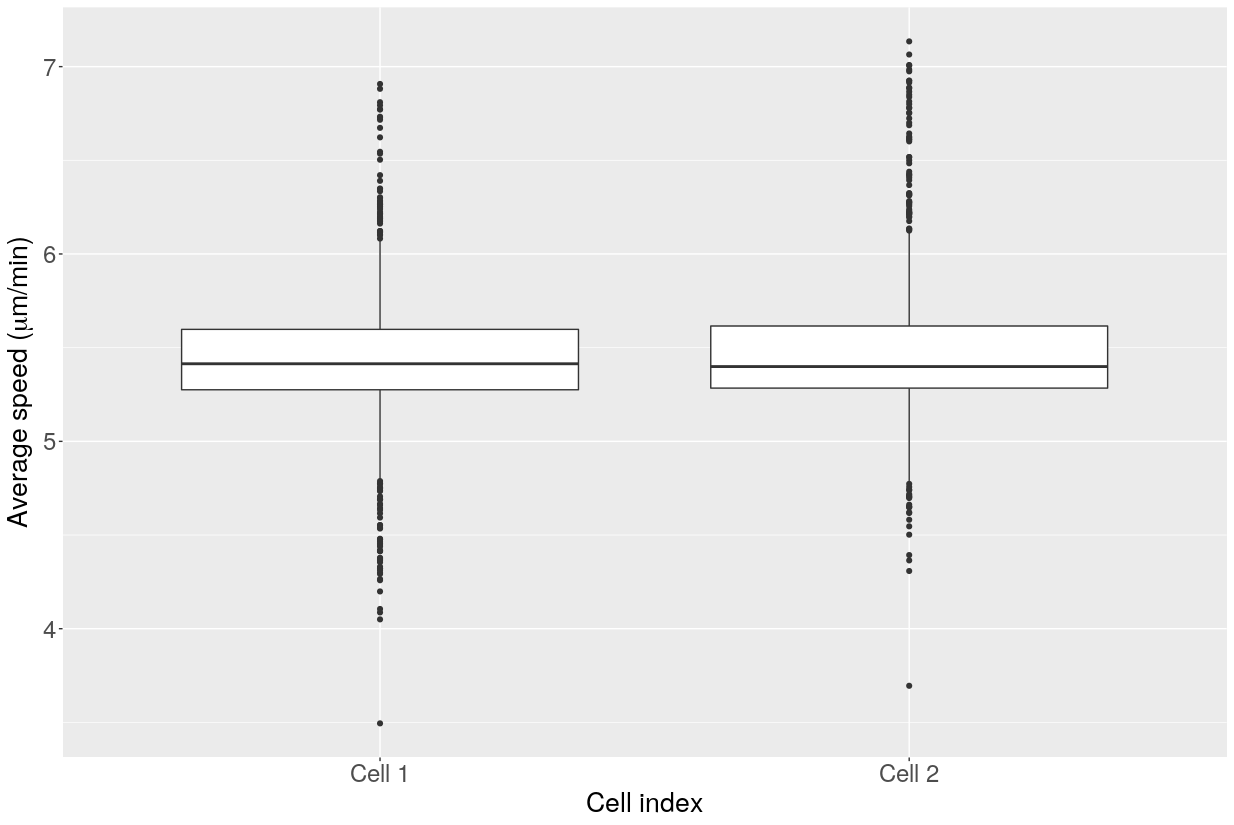}
	\caption{The figure shows the Monte Carlo results of Case (2), in particular, the results of the average speed of the three cells. The box-plot of the average velocity of all three cells, where the median, 25\% (Q1) and 75\% (Q3) percentile (lower and upper bound of the box), the minimum and the maximum of the data excluding the outliers and the outliers are shown.}
	\label{Fig_MC_Case2_vel}
\end{figure}

\noindent
For Case (2), since the leader cell (Cell 1) and the follower cell (Cell 2) are entering the channel at about the same time, the data regarding the channel width after the leader cell exits is less interesting. There are $1044$ samples collected from the Monte Carlo simulations of Case (2) and similarly, we start with analyzing the average speed of the cells. From Figure \ref{Fig_MC_Case2_vel}, there is hardly any difference between the average speed of the cells, which is similar to the preliminary results in the previous section. The Wilcoxon's test verifies that two cells are moving at more or less the same velocity with $\text{p-value}=0.9999$ ; see Table \ref{Tbl_Wilcox_Case2_cell_vel}. In the meantime, there exists a strong positive correlation between the average velocity of two cells ($\corr = 0.86$).

Two cells are moving together through the channel and they will definitely collide, since from Case (1), we have concluded that the follower cell benefits from the expanded channel that is caused by the leader cell. Because of the collisions, even though the follower cell is able  to move faster, the leader cell blocks the way, hence, the average velocity is more or less the same between two cells. 

\subsection{Cell Doublets enhance Cancer Cell Invasiveness and Migration}\label{Subsec_MC_Case12}
\noindent
As mentioned in Section 3, we are interested in knowing whether the model predicts that the leader cell can also benefit from the the follower cell, and vice versa, or whether the follower cells is slowed down by the leader cell. To this extent, we analyze the model setting in Case (2). To achieve this, we collect the data from the Monte Carlo simulation such that every dataset is paired, and each paired dataset has exactly the same input parameters for the Monte Carlo simulations for both cases. Subsequently, the results collected from the both cases are comparable. The distributions of the input parameters are shown in Table \ref{Tbl_MC_all}, whereas the main interesting outputs are the average velocities of Cell 1 and Cell 2 in the both cases. In this section, $1000$ paired samples are collected. 

\begin{figure}
	\centering
	\subfigure[Cell speed]{
	\includegraphics[width=0.48\textwidth]{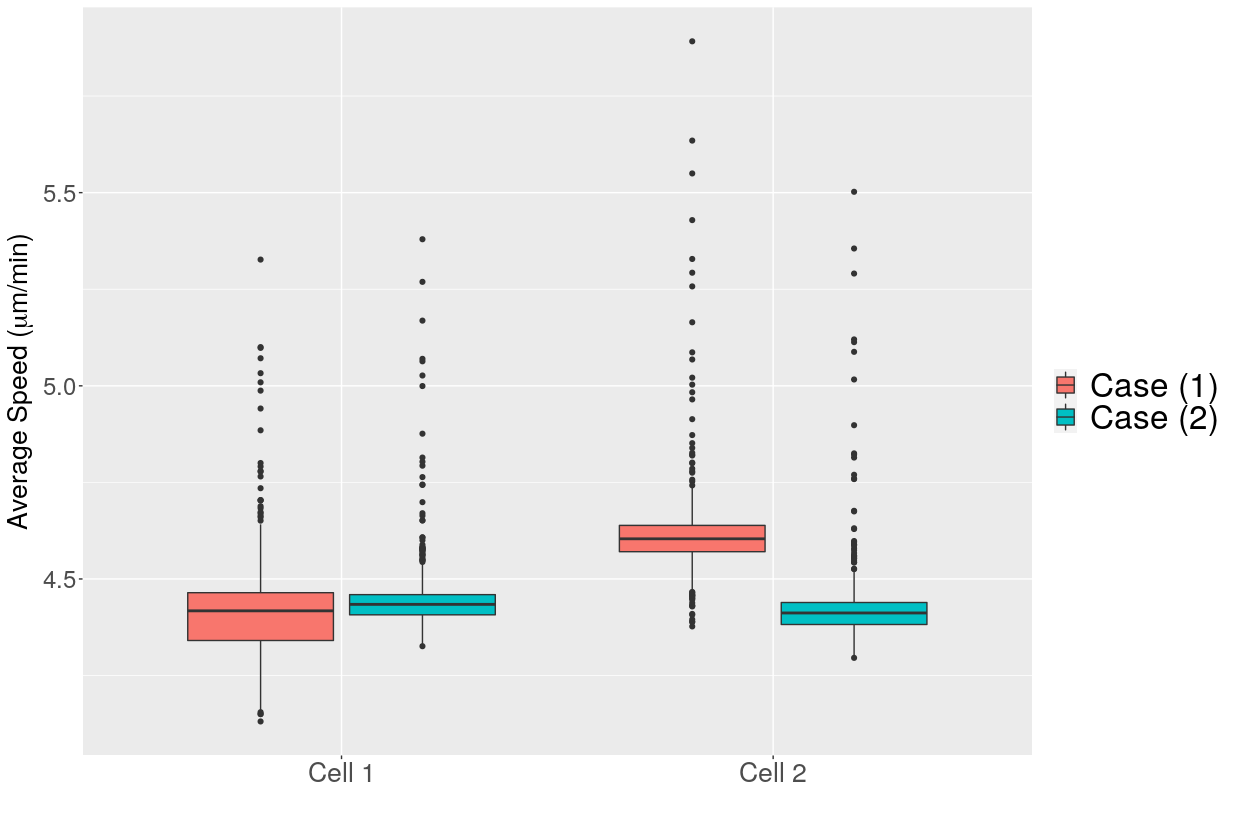}
	\label{fig_Case12_Cell12_vel}}
	\subfigure[Channel width]{
	\includegraphics[width=0.48\textwidth]{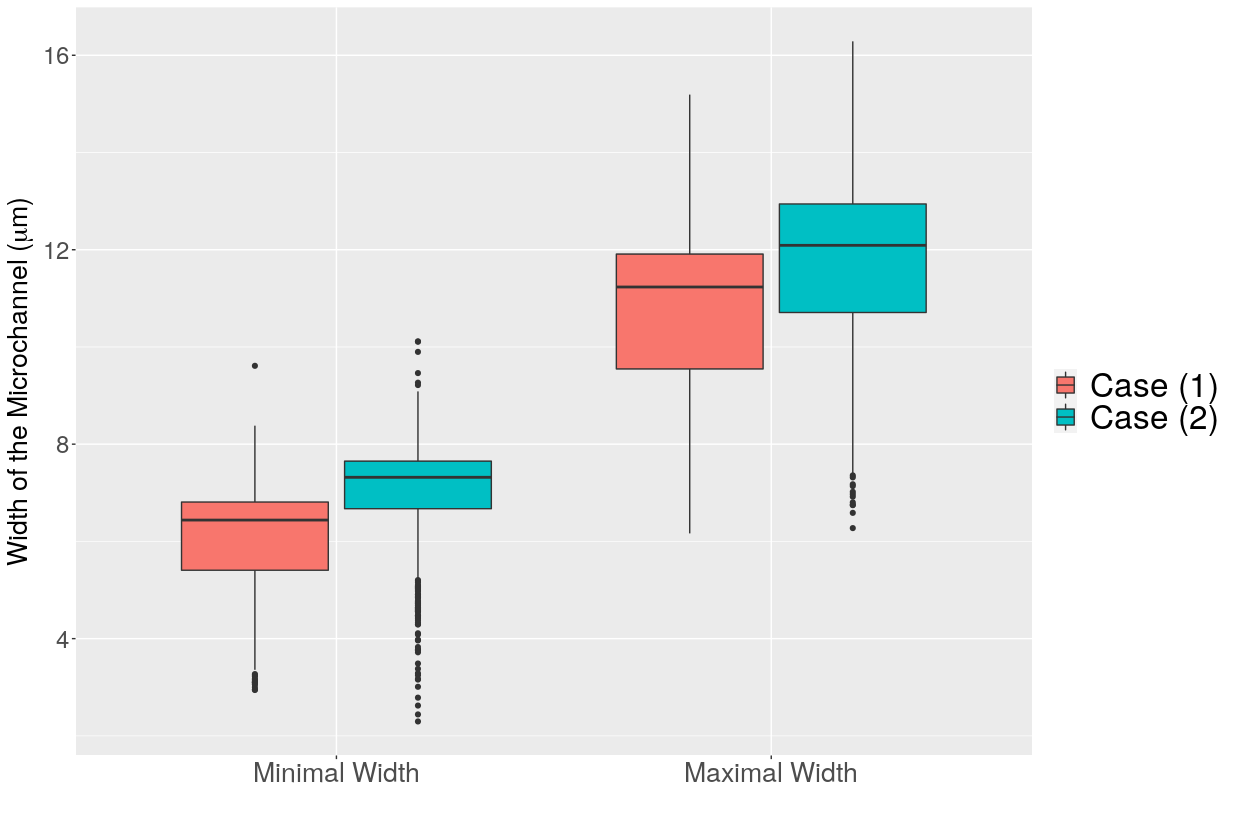}
	\label{fig_Rplot_channel_width_case12}}
	\caption{Using the same input values in the Monte Carlo simulations in both cases, the box-plot show cell speed and the channel width {\it after} the leader cell and the cell doublets exit the channel in both cases. The orange and green box represent Case (1) and Case (2), respectively. \subref{fig_Case12_Cell12_vel} The average speed of Cell 1 and Cell 2 in different cases. \subref{fig_Rplot_channel_width_case12} The maximal and minimal width {\it after} the leader cell (Case (1)) and the cell doublet (Case (2)) exit the channel completely.}
	\label{Fig_Case12}
\end{figure}

In Figure \ref{Fig_Case12}\subref{fig_Case12_Cell12_vel}, we show the box-plot of the average speeds of the two cells in both cases. It is not straightforward to draw any conclusion for Cell 1, while there is a dominant difference in Cell 2 for different cases, that is, the average velocity of Cell 2 in Case (2) is less than the average speed of Cell 2 in Case (1). We conduct the Wilcoxon's test between the paired samples and the results are shown in Table \ref{Tbl_Wilcox_Cell12_Case12_vel}. As we have expected and similar to the conclusion in Section \ref{Sec_pre_results}, Cell 1 transmigrates through the flexible channel significantly faster in Case (2) than in Case (1) (with p-value $1.52\times 10^{-17}$) due to the cellular forces exerted by the follower cell in Case (2). As a result, Cell 2 cannot pass through Cell 1 in Case (2) even though Cell 2 is migrating faster thanks to the expanded channel (otherwise, Cell 2 will no collide with Cell 1 when there is distance between them initially), and Cell 1 also exerts the repelling force on Cell 2, which also slows down the transmigration of Cell 2 significantly ($\text{p-value} = 4.81\times10^{-154}$). In summary, under the setting of collective migration, the leader cell also benefits from the follower cell such that the leader cell transmigrates faster and easier through the narrow channel. Hence, collective cell migration accelerates cell migration and may be responsible for increased metastatic rate.

Figure \ref{Fig_Case12}\subref{fig_Rplot_channel_width_case12} shows the box-plot of the microchannel width after the leader cell (Cell 1) exits the microchannel completely in Case (1) and the cell doublet exits the microchannel completely in Case (2). As each dataset is paired, we conduct the Wilcoxon's test in Table \ref{Tbl_Wilcox_Channel_Width_Case12} and we conclude that both the minimal and maximal width in Case (2) are significantly wider than in Case (1) when all the input parameter values are the same.

\section{Conclusions and Discussions}\label{Sec_conclusions}
\noindent
In this manuscript, we attempt to answer the following research questions: \begin{enumerate}
	\item Can a follower cell benefit in transmigration speed through a narrow flexible from a leading cell that may have widened the channel?
	\item Can cell doublets (i.e. a pair of cells that are migrating close to each other) benefit in transmigration speed by the mechanical repelling forces that they exert on each other?
\end{enumerate}
We designed these questions as two categories of simulations, so-called Case (1) and Case (2), respectively. 

Case (1) answered the question that indeed, the follower cells move significantly faster than the leader cell. However, there exists an optimum regarding the confinement on the cell, such that the cell moves fastest through the channel or the pore. If the confinement is too small, there is barely any polarization of the cell, hence, the cell would prioritize maintaining the equilibrium shape instead of other cellular activities; on the other hand, if the confinement is too large, besides the friction that burdens the migration, most energy is spent on the deformation of the cell, then the cell cannot migrate optimally; furthermore, the cell is under a higher risk of death due to the large mechanical stress \citep{Gefen_2016}. This behavior is verified in the experiment conducted by \citet{Mak2013}, where they show that cells take much longer time to transmigrate through the constrictions and only part of the cell population survive. Another possible explanation is from the perspective of minimizing the energy configuration: to some extent, a smaller deformation costs less energy to deform and to recover to its equilibrium shape, hence, most energy can be spent in the migration; however, if there is hardly any confinement on the cell, the cell prefers to stay in its (nearly) equilibrium shape rather than consume any energy on any cellular activities such as migration or deformation. We remark that this energetic view is speculative since energy is not considered in the current formalism.

Case (2) is developed to mimic collective migration of cells in a simplified setting, where we simulate migration of cell doublets. As we assumed that once a cell is compressed due to encountering any obstacle, a cell exerts repelling forces on the obstacle to obtain more spaces. Therefore, in the simulations, we observed that the leader cell moves significantly faster than in Case (1), whereas the average velocity of the follower cell is much smaller. In fact, the follower cell is moving faster before it touches the leader cell, since the two cells are initially distanced at the entrance of the channel. However, the leader cell blocks the way of the follower cell, which results in a slower average velocity of the follower cell. The experimental results of \citet{Merkher2017-va},  show that invasiveness of cancer cells increases when in close proximity, which likely results from additive and synergistic contributions, see \citep{Tulchinsky2022-vf}. This is qualitatively reproduced by our model (see Figure \ref{Fig_Case12}\subref{fig_Rplot_channel_width_case12}). In summary, adjacent cells do benefit from each other in various aspects.  

There are many possibilities to extend our current model, in terms of parameters considered in the model as well as in mathematical analysis perspectives. The model can be extended, for instance, to include a limit on deformation as large deformations may lead to irreversible cell damage \citep{Gefen_2016} and death. Specifically large deformations may damage the cytoskeleton and the plasma membrane, whereas the springs that maintain the cell intact break. With a smaller deformation, cells may remain in a "softened" structure, which could result in the experimentally observed increased cell speed.  We would like to attempt to answer the question in \citet{Mak2013ASerial}, that after the cell penetrates through the first constriction, why the following-up constrictions cost less time for the cell when invading serial constrictions. On the other hand, instead of splitting the cell membrane into finite line segments, to keep the smoothness of the cell boundary, we can convert it into a free boundary problem that incorporates a PDE for the cell mechanics, which ensures no self-intersecting of the cell membrane, and probably increases the accuracy of the modelling framework for the fluid-structure interaction between cell and fluid. To further consider limits in deformation, the mechanics of the stiff nucleus can be implemented in the model, as it plays an important role in the invasiveness \citep{Tulchinsky2022-vf} and the velocity of cells \citep{Krause2019-dq}. In parallel, the material properties of the channel walls, such as their stiffness, are likely to affect cell migration and can be considered. In addition, to evaluate combined effects of biomechanical and biochemical cell-matrix interactions, biochemical degradation of the cell environment, e.g. by the secretion of matrix metalloproteinase (MMP), can be added.

In summary, the current manuscript demonstrates the bi-directional interactions of cells with their environment and how that is affected by cell proximity. We have extended our earlier cell invasion model to describe transmigration of cells through narrow, flexible channels, such as those existing in various tissues. Furthermore,  we have described the bi-directional impact of cells interactions with adjacent neighbors and with their direct environment through the forces that are exerted by the cells. The cell environment may be a pore (through the cross-linked fibrous structure in the extracellular matrix) or any other narrow channel (a small blood vessel). In order to transmigrate, cancer cells deform and exert forces to widen the channel, and two closely adjacent cells are able to interact to accelerate this process. The current model reveals mechanisms underlying the observations of increased invasiveness and the cancer cell metastatic rate upon collective cell migration, when the deformation of the immediate environment is considered.

\section*{Conflict of Interest}
\noindent
The authors declare that they have no conflict of interest.

\section*{Acknowledgments}    
\noindent
The work was partially supported by the Israeli Ministry of Science and Technology (MOST) Medical Devices Program (Grant no. 3-17427 awarded to Prof. Daphne Weihs) and by the Gerald O. Mann and the Frank and Dolores Corbett Charitable Foundations.

\bibliographystyle{abbrvnat}
\bibliography{ProjectReport}

\begin{thebibliography}{57}
\providecommand{\natexlab}[1]{#1}
\providecommand{\url}[1]{\texttt{#1}}
\expandafter\ifx\csname urlstyle\endcsname\relax
  \providecommand{\doi}[1]{doi: #1}\else
  \providecommand{\doi}{doi: \begingroup \urlstyle{rm}\Url}\fi

\bibitem[Ananthakrishnan and Ehrlicher(2007)]{Ananthakrishnan2007}
R.~Ananthakrishnan and A.~Ehrlicher.
\newblock The forces behind cell movement.
\newblock \emph{International Journal of Biological Sciences}, pages 303--317,
  2007.
\newblock \doi{10.7150/ijbs.3.303}.
\newblock URL \url{https://doi.org/10.7150/ijbs.3.303}.

\bibitem[Angelini et~al.(2012)Angelini, Dunn, Urue{\~{n}}a, Dickrell, Burris,
  and Sawyer]{Angelini2012}
T.~E. Angelini, A.~C. Dunn, J.~M. Urue{\~{n}}a, D.~J. Dickrell, D.~L. Burris,
  and W.~G. Sawyer.
\newblock Cell friction.
\newblock \emph{Faraday Discussions}, 156:\penalty0 31, 2012.
\newblock \doi{10.1039/c2fd00130f}.
\newblock URL \url{https://doi.org/10.1039/c2fd00130f}.

\bibitem[Ben~Amar et~al.(2015)Ben~Amar, Wu, Trejo, and Atlan]{ben2015morpho}
M.~Ben~Amar, M.~Wu, M.~Trejo, and M.~Atlan.
\newblock Morpho-elasticity of inflammatory fibrosis: the case of capsular
  contracture.
\newblock \emph{Journal of the Royal Society Interface}, 12\penalty0
  (111):\penalty0 20150343, 2015.

\bibitem[Bershadsky and Kozlov(2011)]{Bershadsky2011}
A.~D. Bershadsky and M.~M. Kozlov.
\newblock Crawling cell locomotion revisited.
\newblock \emph{Proceedings of the National Academy of Sciences}, 108\penalty0
  (51):\penalty0 20275--20276, Dec. 2011.
\newblock \doi{10.1073/pnas.1116814108}.
\newblock URL \url{https://doi.org/10.1073/pnas.1116814108}.

\bibitem[Chen et~al.(2017)Chen, Weihs, and Vermolen]{chen2017model}
J.~Chen, D.~Weihs, and F.~J. Vermolen.
\newblock A model for cell migration in non-isotropic fibrin networks with an
  application to pancreatic tumor islets.
\newblock \emph{Biomechanics and Modeling in Mechanobiology}, pages 1--20,
  2017.

\bibitem[Chen et~al.(2018)Chen, Weihs, Dijk, and Vermolen]{Chen2018}
J.~Chen, D.~Weihs, M.~V. Dijk, and F.~J. Vermolen.
\newblock A phenomenological model for cell and nucleus deformation during
  cancer metastasis.
\newblock \emph{Biomechanics and Modeling in Mechanobiology}, 17\penalty0
  (5):\penalty0 1429--1450, May 2018.
\newblock \doi{10.1007/s10237-018-1036-5}.
\newblock URL \url{https://doi.org/10.1007/s10237-018-1036-5}.

\bibitem[Cross et~al.(2007)Cross, Jin, Rao, and Gimzewski]{Cross2007}
S.~E. Cross, Y.-S. Jin, J.~Rao, and J.~K. Gimzewski.
\newblock Nanomechanical analysis of cells from cancer patients.
\newblock \emph{Nature Nanotechnology}, 2\penalty0 (12):\penalty0 780--783,
  Dec. 2007.
\newblock \doi{10.1038/nnano.2007.388}.
\newblock URL \url{https://doi.org/10.1038/nnano.2007.388}.

\bibitem[Cusseddu et~al.(2019)Cusseddu, Edelstein-Keshet, Mackenzie, Portet,
  and Madzvamuse]{Cusseddu2019}
D.~Cusseddu, L.~Edelstein-Keshet, J.~Mackenzie, S.~Portet, and A.~Madzvamuse.
\newblock A coupled bulk-surface model for cell polarisation.
\newblock \emph{Journal of Theoretical Biology}, 481:\penalty0 119--135, Nov.
  2019.
\newblock \doi{10.1016/j.jtbi.2018.09.008}.
\newblock URL \url{https://doi.org/10.1016/j.jtbi.2018.09.008}.

\bibitem[Devreotes and Zigmond(1988)]{Devreotes1988}
P.~N. Devreotes and S.~H. Zigmond.
\newblock Chemotaxis in eukaryotic cells: A focus on leukocytes and
  dictyostelium.
\newblock \emph{Annual Review of Cell Biology}, 4\penalty0 (1):\penalty0
  649--686, Nov. 1988.
\newblock \doi{10.1146/annurev.cb.04.110188.003245}.
\newblock URL \url{https://doi.org/10.1146/annurev.cb.04.110188.003245}.

\bibitem[Ebata et~al.(2018)Ebata, Yamamoto, Tsuji, Sasaki, Moriyama, Kuboki,
  and Kidoaki]{Ebata2018}
H.~Ebata, A.~Yamamoto, Y.~Tsuji, S.~Sasaki, K.~Moriyama, T.~Kuboki, and
  S.~Kidoaki.
\newblock Persistent random deformation model of cells crawling on a gel
  surface.
\newblock \emph{Scientific Reports}, 8\penalty0 (1), Mar. 2018.
\newblock \doi{10.1038/s41598-018-23540-x}.
\newblock URL \url{https://doi.org/10.1038/s41598-018-23540-x}.

\bibitem[Gal and Weihs(2012)]{Gal2012}
N.~Gal and D.~Weihs.
\newblock Intracellular mechanics and activity of breast cancer cells correlate
  with metastatic potential.
\newblock \emph{Cell Biochemistry and Biophysics}, 63\penalty0 (3):\penalty0
  199--209, May 2012.
\newblock \doi{10.1007/s12013-012-9356-z}.
\newblock URL \url{https://doi.org/10.1007/s12013-012-9356-z}.

\bibitem[Gefen and Weihs(2016)]{Gefen_2016}
A.~Gefen and D.~Weihs.
\newblock Cytoskeleton and plasma-membrane damage resulting from exposure to
  sustained deformations: A review of the mechanobiology of chronic wounds.
\newblock \emph{Medical Engineering {\&} Physics}, 38\penalty0 (9):\penalty0
  828--833, sep 2016.
\newblock \doi{10.1016/j.medengphy.2016.05.014}.
\newblock URL \url{https://doi.org/10.1016%2Fj.medengphy.2016.05.014}.

\bibitem[Goriely and Moulton(2011)]{goriely2011morphoelasticity}
A.~Goriely and D.~Moulton.
\newblock Morphoelasticity: a theory of elastic growth.
\newblock \emph{New Trends in the Physics and Mechanics of Biological Systems:
  Lecture Notes of the Les Houches Summer School: Volume 92, July 2009},
  92:\penalty0 153, 2011.

\bibitem[Guck et~al.(2005)Guck, Schinkinger, Lincoln, Wottawah, Ebert, Romeyke,
  Lenz, Erickson, Ananthakrishnan, Mitchell, K\"{a}s, Ulvick, and
  Bilby]{Guck2005}
J.~Guck, S.~Schinkinger, B.~Lincoln, F.~Wottawah, S.~Ebert, M.~Romeyke,
  D.~Lenz, H.~M. Erickson, R.~Ananthakrishnan, D.~Mitchell, J.~K\"{a}s,
  S.~Ulvick, and C.~Bilby.
\newblock Optical deformability as an inherent cell marker for testing
  malignant transformation and metastatic competence.
\newblock \emph{Biophysical Journal}, 88\penalty0 (5):\penalty0 3689--3698, May
  2005.
\newblock \doi{10.1529/biophysj.104.045476}.
\newblock URL \url{https://doi.org/10.1529/biophysj.104.045476}.

\bibitem[Haupt and Minc(2018)]{Haupt2018}
A.~Haupt and N.~Minc.
\newblock How cells sense their own shape {\textendash} mechanisms to probe
  cell geometry and their implications in cellular organization and function.
\newblock \emph{Journal of Cell Science}, 131\penalty0 (6):\penalty0 jcs214015,
  Mar. 2018.
\newblock \doi{10.1242/jcs.214015}.
\newblock URL \url{https://doi.org/10.1242/jcs.214015}.

\bibitem[Heuz{\'{e}} et~al.(2011)Heuz{\'{e}}, Collin, Terriac,
  Lennon-Dum{\'{e}}nil, and Piel]{Heuz__2011}
M.~L. Heuz{\'{e}}, O.~Collin, E.~Terriac, A.-M. Lennon-Dum{\'{e}}nil, and
  M.~Piel.
\newblock Cell migration in confinement: A micro-channel-based assay.
\newblock In \emph{Methods in Molecular Biology}, pages 415--434. Humana Press,
  2011.
\newblock \doi{10.1007/978-1-61779-207-6_28}.
\newblock URL \url{https://doi.org/10.1007%2F978-1-61779-207-6_28}.

\bibitem[Ji et~al.(2008)Ji, Lim, and Danuser]{Ji2008}
L.~Ji, J.~Lim, and G.~Danuser.
\newblock Fluctuations of intracellular forces during cell protrusion.
\newblock \emph{Nature Cell Biology}, 10\penalty0 (12):\penalty0 1393--1400,
  Nov. 2008.
\newblock \doi{10.1038/ncb1797}.
\newblock URL \url{https://doi.org/10.1038/ncb1797}.

\bibitem[Keren et~al.(2008)Keren, Pincus, Allen, Barnhart, Marriott, Mogilner,
  and Theriot]{Keren2008}
K.~Keren, Z.~Pincus, G.~M. Allen, E.~L. Barnhart, G.~Marriott, A.~Mogilner, and
  J.~A. Theriot.
\newblock Mechanism of shape determination in motile cells.
\newblock \emph{Nature}, 453\penalty0 (7194):\penalty0 475--480, May 2008.
\newblock \doi{10.1038/nature06952}.
\newblock URL \url{https://doi.org/10.1038/nature06952}.

\bibitem[Koppenol(2017)]{koppenol2017biomedical}
D.~Koppenol.
\newblock Biomedical implications from mathematical models for the simulation
  of dermal wound healing.
\newblock \emph{PhD-thesis at the Delft University of Technology, the
  Netherlands}, 2017.

\bibitem[Kortam et~al.(2021)Kortam, Merkher, Kramer, Metanes, Ad-El, Krausz,
  Har-Shai, and Weihs]{Kortam_2021}
S.~Kortam, Y.~Merkher, A.~Kramer, I.~Metanes, D.~Ad-El, J.~Krausz, Y.~Har-Shai,
  and D.~Weihs.
\newblock Rapid, quantitative prediction of tumor invasiveness in non-melanoma
  skin cancers using mechanobiology-based assay.
\newblock \emph{Biomechanics and Modeling in Mechanobiology}, 20\penalty0
  (5):\penalty0 1767--1774, jun 2021.
\newblock \doi{10.1007/s10237-021-01475-z}.
\newblock URL \url{https://doi.org/10.1007%2Fs10237-021-01475-z}.

\bibitem[Koumoutsakos et~al.(2013)Koumoutsakos, Pivkin, and
  Milde]{Koumoutsakos2013}
P.~Koumoutsakos, I.~Pivkin, and F.~Milde.
\newblock The fluid mechanics of cancer and its therapy.
\newblock \emph{Annual Review of Fluid Mechanics}, 45\penalty0 (1):\penalty0
  325--355, Jan. 2013.
\newblock \doi{10.1146/annurev-fluid-120710-101102}.
\newblock URL \url{https://doi.org/10.1146/annurev-fluid-120710-101102}.

\bibitem[Krause et~al.(2019)Krause, Yang, Te~Lindert, Isermann, Schepens, Maas,
  Venkataraman, Lammerding, Madzvamuse, Hendriks, Te~Riet, and
  Wolf]{Krause2019-dq}
M.~Krause, F.~W. Yang, M.~Te~Lindert, P.~Isermann, J.~Schepens, R.~J.~A. Maas,
  C.~Venkataraman, J.~Lammerding, A.~Madzvamuse, W.~Hendriks, J.~Te~Riet, and
  K.~Wolf.
\newblock Cell migration through three-dimensional confining pores: speed
  accelerations by deformation and recoil of the nucleus.
\newblock \emph{Philos. Trans. R. Soc. Lond. B Biol. Sci.}, 374\penalty0
  (1779):\penalty0 20180225, Aug. 2019.

\bibitem[Köppl and Wohlmuth(2014)]{K_ppl_2014}
T.~Köppl and B.~Wohlmuth.
\newblock Optimal a priori error estimates for an elliptic problem with dirac
  right-hand side.
\newblock \emph{{SIAM} Journal on Numerical Analysis}, 52\penalty0
  (4):\penalty0 1753--1769, jan 2014.
\newblock \doi{10.1137/130927619}.
\newblock URL \url{https://doi.org/10.1137%2F130927619}.

\bibitem[Ladoux et~al.(2016)Ladoux, M{\`{e}}ge, and Trepat]{Ladoux2016}
B.~Ladoux, R.-M. M{\`{e}}ge, and X.~Trepat.
\newblock Front{\textendash}rear polarization by mechanical cues: From single
  cells to tissues.
\newblock \emph{Trends in Cell Biology}, 26\penalty0 (6):\penalty0 420--433,
  June 2016.
\newblock \doi{10.1016/j.tcb.2016.02.002}.
\newblock URL \url{https://doi.org/10.1016/j.tcb.2016.02.002}.

\bibitem[Llense and Etienne-Manneville(2015)]{Llense2015}
F.~Llense and S.~Etienne-Manneville.
\newblock Front-to-rear polarity in migrating cells.
\newblock In \emph{Cell Polarity 1}, pages 115--146. Springer International
  Publishing, 2015.
\newblock \doi{10.1007/978-3-319-14463-4_5}.
\newblock URL \url{https://doi.org/10.1007/978-3-319-14463-4_5}.

\bibitem[Mak and Erickson(2013)]{Mak2013ASerial}
M.~Mak and D.~Erickson.
\newblock A serial micropipette microfluidic device with applications to cancer
  cell repeated deformation studies.
\newblock \emph{Integrative Biology}, 5\penalty0 (11):\penalty0 1374--1384,
  Sept. 2013.
\newblock \doi{10.1039/c3ib40128f}.
\newblock URL \url{https://doi.org/10.1039/c3ib40128f}.

\bibitem[Mak et~al.(2013)Mak, Reinhart-King, and Erickson]{Mak2013}
M.~Mak, C.~A. Reinhart-King, and D.~Erickson.
\newblock Elucidating mechanical transition effects of invading cancer cells
  with a subnucleus-scaled microfluidic serial dimensional modulation device.
\newblock \emph{Lab Chip}, 13\penalty0 (3):\penalty0 340--348, 2013.
\newblock \doi{10.1039/c2lc41117b}.
\newblock URL \url{https://doi.org/10.1039/c2lc41117b}.

\bibitem[Massalha and Weihs(2016)]{Massalha2016}
S.~Massalha and D.~Weihs.
\newblock Metastatic breast cancer cells adhere strongly on varying stiffness
  substrates, initially without adjusting their morphology.
\newblock \emph{Biomechanics and Modeling in Mechanobiology}, 16\penalty0
  (3):\penalty0 961--970, Dec. 2016.
\newblock \doi{10.1007/s10237-016-0864-4}.
\newblock URL \url{https://doi.org/10.1007/s10237-016-0864-4}.

\bibitem[Merkher and Weihs(2017)]{Merkher2017-va}
Y.~Merkher and D.~Weihs.
\newblock Proximity of metastatic cells enhances their mechanobiological
  invasiveness.
\newblock \emph{Ann. Biomed. Eng.}, 45\penalty0 (6):\penalty0 1399--1406, June
  2017.

\bibitem[Merkher et~al.(2020)Merkher, Horesh, Abramov, Shleifer, Ben-Ishay,
  Kluger, and Weihs]{Merkher_2020}
Y.~Merkher, Y.~Horesh, Z.~Abramov, G.~Shleifer, O.~Ben-Ishay, Y.~Kluger, and
  D.~Weihs.
\newblock Rapid cancer diagnosis and early prognosis of metastatic risk based
  on mechanical invasiveness of sampled cells.
\newblock \emph{Annals of Biomedical Engineering}, 48\penalty0 (12):\penalty0
  2846--2858, jun 2020.
\newblock \doi{10.1007/s10439-020-02547-4}.
\newblock URL \url{https://doi.org/10.1007%2Fs10439-020-02547-4}.

\bibitem[Mitchison and Cramer(1996)]{Mitchison1996}
T.~Mitchison and L.~Cramer.
\newblock Actin-based cell motility and cell locomotion.
\newblock \emph{Cell}, 84\penalty0 (3):\penalty0 371--379, Feb. 1996.
\newblock \doi{10.1016/s0092-8674(00)81281-7}.
\newblock URL \url{https://doi.org/10.1016/s0092-8674(00)81281-7}.

\bibitem[Mogilner and Keren(2009)]{Mogilner2009}
A.~Mogilner and K.~Keren.
\newblock The shape of motile cells.
\newblock \emph{Current Biology}, 19\penalty0 (17):\penalty0 R762--R771, Sept.
  2009.
\newblock \doi{10.1016/j.cub.2009.06.053}.
\newblock URL \url{https://doi.org/10.1016/j.cub.2009.06.053}.

\bibitem[Paluch and Heisenberg(2009)]{Paluch2009}
E.~Paluch and C.-P. Heisenberg.
\newblock Biology and physics of cell shape changes in development.
\newblock \emph{Current Biology}, 19\penalty0 (17):\penalty0 R790--R799, Sept.
  2009.
\newblock \doi{10.1016/j.cub.2009.07.029}.
\newblock URL \url{https://doi.org/10.1016/j.cub.2009.07.029}.

\bibitem[Peng and Vermolen(2020)]{Peng2020}
Q.~Peng and F.~Vermolen.
\newblock Agent-based modelling and parameter sensitivity analysis with a
  finite-element method for skin contraction.
\newblock \emph{Biomechanics and Modeling in Mechanobiology}, 19\penalty0
  (6):\penalty0 2525--2551, July 2020.
\newblock \doi{10.1007/s10237-020-01354-z}.
\newblock URL \url{https://doi.org/10.1007/s10237-020-01354-z}.

\bibitem[Peng and Vermolen(2022)]{Peng_2022_JCAM}
Q.~Peng and F.~Vermolen.
\newblock Numerical methods to compute stresses and displacements from cellular
  forces: Application to the contraction of tissue.
\newblock \emph{Journal of Computational and Applied Mathematics},
  404:\penalty0 113892, apr 2022.
\newblock \doi{10.1016/j.cam.2021.113892}.
\newblock URL \url{https://doi.org/10.1016%2Fj.cam.2021.113892}.

\bibitem[Peng et~al.(2021)Peng, Vermolen, and Weihs]{Peng2021}
Q.~Peng, F.~J. Vermolen, and D.~Weihs.
\newblock A formalism for modelling traction forces and cell shape evolution
  during cell migration in various biomedical processes.
\newblock \emph{Biomechanics and Modeling in Mechanobiology}, Apr. 2021.
\newblock \doi{10.1007/s10237-021-01456-2}.
\newblock URL \url{https://doi.org/10.1007/s10237-021-01456-2}.

\bibitem[Peng et~al.(2022)Peng, Gorter, and Vermolen]{PengGorter2022}
Q.~Peng, W.~S. Gorter, and F.~J. Vermolen.
\newblock Comparison between a phenomenological approach and a morphoelasticity
  approach regarding the displacement of extracellular matrix.
\newblock \emph{Biomechanics and Modeling in Mechanobiology}, 21\penalty0
  (3):\penalty0 919--935, Apr. 2022.
\newblock \doi{10.1007/s10237-022-01568-3}.
\newblock URL \url{https://doi.org/10.1007/s10237-022-01568-3}.

\bibitem[Popov et~al.(2019)Popov, He{\ss}, and Willert]{popov2019handbook}
V.~L. Popov, M.~He{\ss}, and E.~Willert.
\newblock \emph{Handbook of contact mechanics: exact solutions of axisymmetric
  contact problems}.
\newblock Springer Nature, 2019.

\bibitem[Rappel and Edelstein-Keshet(2017)]{Rappel2017}
W.-J. Rappel and L.~Edelstein-Keshet.
\newblock Mechanisms of cell polarization.
\newblock \emph{Current Opinion in Systems Biology}, 3:\penalty0 43--53, June
  2017.
\newblock \doi{10.1016/j.coisb.2017.03.005}.
\newblock URL \url{https://doi.org/10.1016/j.coisb.2017.03.005}.

\bibitem[Rens and Edelstein-Keshet(2019)]{Rens2019}
E.~G. Rens and L.~Edelstein-Keshet.
\newblock From energy to cellular forces in the cellular potts model: An
  algorithmic approach.
\newblock \emph{{PLOS} Computational Biology}, 15\penalty0 (12):\penalty0
  e1007459, Dec. 2019.
\newblock \doi{10.1371/journal.pcbi.1007459}.
\newblock URL \url{https://doi.org/10.1371/journal.pcbi.1007459}.

\bibitem[Roussos et~al.(2011)Roussos, Condeelis, and Patsialou]{Roussos2011}
E.~T. Roussos, J.~S. Condeelis, and A.~Patsialou.
\newblock Chemotaxis in cancer.
\newblock \emph{Nature Reviews Cancer}, 11\penalty0 (8):\penalty0 573--587,
  July 2011.
\newblock \doi{10.1038/nrc3078}.
\newblock URL \url{https://doi.org/10.1038/nrc3078}.

\bibitem[Rudraraju et~al.(2019)Rudraraju, Moulton, Chirat, Goriely, and
  Garikipati]{rudraraju2019computational}
S.~Rudraraju, D.~E. Moulton, R.~Chirat, A.~Goriely, and K.~Garikipati.
\newblock A computational framework for the morpho-elastic development of
  molluskan shells by surface and volume growth.
\newblock \emph{PLoS computational biology}, 15\penalty0 (7):\penalty0
  e1007213, 2019.

\bibitem[Selmeczi et~al.(2005)Selmeczi, Mosler, Hagedorn, Larsen, and
  Flyvbjerg]{Selmeczi2005}
D.~Selmeczi, S.~Mosler, P.~H. Hagedorn, N.~B. Larsen, and H.~Flyvbjerg.
\newblock Cell motility as persistent random motion: Theories from experiments.
\newblock \emph{Biophysical Journal}, 89\penalty0 (2):\penalty0 912--931, Aug.
  2005.
\newblock \doi{10.1529/biophysj.105.061150}.
\newblock URL \url{https://doi.org/10.1529/biophysj.105.061150}.

\bibitem[Shapiro(1990)]{shapiro1990test}
S.~S. Shapiro.
\newblock How to test normality and other distributional assumptions.
\newblock Technical report, 1990.

\bibitem[Swaminathan et~al.(2011)Swaminathan, Mythreye,
  O{\textquotesingle}Brien, Berchuck, Blobe, and Superfine]{Swaminathan2011}
V.~Swaminathan, K.~Mythreye, E.~T. O{\textquotesingle}Brien, A.~Berchuck, G.~C.
  Blobe, and R.~Superfine.
\newblock Mechanical stiffness grades metastatic potential in patient tumor
  cells and in cancer cell lines.
\newblock \emph{Cancer Research}, 71\penalty0 (15):\penalty0 5075--5080, June
  2011.
\newblock \doi{10.1158/0008-5472.can-11-0247}.
\newblock URL \url{https://doi.org/10.1158/0008-5472.can-11-0247}.

\bibitem[Takagi et~al.(2008)Takagi, Sato, Yanagida, and Ueda]{Takagi2008}
H.~Takagi, M.~J. Sato, T.~Yanagida, and M.~Ueda.
\newblock Functional analysis of spontaneous cell movement under different
  physiological conditions.
\newblock \emph{{PLoS} {ONE}}, 3\penalty0 (7):\penalty0 e2648, July 2008.
\newblock \doi{10.1371/journal.pone.0002648}.
\newblock URL \url{https://doi.org/10.1371/journal.pone.0002648}.

\bibitem[Trickey et~al.(2006)Trickey, Baaijens, Laursen, Alexopoulos, and
  Guilak]{Trickey2006}
W.~R. Trickey, F.~P. Baaijens, T.~A. Laursen, L.~G. Alexopoulos, and F.~Guilak.
\newblock Determination of the poisson{\textquotesingle}s ratio of the cell:
  recovery properties of chondrocytes after release from complete micropipette
  aspiration.
\newblock \emph{Journal of Biomechanics}, 39\penalty0 (1):\penalty0 78--87,
  Jan. 2006.
\newblock \doi{10.1016/j.jbiomech.2004.11.006}.
\newblock URL \url{https://doi.org/10.1016/j.jbiomech.2004.11.006}.

\bibitem[Tulchinsky and Weihs(2022)]{Tulchinsky2022-vf}
M.~Tulchinsky and D.~Weihs.
\newblock Computational modeling reveals a vital role for proximity-driven
  additive and synergistic cell-cell interactions in increasing cancer
  invasiveness.
\newblock \emph{Acta Biomater.}, Mar. 2022.

\bibitem[Verkhovsky et~al.(1999)Verkhovsky, Svitkina, and
  Borisy]{Verkhovsky1999}
A.~B. Verkhovsky, T.~M. Svitkina, and G.~G. Borisy.
\newblock Self-polarization and directional motility of cytoplasm.
\newblock \emph{Current Biology}, 9\penalty0 (1):\penalty0 11--S1, Jan. 1999.
\newblock \doi{10.1016/s0960-9822(99)80042-6}.
\newblock URL \url{https://doi.org/10.1016/s0960-9822(99)80042-6}.

\bibitem[Vermolen and Gefen(2012)]{Vermolen2012cellshape}
F.~J. Vermolen and A.~Gefen.
\newblock A phenomenological model for chemico-mechanically induced cell shape
  changes during migration and cell{\textendash}cell contacts.
\newblock \emph{Biomechanics and Modeling in Mechanobiology}, 12\penalty0
  (2):\penalty0 301--323, May 2012.
\newblock \doi{10.1007/s10237-012-0400-0}.
\newblock URL \url{https://doi.org/10.1007/s10237-012-0400-0}.

\bibitem[Wang(2009)]{Wang2009}
F.~Wang.
\newblock The signaling mechanisms underlying cell polarity and chemotaxis.
\newblock \emph{Cold Spring Harbor Perspectives in Biology}, 1\penalty0
  (4):\penalty0 a002980--a002980, Sept. 2009.
\newblock \doi{10.1101/cshperspect.a002980}.
\newblock URL \url{https://doi.org/10.1101/cshperspect.a002980}.

\bibitem[Weaver et~al.(2017)Weaver, Morales, Dunn, Godde, and
  Weaver]{Weaver2017}
K.~F. Weaver, V.~Morales, S.~L. Dunn, K.~Godde, and P.~F. Weaver.
\newblock \emph{An Introduction to Statistical Analysis in Research}.
\newblock John Wiley {\&} Sons, Inc., July 2017.
\newblock \doi{10.1002/9781119454205}.
\newblock URL \url{https://doi.org/10.1002/9781119454205}.

\bibitem[Yeung et~al.(2004)Yeung, Georges, Flanagan, Marg, Ortiz, Funaki,
  Zahir, Ming, Weaver, and Janmey]{Yeung_2004}
T.~Yeung, P.~C. Georges, L.~A. Flanagan, B.~Marg, M.~Ortiz, M.~Funaki,
  N.~Zahir, W.~Ming, V.~Weaver, and P.~A. Janmey.
\newblock Effects of substrate stiffness on cell morphology, cytoskeletal
  structure, and adhesion.
\newblock \emph{Cell Motility and the Cytoskeleton}, 60\penalty0 (1):\penalty0
  24--34, 2004.
\newblock \doi{10.1002/cm.20041}.
\newblock URL \url{https://doi.org/10.1002%2Fcm.20041}.

\bibitem[Zemel et~al.(2010)Zemel, Rehfeldt, Brown, Discher, and
  Safran]{Zemel2010}
A.~Zemel, F.~Rehfeldt, A.~E.~X. Brown, D.~E. Discher, and S.~A. Safran.
\newblock Optimal matrix rigidity for stress-fibre polarization in stem cells.
\newblock \emph{Nature Physics}, 6\penalty0 (6):\penalty0 468--473, Mar. 2010.
\newblock \doi{10.1038/nphys1613}.
\newblock URL \url{https://doi.org/10.1038/nphys1613}.

\bibitem[Zhang et~al.(2021)Zhang, Chan, and Mak]{Zhang2021-yu}
X.~Zhang, T.~Chan, and M.~Mak.
\newblock Morphodynamic signatures of {MDA-MB-231} single cells and cell
  doublets undergoing invasion in confined microenvironments.
\newblock \emph{Sci. Rep.}, 11\penalty0 (1):\penalty0 6529, Mar. 2021.

\bibitem[Zhao et~al.(2017)Zhao, Cao, DiPietro, and Liang]{Zhao2017}
J.~Zhao, Y.~Cao, L.~A. DiPietro, and J.~Liang.
\newblock Dynamic cellular finite-element method for modelling large-scale cell
  migration and proliferation under the control of mechanical and biochemical
  cues: a study of re-epithelialization.
\newblock \emph{Journal of The Royal Society Interface}, 14\penalty0
  (129):\penalty0 20160959, Apr. 2017.
\newblock \doi{10.1098/rsif.2016.0959}.
\newblock URL \url{https://doi.org/10.1098/rsif.2016.0959}.

\bibitem[Zhao et~al.(2020)Zhao, Manuchehrfar, and Liang]{Zhao2020}
J.~Zhao, F.~Manuchehrfar, and J.~Liang.
\newblock Cell{\textendash}substrate mechanics guide collective cell migration
  through intercellular adhesion: a dynamic finite element cellular model.
\newblock \emph{Biomechanics and Modeling in Mechanobiology}, 19\penalty0
  (5):\penalty0 1781--1796, Feb. 2020.
\newblock \doi{10.1007/s10237-020-01308-5}.
\newblock URL \url{https://doi.org/10.1007/s10237-020-01308-5}.

\end{thebibliography}



\section*{Supplementary material}

# Supplementary Material

Q. Peng, F.J. Vermolen, D. Weihs

# S1 Supplementary Video

- **Supplementary Video 1:** An animation of the simulation for Case (1), which is corresponding to Section 3.1: the next cell is only allowed to enter the microchannel once the previous cell has left the microchannel completely.

- **Supplementary Video 2:** An animation of the simulation for Case (2), which is corresponding to Section 3.2: two cells enter the microchannel together as a cell doublet.

# S2 Wilcoxon's Test for Monte Carlo Simulations

## S2.1 Case (1): Single Cell Migration

In Case (1), we are interested in whether the leader cell expanding the flexible channel is beneficial to the follower cells and whether there is an optimum of the width of the channel such that the cell moves fastest. This section shows the statistical results from the Wilcoxon's test corresponding to the Monte Carlo results in Section 4.1. There are 1019 samples. From Table S1, all the p-value of the hypothesis tests are small, that is, Cell 1 expanding the channel helps the transmigration speed of Cell 2 and Cell 3. However, Cell 2 is the fastest cell among all the three cells, which suggests an optimum of the width of the channel to maximize the cell speed. Further, we show the scatter plots between the cell and the minimal width of the channel *before* the cell enters the channel in Figure S1. When the width of the channel is below a threshold, a wider channel increases the speed of the cell (see Figure S1(a)); however, when the width of the channel is above the threshold, a wider channel reduces the compression on the cell, and subsequently, the cell moves more slowly (see Figure S1(c)).

*Table S1: Wilcoxon's test for the average speed of cells between Cell 1 and Cell 2 , Cell 2 and Cell 3, and Cell 1 and Cell 3, respectively, in Case (1)*

| **Wilcoxon's text between Cell 1 and Cell 2 regarding the average speed** | |
| --- | --- |
| **Null Hypothesis** ($H_0$: $\bar{v}^1 = \bar{v}^2$) | There is no difference in the average speed of Cell 1 and Cell 2. |
| **Alternative Hypothesis** ($H_1$: $\bar{v}^1 < \bar{v}^2$) | The average speed of Cell 2 is larger than Cell 1. |
| $p-value = \mathbb{P}(Data|H_0) = 2.244267 \times 10^{-68}$ | |
| **Wilcoxon's text between Cell 2 and Cell 3 regarding the average speed** | |
| **Null Hypothesis** ($H_0$: $\bar{v}^2 = \bar{v}^3$) | There is no difference in the average speed of Cell 2 and Cell 3. |
| **Alternative Hypothesis** ($H_1$: $\bar{v}^2 > \bar{v}^3$) | The average speed of Cell 3 is *less* than Cell 2. |
| $p-value = \mathbb{P}(Data|H_0) = 5.797204 \times 10^{-09}$ | |
| **Wilcoxon's text between Cell 1 and Cell 3 regarding the average speed** | |
| **Null Hypothesis** ($H_0$: $\bar{v}^1 = \bar{v}^3$) | There is no difference in the average speed of Cell 1 and Cell 3. |
| **Alternative Hypothesis** ($H_1$: $\bar{v}^1 < \bar{v}^3$) | The average speed of Cell 3 is larger than Cell 1. |
| $p-value = \mathbb{P}(Data|H_0) = 8.279773 \times 10^{-47}$ | |

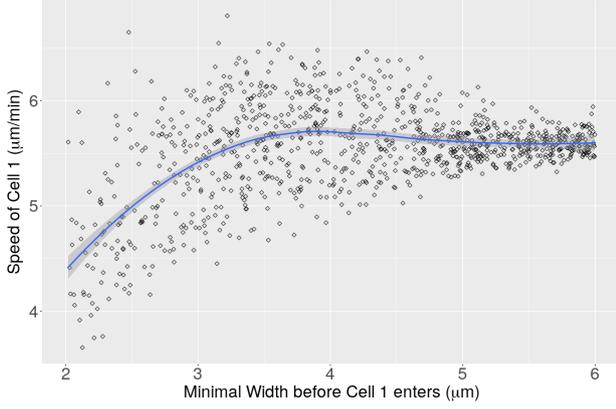

(a) Scatter plot between the average speed of Cell 1 and the minimal width of the channel before it enters

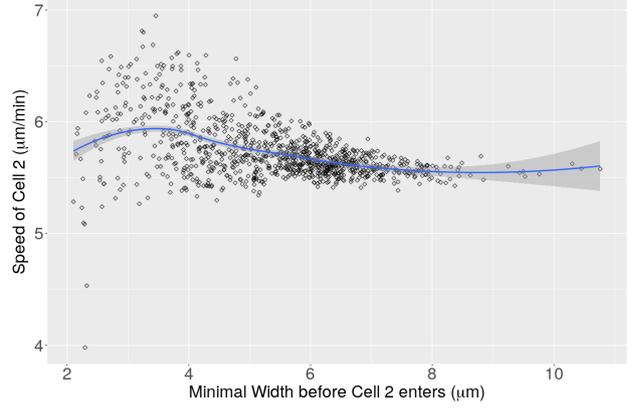

(b) Scatter plot between the average speed of Cell 2 and the minimal width of the channel before it enters

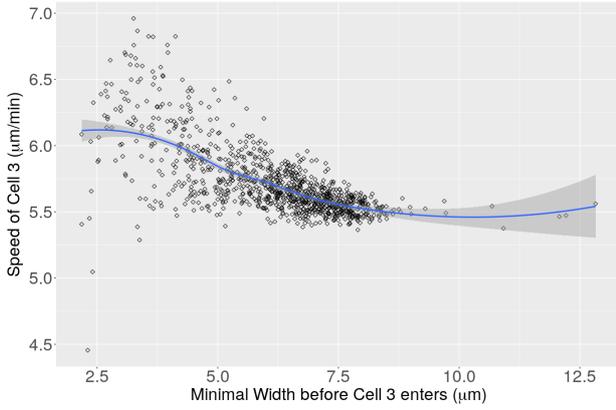

(c) Scatter plot between the average speed of Cell 3 and the minimal width of the channel before it enters

*Figure S1: (a) The scatter plot between the average speed of Cell 1 and the minimal width of the flexible channel before it enters the channel, where the blue line indicates the linear fitted value and the gray envelop represents the 95% confidence interval. (b) The scatter plot between the average speed of Cell 2 and the minimal width of the flexible channel before it enters the channel, where the blue line indicates the linear fitted value and the gray envelop represents the 95% confidence interval. (c) The scatter plot between the average speed of Cell 3 and the minimal width of the flexible channel before it enters the channel, where the blue line indicates the linear fitted value and the gray envelop represents the 95% confidence interval.*

## S2.2 Case (2): Cell Doublet Migration

There are 1044 samples collected. When cells move as a doublet, their speed is more or less the same, which is verified in Table S2. This subsection is corresponding to Section 4.2 in the main manuscript.

*Table S2: Wilcoxon's test for the average speed of cells between Cell 1 and Cell 2 in Case (2)*

| **Wilcoxon's text between Cell 1 and Cell 2 regarding the average speed** | |
| --- | --- |
| **Null Hypothesis** ($H_0$: $\bar{v}^1 = \bar{v}^2$) | There is no difference in the average speed of Cell 1 and Cell 2. |
| **Alternative Hypothesis** ($H_1$: $\bar{v}^1 < \bar{v}^2$) | The average speed of Cell 2 is larger than Cell 1. |
| $p-value = \mathbb{P}(Data|H_0) = 0.9999255$ | |

## S2.3 Comparison between Case (1) and Case (2)

In the main manuscript, we have shown that the cell doublets accelerate the transmigration of the leader cell (see Section 3.3 and 4.3). On the other hand, the cell doublets (Case (2)) also expand the microchannel more compared with single cell (Case (1)). We use the same data as Section 4.3 in the main manuscript, that is, there are 1000 paired samples.

Table S3 shows that the cell doublets transmigrates significantly faster than the single cell migration, since the p-value is very small. In Table S4, we show that there is a significant difference between the minimal width of the channel *after* the single cell (Case (1)) and the cell doublet (Case (2)) have left the channel completely, which implies that cell doublets are more invasive comparing with the single cell, since they both exert forces on the channel under the compression within a relative small distance.

*Table S3: Wilcoxon's test for the average speed of both cells between different cases*

| **Wilcoxon's test for the average speed of Cell 1 in Case (1) and Case (2)** | |
|---|---|
| **Null Hypothesis $H_0$** | There is no difference in the average speed of Cell 1 in Case (1) and Case (2). |
| **Alternative Hypothesis $H_1$** | The average speed of Cell 1 in Case (2) is larger than in Case (1). |
| $p - value = \mathbb{P}(Data\|H_0) = 1.52139085783176 \times 10^{-17}$ | |
| **Wilcoxon's test for the average speed of Cell 2 in Case (1) and Case (2)** | |
| **Null Hypothesis $H_0$** | There is no difference in the average speed of Cell 2 in Case (1) and Case (2). |
| **Alternative Hypothesis $H_1$** | The average speed of Cell 2 in Case (1) is larger than in Case (2). |
| $p - value = \mathbb{P}(Data\|H_0) = 4.80941709268107 \times 10^{-154}$ | |

*Table S4: Wilcoxon's test for the width of deformed microchannel between different cases*

| **Wilcoxon's test for the minimal width of the deformed microchannel in both cases** | |
|---|---|
| **Null Hypothesis $H_0$** | There is no difference in the minimal width of deformed microchannel 1 in Case (1) and Case (2). |
| **Alternative Hypothesis $H_1$** | The minimal width of deformed microchannel in Case (2) is larger than in Case (1). |
| $p - value = \mathbb{P}(Data\|H_0) = 2.63633038451654 \times 10^{-80}$ | |
| **Wilcoxon's test for the maximal width of the deformed microchannel in both cases** | |
| **Null Hypothesis $H_0$** | There is no difference in the maximal width of deformed microchannel 1 in Case (1) and Case (2). |
| **Alternative Hypothesis $H_1$** | The maximal width of deformed microchannel in Case (2) is larger than in Case (1). |
| $p - value = \mathbb{P}(Data\|H_0) = 6.65532947544036 \times 10^{-46}$ | |



\end{document}